\documentclass{jpp}

\usepackage{amsmath}
\usepackage{amsbsy}
\usepackage{amssymb}
\usepackage{graphicx}
\usepackage{bm,ulem,color}
\usepackage{natbib}
\providecommand{\Or}[1]{\ensuremath{\mbox{\textit{O}}#1}}
\providecommand{\ianabel}{I.\ns G.\ns A\ls B\ls E\ls L}

\providecommand{\email}{\thanks{Email address for correspondence: iga@physics.org}}

\providecommand{\axel}{A.\ns H\ls A\ls L\ls L\ls E\ls N\ls B\ls E\ls R\ls T}

\providecommand{\dv}{\bm{\nabla}\cdot}
\providecommand{\curl}{\bm{\nabla}\times}

\providecommand{\tensor}[1]{ \underline{\bm{\mathsf{#1}}}}

\newcommand{\collop}[1][f_s]{\ensuremath{{C}\hspace{-0.5mm}\left[#1\right]}}
\newcommand{\vth}[1][s]{\ensuremath{v_{\mathrm{th}_#1}}}

\newcommand{\pd}[2]{\ensuremath{ \frac{\partial #1} {\partial #2} } }
\newcommand{\inpd}[2]{\ensuremath{ \infrac{\partial #1} {\partial #2} } }
\newcommand{\gyror}[1]{\ensuremath{ {\left< #1 \right>}_{\bm{r}}}}
\newcommand{\ensav}[1]{\ensuremath{ {\left< #1 \right>}_{\mathrm{turb}}}}  
\newcommand{\fav}[1]{\ensuremath{\left< #1 \right>_{\psi}}} 
\newcommand{\gyroR}[1]{\ensuremath{{\left< #1 \right>}_{\bm{R}}}}

\providecommand{\eqref}[1]{Eq.\ (\ref{#1})}

\newcommand{\Secref}[1]{Section\ \ref{#1}}

\providecommand{\Apref}[1]{Appendix\ \ref{#1}}

\providecommand{\Or}[1]{\mathcal{O}#1}

\providecommand{\eref}[1]{(\ref{#1})}

\newcommand{\dg}{\cdot\nabla}      
\providecommand{\dv}{\nabla\cdot}
\providecommand{\curl}{\nabla\times}
\newcommand{\pb}[2]{\left\{ {#1} \middle., {#2} \right\}} 

\newcommand{\gyr}{\vartheta} 
\newcommand{\pot}{\varphi} 
\newcommand{\pol}{\theta} 
\newcommand{\gkeps}{\epsilon} 
\newcommand{\energy}[1][s]{{{\varepsilon}_#1}} 
\newcommand{\denergy}[1][s]{{\dot{\varepsilon}}_{#1}}
\newcommand{\source}[1][s]{\ensuremath{{{{S}}_{#1}}}} 

\newcommand{\gkpot}{\chi}

\newcommand{\magmom}[1][s]{\mu_#1}

\newcommand{\ddR}[2][s]{\pd{#2}{\bm{R}_{#1}}}
\newcommand{\dgR}[2][s]{\cdot\ddR[#1]{#2}}

\providecommand{\tensor}[1]{ {\bm{\mathsf{#1}}}}

\newcommand{\idmat}{\tensor{I}}
\newcommand{\ParticleFlux}[1][s]{\Gamma_{#1}}
\newcommand{\HeatFlux}[1][s]{q_{#1}}

\newcommand{\daccel}[1][s]{\delta \bm{a}_{#1}}
\newcommand{\ddtpsi}{\left.\pd{}{t}\right|_{\psi}}

\newcommand{\pcts}{
Princeton Center for Theoretical Science, Princeton University, Princeton, NJ 08544, USA
}

\newcommand{\chalmers}{
Department of Physics, Chalmers University of Technology, G\"oteborg, SE-41296, Sweden
}

\newcommand{\wint}{\int\hspace{-1.25mm} d^3 \bm{w}}

\newcommand{\Esource}[1][s]{S^{({E})}_{#1}}
\newcommand{\Psource}[1][s]{S^{({n})}_{#1}}

\newcommand{\infrac}[2]{ \left.{#1}\middle/{#2}\right. }

\newcommand{\CompHeat}[1][s]{P^{\mathrm{comp}}_#1}
\newcommand{\TurbInj}[1][s]{P^{\mathrm{drive}}_#1}
\newcommand{\TurbColl}[1][s]{P^{\mathrm{diss}}_#1}
\newcommand{\TurbPow}[1][s]{P^{\mathrm{turb}}_#1}

\newcommand{\vchi}{\bm{V}_\chi}
\newcommand{\vchiR}{\gyroR{\vchi}}
\newcommand{\vdrift}[1][s]{\bm{V}_{\mathrm{D}#1}}

\newcommand{\delAp}{{\ensuremath{\delta A_\parallel}}}

\newcommand{\delpot}{{\delta\pot}}
 \newcommand{\bhat}{{\widetilde{\bm{b}}}}

\newcommand{\MeanMagB}{{B}}
\newcommand{\Meanb}{{\bm{b}}}

\newcommand{\Bfield}{{\widetilde{\bm{B}}}}

\newcommand{\current}{{\widetilde{\bm{j}}}}

\newcommand{\upar}[1][e]{\delta u_{\parallel #1}}

\newcommand{\qpar}{\delta q_{\parallel e}}

\newcommand{\cycfreq}[1][s]{\Omega_{#1}}

\newcommand{\vA}{v_{\mathrm{A}}} 

\providecommand{\omstar}[1][]{\ensuremath{\omega_{*#1}}}

\newcommand{\overbar}[1]{\mkern 1.5mu\overline{\mkern-1.5mu#1\mkern-1.5mu}\mkern 1.5mu}
\newcommand{\nustar}[1][e]{\nu^{*}_{#1}}

\newcommand{\lincol}[1][\cdot]{\ensuremath{{C}_{L}\hspace{-0.5mm}\left[#1\right]}}

\RequirePackage{amsmath}
\makeatletter
\newcommand*\rel@kern[1]{\kern#1\dimexpr\macc@kerna}
\newcommand*\widebar[1]{%
  \begingroup
  \def\mathaccent##1##2{%
    \rel@kern{0.8}%
    \overline{\rel@kern{-0.8}\macc@nucleus\rel@kern{0.2}}%
    \rel@kern{-0.2}%
  }%
  \macc@depth\@ne
  \let\math@bgroup\@empty \let\math@egroup\macc@set@skewchar
  \mathsurround\z@ \frozen@everymath{\mathgroup\macc@group\relax}%
  \macc@set@skewchar\relax
  \let\mathaccentV\macc@nested@a
  \macc@nested@a\relax111{#1}%
  \endgroup
}
\makeatother

\usepackage[colorlinks=true]{hyperref}
\newcommand{\Ptens}[1][s]{\tensor{P}_{#1}}
\newcommand{\pitens}[1][s]{\tensor{\pi}_{#1}}
\renewcommand{\bhat}{\bm{b}}
\newcommand{\vint}{\int\mathrm{d}^3\bm{v}\,}

\newcommand{\uE}{\bm{u}_{\bm{E}}}
\renewcommand{\upar}[1][s]{u_{\parallel #1}}
\providecommand{\turbf}[1][s]{\delta {f}_#1}
\providecommand{\turbPot}{\delta {\pot}}
\providecommand{\turbB}{\delta {\bm{B}}}
\providecommand{\turbE}{\delta {\bm{E}}}
\providecommand{\turbChi}{\delta\chi}

\providecommand{\turbApar}{\delta A_\parallel}
\providecommand{\turbBpar}{\delta B_\parallel}
\providecommand{\exactf}[1][s]{\widetilde{f}_{#1}}
\providecommand{\exactE}{\widetilde{\bm{E}}}
\providecommand{\exactB}{\widetilde{\bm{B}}}

\providecommand{\elmB}{\bm{B}_1}

\newcommand{\feN}[1]{{\ensuremath{f_e^{(#1)}}}}
\providecommand{\grad}{\bm{\nabla}}
\renewcommand{\dg}{\cdot\grad}
\newcommand{\notnow}[1]{ }
\providecommand{\apSmall}{\xi}
\providecommand{\realGKeps}{\epsilon_{\mathrm{\scriptscriptstyle GK}}}
\newcommand{\uD}[1][s]{\bm{u}_{\mathrm{d} #1}}
\renewcommand{\qpar}{q_{\parallel s}}

\begin{document}

\date{\today}

\title{Multiscale Modelling for Tokamak Pedestals}
\author[Abel and Hallenbert]{\ianabel$^{1,2}$\email and \axel$^{1}$}
\affiliation{$^1$\chalmers\\$^2$\pcts}

\pubyear{2017}
\volume{653}
\pagerange{1--10}
\date{?; revised ?; accepted ?. - To be entered by editorial office}

\maketitle
\begin{abstract}
Pedestal modelling is crucial to predict the performance of future fusion devices. Current modelling efforts suffer either from a lack of 
kinetic physics, or an excess of computational complexity.  To ameliorate these problems, we take a first-principles multiscale approach to the pedestal.

We will present three separate sets of equations, covering the dynamics of Edge Localised Modes (ELMs), the inter-ELM pedestal, and pedestal turbulence, respectively. 
Precisely how these equations should be coupled to each other are covered in detail.  This framework is completely self-consistent; it is derived
from first principles by means of an asymptotic expansion of the fundamental Vlasov-Landau-Maxwell system in appropriate small parameters. The derivation exploits the 
narrowness of the pedestal region, the smallness of the thermal gyroradius, and the low plasma $\beta$ (the ratio of thermal to magnetic pressures) typical of current pedestal operation to achieve its simplifications.

The relationship between this framework and gyrokinetics is analysed, and possibilities to directly match our systems of equations onto multiscale gyrokinetics are explored. 
A detailed comparison between our model and other models in the literature is performed. 
Finally, the potential for matching this framework onto an open-field-line region is briefly discussed.
\end{abstract}
\section{Introduction}

The pedestal region of current fusion devices is both a boon and a serious obstacle to achieving sustained fusion power. Discovered by pioneering work on the ASDEX tokamak in Germany~\citep{wagner1982hmode,ASDEXHMode}, 
 the ``$H$-mode edge pedestal'' or ``Edge Transport Barrier'' is a narrow region at the edge of the tokamak where extremely steep gradients can be maintained (thus improving confinement) but 
 which intermittently relaxes through the release of large amounts of hot plasma, known as ``Edge Localised Modes''~\citep{zohm1996elms} (henceforth ELMs). The dichotomy between the confinement that can make a tokamak into a smaller and more efficient reactor and the ELMs that can severely damage that reactor has lead to a vast amount of study of pedestals since their discovery.

Enormous strides have been made, including understanding the fundamental instabilities that are responsible for ELMs~\citep{connor-edge-stability,snyder2002eped} and incorporating this knowledge into successful analysis of experiments~\citep{snyder2007qhmode,snyder2011eped}.
Experimentally, analyses have been made from data acquired on multiple machines~\citep{beurskens2011pedestals}, but the narrowness of the pedestal and the difficulty of making these measurements has hampered making predictions for future devices.
Indeed, recent results and analyses~\citep{mikek2017pedestalfail,hatch2017jetilwped} suggest that the scaling of the pedestal width with fundamental plasma parameters from current machines to ITER may be in doubt.
More detailed modelling is thus essential.

Edge modelling comprises both first principles analysis and a large and established effort in trying to understand existing devices through interpretative modelling.
Many difficulties arise in modelling this region of the plasma; any model must be able to handle large, rapidly-evolving perturbations associated with ELMs as well as small-scale fluctuations that may regulate the residual transport through the inter-ELM pedestal. In addition, the collisionality of the plasma may vary by orders of magnitude across the pedestal.
Taking data from a multi-machine comparison~\cite{militello2011multimachine}, the collisionality at the pedestal top for the Joint European Torus (JET) is $\nustar[e] \approx 4 \times 10^{-3}$ climbing to $\nustar[e] \approx 0.13$ at the separatrix. Using the predictions for ITER in the same paper, the collisionality at the pedestal top is now $\nustar[e] \approx 4 \times 10^{-4}$ and still $\nustar[e] \sim 0.2$ at the separatrix. This indicates a trend towards increasingly collisionless pedestals as machines get larger, driven in part by divertor heat load requirements preventing a high separatrix temperature.
Suitable modelling must therefore be able to transition from collisionless to collisional physics across the pedestal region. 

Several main branches of research into near-edge (i.e. regions at the edge of the plasma, but still inside the closed-field-line region) modelling currently exist. 
The most experimentally-successful approach has been that taken by the \verb#EPED# collaboration~\citep{snyder2011eped,snyder2007qhmode}. 
This consists of linear magnetohydrodynamic (MHD) modelling (with the \verb#ELITE# code) of the peeling-ballooning mode to determine the onset of an ELM, the use of the infinite-$n$ ballooning stability criterion as a proxy for 
the stability of the inter-ELM pedestal to microturbulence, and an assumption that the pedestal is pinned to this stability threshold between ELMs.
Despite the success of this model, the \textit{ad hoc} marginal stability criterion is required to constrain the width of pedestals, the linear stability calculation may neglect important kinetic effects, and the model has no capacity for handling time-dependent inter-ELM dynamics.
The dominant paradigm for first-principles edge modelling is that of ``drift-reduced'' or other anisotropic fluid equations~\citep{zeiler1997driftbraginskii}. These are the basis of many current simulation codes, such as \verb#GBS#~\citep{ricci2012gbs}, and also older 2D codes including \verb#SOLPS# and others~\citep{solps2006}, and have been used to study the pedestal directly~\citep{rasmussen2016lhtransition}. Gyrofluid equations, originally developed for the core~\citep{hammett1993developments} have also found much use in edge modelling~\citep{scott2000dalf,scott2007edge}. Gyrofluid models are derived by taking moments of gyrokinetic equations and then applying one of a variety of closure conditions.
These models have the advantage that a limited number of fluid moments is numerically tractable even in complex edge geometries. However, the strong requirements on collisionality for Braginskii equations to be applicable are often not satisfied in high-performance pedestals. Gyrofluid equations at low collisionality become ever-more plagued by questions of the validity of the chosen closure scheme, and lack a first-principles justification for the closures used. Thus, to handle a potentially trans-collisional pedestal region, we will take a kinetic approach. 

Recently, kinetic simulations of the plasma edge have been performed. The first category of these are the global gyrokinetic simulations~\citep{chang2017divertor,ku2006gkped,shi2017gksol}. These endeavour to capture all the dynamics of the edge region, from the macro scales down to the ion gyroradius scale. The computational expense is titanic, and thus parameter scans, and predictive modelling are currently out of reach for most applications. The second category of kinetic simulations only seeks to tackle turbulence in the edge, and have shed some light upon the inter-ELM pedestal~\citep{dickinson2012prl,dickinson2013microtearing,hatch2017jetilwped,mikek2017pedestalfail}. These simulations have been conducted despite the fact that the levels of $\bm{E}\times\bm{B}$ shear and flow velocities in the pedestal are not consistent with the assumptions under which gyrokinetics is usually derived~\citep{flowtome1}. Thus, we seek to embed gyrokinetics within a more complete pedestal model.

Bridging the gap between fluid models that cannot fully capture the kinetic physics, and gyrokinetic modelling that is extremely expensive and hard to interpret, we instead use a first-principles multiscale approach.
Our formulation neatly separates the physics of ELMs, coherent oscillations, and inter-ELM turbulence. This permits models for each component that are only as complex as needed, rather than a model that endeavours to rigorously include all possible physics. 
This conceptual simplification should enable more detailed studies of each separate part of the pedestal physics than is currently possible.
However, in this paper we only present a new model for the pedestal, which, in this work, we consider to be the steep gradient part of the closed-field-line region -- we leave the issues of extending this to an open-field-line region at the top of the scrape-off-layer to future work. 

The structure of the remainder of the paper, which lays this model out in detail, is as follows. 
In \Secref{SecELM} we discuss the physics needed to capture ELMs. Our physical requirements are translated into formal ordering assumptions in \Secref{SecELMOrdering}. In \Secref{SecELMEquations} we follow the well-trodden paths of asymptotic theory to turn our ordering assumptions into a closed set of equations for ELM dynamics.

Proceeding to the longer, inter-ELM, timescale in \Secref{SecInterELM} we follow the same procedure, presenting our orderings in \Secref{SecIntELMOrd} and the equations for the inter-ELM evolution in \Secref{SecSlowPed}. However, in this procedure we have to introduce an explicit multiple-scales approach to handle inter-ELM turbulence -- governed in turn by orderings and equations that are detailed in \Secref{SecPedTurb}.

Finally, in \Secref{SecBoundaries} we make detailed contact between our model and existing models of the pedestal. In \Secref{SecBCs} we describe how our model can be integrated with existing validated models for the core and scrape-off-layer, acting as a boundary condition for both. Then, in \Secref{SecExisting}, we compare our model to the models currently used to study the pedestal.
Ultimately, we end the paper with a summary of our results in \Secref{SecSummary} and some discussion of exploration of new avenues this model may open up.

\section{Fast and Furious: the physics of Edge Localised Modes}
\label{SecELM}

As alluded to in the introduction, ELMs are the large-amplitude convulsions of the edge pedestal that pose a great threat to current fusion devices. For this reason, ELM stability must be comprehensively addressed in any plausible pedestal model. To this end, in \Secref{SecELMOrdering} we codify the physics of ELMs in terms of temporal and spatial scales of interest, and relations between plasma parameters. These are then applied in \Secref{SecELMEquations} to obtain the dynamical equations governing ELMs.

\subsection{Orderings For ELM Dynamics}
\label{SecELMOrdering}
ELMs are now well-understood to be nonlinear peeling-ballooning modes~\citep{connor-edge-stability}. These modes are driven by the combination of edge current density, curvature, and pressure gradients. They naturally occur on an Alfv\'enic timescale, that is one which is comparable to $L_\parallel / \vA$, where $L_\parallel$ is the typical length-scale of the instability along a field line and $\vA$ is the Alfv\'en velocity. 
In this work we will use Gaussian units throughout and so, explicitly, $\vA = B / \sqrt{4\pi \sum_s m_s n_s}$ where $B$ is the magnetic field strength, and the plasma is composed of several species with masses, $m_s$, and densities, $n_s$.
The parallel length scale $L_\parallel$ will be taken to be the longest length scale in our system, and comparable to the connection length of the magnetic field. In general this means that the defining relationship is $L_\parallel \sim \left( \bm{b}\dg \ln B \right)^{-1}$, where as $L_\parallel$ is only ever used in our asymptotic orderings prefactors of order unity do not matter and will be neglected.

In order to cause such violent instabilities, the pressure gradients must be large, with the pressure varying over a distance $L_\perp \ll L_\parallel$ perpendicular to the field lines. Indeed, pedestals are observed to be extremely narrow structures in current experiments~\citep{beurskens2011pedestals}. Hence we adopt
\begin{equation}
\gkeps = \frac{L_\perp}{L_\parallel} \ll 1,
\label{epsDef}
\end{equation}
as our fundamental small parameter.

The most basic stability limit for the ballooning part of the peeling-ballooning mode is a limit on $\beta = 8\pi \sum_s n_s T_s / B^2$, with $T_s$ the temperature of species $s$.\footnote{We will later discover that our orderings do not constrain all species to be in local thermal equilibrium. Thus $\beta$ should really be defined instead in terms of the pressure of species $s$ perpendicular to field lines. As we are only interested in asymptotic orderings in this section, we will retain $T_s$ as a useful placeholder, that can be thought of as a mean kinetic energy per particle if species $s$ is not in thermal equilibrium.} This limit derives from the constraint that the energy required to bend field lines must be comparable to the energy released by relaxing the pressure gradient~\citep{IdealMHD}. 
Estimating this balance, we obtain
\begin{equation}
\beta \sim \frac{L_\perp}{L_\parallel},
\label{betaOrdering}
\end{equation}
which immediately implies that $\beta$ must be small in these systems. N.B. in our slightly unconventional usage $a \sim b$ means that $a$ and $b$ are of the same asymptotic order as $\gkeps\rightarrow 0$, but may differ by numerical prefactors of order unity. It is also important at this stage to note that we are expanding in the ratio of the pedestal width to the magnetic scale length, not in the aspect ratio of the magnetic surfaces.\footnote{Hence, our ordering for $\beta$ is compatible with the usual tokamak estimate $\beta \lesssim (r/q^2 R) d p / dr$ as $r/R$ and $q$ are taken to be $\Or(1)$.}

At this point, we shall not distinguish between profiles and fluctuations in terms of frequencies and length scales. Thus, all gradients will be ordered as $L_\perp^{-1}$ or $L_\parallel^{-1}$. 
This is consistent with our next assumption: that the kinetic plasma variables (e.g. density, pressure) will have $\Or(1)$ variation. This is physically reasonable as ELMs are observed to result in filaments of hot dense plasma erupting into much colder, lower density regions. To support finite density variations on times comparable to the electron transit time $L_\parallel / \vth[e]$, we require that the electrostatic potential also support large variations
\begin{equation}
\frac{e\pot}{T_e} \sim 1.
\label{largePot}
\end{equation}

Strong radial electric fields are known to play a role in pedestals, and many empirical measurements find these 
electric fields to be present in the absence of large plasma flows. The simplest and most plausible mechanism for
this to occur is that diamagnetic flows, for the bulk ion species, are comparable to, and may sometimes cancel, the $\bm{E}\times\bm{B}$ flow in the pedestal.
To permit such cancellation we require the diamagnetic flow to be of the same order as the $\bm{E}\times\bm{B}$ flow:
\begin{equation}
\frac{T_s}{m_s \cycfreq[s] L_\perp} \sim \uE,
	\label{diamagImportant}
\end{equation}
with $\cycfreq[s] = Z_s e B / m_s c$ the cyclotron frequency of species $s$, $c$ the speed of light, and $Z_s e$ the charge of species $s$ in terms of the unit charge $e$.
Note that this is also the ordering that follows from \eref{largePot} and assuming that $\pot$ varies on the $L_\perp$ scale.
In this, we assume $Z_s \sim 1$ with respect to the $\gkeps \ll 1$ expansion. Thus, this ordering does not describe the behaviour of an impurity species that is so highly charged that $Z_s \sim \gkeps^{-1}$ or even greater. We also will assume that all temperatures are comparable, in particular $T_e \sim T_i$, and will not keep track of factors of $T_i/T_e$ in our orderings.

Next, requiring that the system be fully nonlinear, the typical decorrelation rate due to the $\bm{E}\times\bm{B}$ nonlinearity should be comparable to the frequency of interest, $\omega$:
\begin{equation}
\frac{\uE}{L_\perp} \sim \omega,
\end{equation}
Hence our frequencies are comparable to the diamagnetic drift frequency $\omstar$
\begin{equation}
\omega \sim \omstar = \frac{cT_e}{e B} \frac{1}{L_\perp^2}.
\label{drift-req}
\end{equation}
Using our assumption of Alfv\'enic timescales
\begin{equation}
\omega \sim \frac{\vA}{L_\parallel},
\end{equation}
and our ordering for $\beta$, we have
\begin{equation}
\omstar \sim \frac{\vA}{L_\parallel} \sim \sqrt{\gkeps} \frac{\vth[i]}{L_\perp},
\end{equation}
which immediately gives
\begin{equation}
\frac{\rho_i}{L_\perp} \sim \sqrt{\gkeps}.
\label{rhoEpsDef}
\end{equation}
Similarly, we see that our frequencies are small compared to the cyclotron frequency 
\begin{equation}
\frac{\omega}{\cycfreq[s]} \sim \frac{\rho_i^2}{L_\perp^2} \sim \gkeps.
\label{omCycOrd}
\end{equation}

Substituting the definition of $\gkeps$ from \eref{epsDef} into \eref{rhoEpsDef} gives an estimate for the typical pedestal scale $L_\perp$ in terms of $L_\parallel \sim R$ (a typical tokamak estimate, where $R$ is the major radius of the device) and $\rho_i$:
\begin{equation}
L_\perp \sim \sqrt[3]{\rho_i^2 R}.
\label{PedestalWidth}
\end{equation}
With this ordering in hand, we can also relate our $\gkeps$ to the ratio of the gyroradius to the system size:
\begin{equation}
\gkeps \sim \left(\frac{\rho_i}{R}\right)^{2/3}.
\end{equation}

Let us discuss the impact of the scaling~\eref{PedestalWidth}. Firstly, we note that our orderings have given us pedestal width where diamagnetic stabilization of the
ballooning mode may be important for ELM stability~\citep{rogers1999diamagEdge}.
This scaling is more optimistic for future devices than a na\"ive neoclassical scaling which would predict a pedestal width $L_\perp$ comparable to the poloidal Larmor radius $\rho_{\pol}$~\footnote{This being the width required to increase ion neoclassical transport to the level where it can provide transport on the $\vth[i] / L_\parallel$ timescale. This is necessary if one wishes to match to an open field line region, where parallel losses to material surfaces occur on this timescale. Such a scaling is also predicted for the near-plasma scrape off layer from drift-orbit estimates in \citet{goldstonScaling}.
The full neoclassical theory for such a pedestal is detailed in \citet{kagan2008apg,kagan2010neoclassical} and related publications.
} -- in our orderings the poloidal Larmor radius is equivalent to $\rho_i$ -- but much less optimistic than the experimental results which typically show remarkably little dependence upon $\rho_i$~\citep{beurskens2011pedestals}.

One might take this experimental data as discouragement from believing our orderings. At the same time some pedestals appear to be dominated by neoclassical transport. These two can be reconciled by noting that
in current devices $\rho_\pol$ is not much smaller than the pedestal width. In addition $\rho_i^{2/3} qR$ is not much larger than $\rho_\pol$. Both of these relationships need to be well satisfied for our theory to be valid.
Hence, it is very likely that current machines do not operate at a small enough value of $\gkeps$ to see the asymptotic scaling in \eref{PedestalWidth}. 
Thus, \eref{PedestalWidth} is a prediction for machines where $\rho_i / R$ is smaller, with a prefactor that could be determined by simulation of the equations presented herein.

We still have a couple of parameters to order with respect to $\gkeps$. Firstly, we choose to retain electron parallel kinetic effects (Landau resonances etc.) upon the ELMs and so order
\begin{equation}
\frac{m_e}{m_i} \sim \beta \sim \gkeps,
\end{equation}
resulting in $\vth[e]\sim\vA$. 
Secondly, in order to straddle the divide between collisional and collisionless physics, we make the maximal ordering of 
\begin{equation}
\nu_{ee} \sim \frac{\vth[e]}{L_\parallel} \sim \omega,
\end{equation}
which naturally results in $\nu_{ii} \sim \vth[i] / L_\parallel$.
This concludes the ordering for the ELMs. Before moving on to turning these orderings into dynamical equations, we must consider
the possibility of the existence of small scale turbulence being present whilst an ELM is occurring.

\subsubsection{Orderings for Pedestal Turbulence}
\label{SecForeshadow}
Foreshadowing the results of \Secref{SecPedTurb}, it will turn out that our pedestal supports turbulent fluctuations, and the inter-ELM transport is dominated by them\footnote{Whilst it is often presumed that all turbulence is suppressed after an L-H transition, this is only possible for a pedestal that is on the scale of the (poloidal) gyroradius. For a pedestal that is asymptotically wider than this, turbulence is required to carry a finite heat flux through the pedestal. See the discussion at the end of Section 2}. The effects of the turbulence are, \textit{a priori}, strong enough that they could influence ELM evolution. In order to consistently include these effects, we will state here the assumptions we make about such turbulence. These assumptions will be justified in \Secref{SecPedTurb}.

The fluctuations are small, with
\begin{equation}
\frac{\turbf}{f_s} \sim \frac{e \turbPot}{T_e} \sim \sqrt{\gkeps},
	\label{firstAssumption}
\end{equation}
and
\begin{equation}
\frac{\turbB}{B} \sim \gkeps.
\end{equation}

The fluctuations will have typical length scales, perpendicular to the field, given by
\begin{equation}
\nabla_\perp \ln \turbf \sim \rho_i^{-1},
\end{equation}
and along the field given by
\begin{equation}
\bm{b}\dg \ln \turbf \sim \left( L_\perp L_\parallel \right)^{-\infrac{1}{2}}.
\end{equation}
One might worry that the large drive in the pedestal will cause turbulence from $k_\perp \rho_i \sim 1$ to reach larger scales. However, in our ordering, this has already occurred. These fluctuations are naturally electron-scale fluctuations, forced out to ion scales by the strong drive. 
Indeed, the timescale of these fluctuations is assumed to be
\begin{equation}
\pd{ }{t} \ln \turbf \sim \frac{\vth[e]}{\sqrt{L_\perp L_\parallel}}.
\label{lastAssumption}
\end{equation}

As these fluctuations are faster than the ELMs and on a smaller scale, by the usual method of introducing intermediate temporal and spatial length scales we can 
assume the existence of an averaging operation $\ensav{\cdot}$ such that
\begin{equation}
\turbf = \exactf - \ensav{\exactf},
	\label{deltaDef}
\end{equation}
where $\exactf$ is the exact (inclusive of fluctuations) distribution function.
The averaged quantities are assumed to obey the orderings given above, and so we will use an undecorated notation for the averaged quantities, e.g.
\begin{equation}
f_s = \ensav{\exactf}.
\end{equation}

We note that, in these orderings, $\sqrt{\gkeps} \sim \rho_i / L_\perp$, and that $L_\perp$  is the appropriate local scale length for the physics that drives the turbulence. Thus, our orderings say that the fluctuations are small in $\rho^* \equiv \rho_i/L_\perp$. It is this relationship that will give rise to the similarity between the equations governing the turbulence and gyrokinetics.
\subsection{Dynamical Equations}
\label{SecELMEquations}
Having argued for our consistent orderings for the ELM dynamics, we now apply them to the governing equations of the plasma to obtain our reduced model.
The fundamental equations we will work from are the Vlasov-Landau equation for the exact distribution function $\exactf$, in terms of the exact fields $\exactE$ and $\exactB$:
\begin{equation}
\pd{\exactf}{t} + \bm{v}\dg\exactf + \frac{Z_s e}{m_s} \left( \exactE + \frac{1}{c} \bm{v}\times\exactB \right) \cdot\pd{ \exactf}{\bm{v}} = \collop[\exactf] + \source,
\label{vfp0}
\end{equation}
where $\bm{v}$ is the particle velocity and the collision operator on the right-hand side is the Landau operator. We have also added an arbitrary source $\source$ that we will keep in our inter-ELM equations (using the maximal ordering $\source \sim f_s \vth[i] / L_\parallel$ so that the source can compete with inter-ELM transport).
We will assume a non-relativistic plasma and also that all frequencies are low compared to the plasma frequency and 
all lengths long compared to the Debye length.
Thus, to close the system for the fields, we have the quasineutrality constraint
\begin{equation}
\label{qn}
\sum_s Z_s e n_s= 0,
\end{equation}
and  the pre-Maxwell Amp\`ere's law:
\begin{equation}
\label{ampere}
\curl\Bfield = \frac{4\pi}{c} \current.
\end{equation}

Averaging these equations over any possible fast fluctuations, we have as our fundamental kinetic equation:
\begin{equation}
\begin{split}
\pd{f_s}{t} + &\bm{v}\dg f_s + \frac{Z_s e}{m_s} \left( \bm{E} + \frac{1}{c} \bm{v}\times\bm{B} \right) \cdot\pd{ f_s}{\bm{v}} = \collop[f_s]\\
		&- \frac{Z_se}{m_s}\ensav{ \left(\turbE + \frac{1}{c} \bm{v}\times\turbB\right)\cdot\pd{\turbf}{\bm{v}}} + \ensav{\collop[\turbf]},
	\end{split}
\label{vfp}
\end{equation}
and as the field equations are linear, they apply separately to the averaged and fluctuating fields. 

\subsubsection{The Magnetic Field}
The first problem we tackle is that of the structure of the magnetic field.
As $\beta \sim \infrac{L_\perp}{L_\parallel} \ll 1$, we naturally expect the time-dependent piece of the magnetic field to be small compared to the background, confining, field. 
To investigate this, we first take the $m_s\bm{v}$ moment of \eref{vfp}, and sum over all species to obtain
\begin{equation}
\begin{split}
\pd{}{t} \sum_s n_s m_s \bm{u}_s + \dv\sum_s\Ptens   = -\nabla\left( \frac{B^2}{8\pi} \right) + \frac{1}{4\pi} \bm{B}\dg\bm{B} + \dv\ensav{ \frac{\turbB\turbB}{4\pi} - \idmat \left|\turbB\right|^2},
\end{split}
	\label{MomEq1}
\end{equation}
where 
\begin{equation}
\Ptens = \vint m_s \bm{v}\bm{v} f_s,
\end{equation}
is the full pressure tensor of species $s$.

Estimating the size of each term in this equation, the magnetic pressure force is larger than all other terms by one power of $\gkeps$, due to the small $\beta$.
Thus, we conclude that the time-dependent piece of the magnetic field must be small compared to the background, confining, field, and so we make the split 
\begin{equation}
\bm{B} = \bm{B}_0 + \elmB
\end{equation}
where $\elmB \sim \gkeps \bm{B}_0$. Note that, despite this decomposition, we will define parallel and perpendicular components of vectors with respect to the total field~$\bm{B}$.

With this ordering in hand, we can solve the lowest order equation
\begin{equation}
\nabla_\perp B_0^2 = \Or(\gkeps),
\end{equation}
by insisting that $B_0$ vary only on the long $L_\parallel$ scale.

We note that the size estimate for $\elmB$ is the same as what one would obtain by assuming the field follows the fluid flow and so can be displaced at most $L_\perp$ in the perpendicular direction, and that this displacement occurs after following the perturbed field for a distance $L_\parallel$:
\begin{equation}
\elmB \sim \frac{L_\perp}{L_\parallel} \bm{B}_0.
\end{equation}
This is also the estimate one would obtain by insisting that $\elmB \cdot\bm{\nabla}_\perp \sim \bm{B}_0\dg$.

Given that $\bm{b}\dg\bm{b} \sim L_\parallel^{-1}$ by definition of the parallel scale, one might be tempted to conclude that $\bm{B}_0$ only varies on the $L_\parallel$ scale. However, we can still admit a large
magnetic shear. That is, the tensor $\bm{\nabla}_\perp \bm{b}_0$ may have large components. As will be argued later, these cannot be as large as $\Or(L_\perp^{-1})$ because it would entail an infeasibly strong electron current. However we will retain the possibility that the shear length $L_s$ of $\bm{B}_0$ is as short as $\Or(\sqrt{L_\perp L_\parallel})$.

Given this decomposition of the magnetic field into fixed and time-dependent pieces, we introduce the vector potential $\bm{A}$ for $\elmB$ only:
\begin{equation}
\elmB = \curl\bm{A},
	\label{elmBdef}
\end{equation}
and we will work in Coulomb gauge $\dv\bm{A} = 0$.

The piece of $\elmB$ that will turn out to be important is the change in the direction of the magnetic field, thus we make the natural decomposition $\bm{A} = A_\parallel \bm{b} + \bm{A}_\perp$ to find
\begin{equation}
\elmB = \nabla A_\parallel \times \bm{b} + \elmB \cdot\bm{b}\bm{b}.
\end{equation}

From hereon in, we will define perpendicular and parallel components of vectors with respect to the total large-scale field $\bm{B}$, inclusive of $\bm{B}_1$ but not of any possible turbulent magnetic field.
If we at any point we only wish to consider the static field $\bm{B}_0$ we will denote that explicitly with a subscript $0$ on the appropriate terms.
\subsubsection{Kinetic Variables}

It will be simplest, despite not retaining finite gyroradius effects, to follow the gyrokinetic literature and transform to guiding centre variables for our derivations.
To simplify later derivations, we will also use a peculiar velocity relative to the mean $\bm{E}\times\bm{B}$ velocity $\bm{w} = \bm{v} - \uE$. 
It is important to note that $\uE = c \bm{E}\times\bm{b} / B$ is defined using the total mean $\bm{B}$ including $\bm{B}_1$. Hence, $w_\parallel = v_\parallel$ exactly.
Thus, we move from $(\bm{r},\bm{v})$ to the Catto-transformed variables $(\bm{R}_{s},\energy,\magmom,\gyr,\sigma)$:
\begin{align}
\label{Rdef}
\bm{R}_{s} &= \bm{r} - \frac{1}{\cycfreq[s]} \Meanb\times\bm{w},\\
			 \label{energydef}
\energy &= \frac{1}{2} m_{s} w^2\\
			  \label{mudef}
\magmom &= \frac{m_{s} w_\perp^2}{2\MeanMagB}+\Or({\gkeps}),
\end{align}
where $\gyr$ is the gyrophase, $\sigma$ is the sign of the parallel velocity, and perpendicular and parallel components are taken with respect to the total mean magnetic field $\bm{B}$.
We take the magnetic moment $\magmom$ to be exact adiabatic invariant of the Larmor motion that is conserved to all orders in $\gkeps$~\citep{kruskal1958gyration,berkowitz1959asymptoticparticle}~\footnote{The excessively cautious reader may wonder if the fact that the $\gkeps$ of this paper and the small parameter of \citet{kruskal1958gyration} differ will give rise to a problem in this argument.
Thankfully, it is obvious that if a term is small in our ordering then it will also be small in Kruskal's -- this follows from the fact that all of our time variation is slow compared to the cyclotron frequency, and all spatial variation (at this point) is long compared to the gyroradius. Extending these results to include a small component of the fields that has $\rho_i$-scale variation can be done order by order. The magnetic moment including first-order gyrokinetic corrections for electrostatic turbulence can be found in Appendix A of \citet{parra2008lgt}.}
	.

Finally, the gyrophase $\gyr$ is defined by
\begin{equation}
\bm{w} = w_\parallel \bm{b} + w_\perp \left(\cos \gyr\,\bm{e}_2 - \sin\gyr\,\bm{e}_1 \right),
\label{vDef}
\end{equation}
where $\bm{e}_1$ and $\bm{e}_2$ are mutually orthogonal basis vectors that satisfy $\bm{b} = \bm{e}_2 \times\bm{e}_1$

We will find it convenient to average over this gyrophase at various points in this work.
All of these gyroaverages are taken at constant $\bm{R}_s$, $\energy$, and $\magmom$; we denote the gyroaverage by
\begin{equation}
\gyroR{\dots} = \frac{1}{2\pi} \int_0^{2\pi} \left( \dots \right) d\gyr.
\end{equation}
We will also occasionally need a gyroaverage at fixed $\bm{r}$, $w_\parallel$, $w_\perp$, which is denoted by
\begin{equation}
\gyror{\dots} =\frac{1}{2\pi} \int_0^{2\pi} \left( \dots \right) \left.d\gyr\right|_{\bm{r},w_\parallel,w_\perp}.
\end{equation}

The gyroaveraged (at fixed $\bm{R}_s$) time derivatives of the gyrokinetic variables are calculated in \Apref{APDdt}, and are found to be (c.f. \eref{APRDot} and \eref{APeDot})
\begin{align}
\label{eqRdot}
\gyroR{\dot{\bm{R}}_{s}} &= w_\parallel \bm{b} + \frac{c}{B} \bm{b}\times\nabla\pot + \Or\left(\gkeps^{3/2} \vth[i]\right),\\
			 \label{eqEdot}
\gyroR{\denergy} &= -Z_s e w_\parallel \left( \bm{b}\dg \pot + \frac{1}{c} \pd{A_\parallel}{t}\right) + \Or\left( \sqrt{\gkeps} \frac{\vth[i]}{L_\parallel} T_e\right)\\
\gyroR{\dot{\gyr}} &= \cycfreq[s] + \Or(\gkeps^{1/2} \cycfreq[i]).
\end{align}

\subsubsection{Kinetic Equations}
Rewriting \eref{vfp} in the above variables, we obtain
\begin{equation}
\begin{split}
\pd{f_s}{t} + &\bm{\dot{R}}_s \dgR{f_s} + \denergy \pd{f_s}{\energy} + \dot{\gyr} \pd{f_s}{\gyr} = \collop[f_s]\\
		&- \frac{Z_se}{m_s}\ensav{ \left(\turbE + \frac{1}{c} \bm{v}\times\turbB\right)\cdot\pd{\turbf}{\bm{v}}} + \ensav{\collop[\turbf]}.
	\end{split}
\label{vfpCatto}
\end{equation}
Applying our ordering for the ELM dynamics to this equation, we immediately see that the lowest order equation is
\begin{equation}
\cycfreq \pd{f_s}{\gyr} = 0.
\end{equation}
Thus, both ions and electrons are immediately found to be gyrotropic at fixed $\bm{R}_s$.
The size of the gyrophase-dependent corrections differs between species and so the accuracy of this statement is
\begin{equation}
\pd{f_i}{\gyr} = \Or(\gkeps f_i) \qquad \mathrm{ and }\qquad \pd{f_e}{\gyr} = \Or(\gkeps^2 f_e).
\label{howSmallGyr}
\end{equation}

Continuing to higher order in our expansion of \eref{vfpCatto}, we find the kinetic equation for the electrons
\begin{equation}
\pd{f_e}{t} + \left( w_\parallel \bm{b} + \uE \right) \dgR[e]{f_e} + ew_\parallel \left( \bm{b}\dg\pot + \frac{1}{c} \pd{A_\parallel}{t}\right) \pd{f_e}{\energy[e]} = \collop[f_e],
	\label{eKE-1}
\end{equation}
where we note that the fluctuations from the pedestal turbulence are not strong enough to affect $f_e$ on this timescale. This kinetic equation can, in principle, have solutions with a non-zero parallel flow. This would, naturally, be of order $n_e\vth[e]$. Such a strong parallel flow is prohibited for two reasons. Firstly, the parallel current, estimated from Amp\`ere's law and using the scaling for the magnetic shear posited above, results in
\begin{equation}
j_\parallel \sim \frac{c}{4\pi} \frac{B}{L_s} \sim \frac{c m_e B^2}{e B n_e  T_e} \frac{\vth[e]}{L_s} e n_e \vth[e]\sim  \gkeps e n_e \vth[e],
\end{equation}
which is inconsistent with a strong electron flow (because $j_\parallel \sim e n_e \upar[e]$). Secondly, electron and ion velocities which have a relative drift comparable to the electron thermal speed are strongly unstable to Debye-scale fluctuations~\citep{jackson1960twostream}.
Thus, we limit ourselves to only considering solutions to \eref{eKE-1} that have no parallel flow (as this equation is correct only to lowest order in $\sqrt{\gkeps}$ the vanishing of the flow to this order simply means that the electron flow is at most comparable to the ion thermal speed). The constraint equation is simply the $w_\parallel$ moment of \eref{eKE-1} and can be written as an equation for the parallel vector potential $A_\parallel$:
\begin{equation}
e n_e \left( \bm{b}\dg \pot + \frac{1}{c} \pd{{A}_\parallel}{t}\right) = \bm{b}\dg p_{\parallel e} - \left( \bm{b}\dg\ln B \right) \left(p_{\parallel e} - p_{\perp e}\right).
\label{ELMAparEq}
\end{equation}
Note that we can neglect the time derivative of $f_e$ because the electron momentum is negligible (due to both the small electron mass and the absence of sonic electron flows) and the integral of the collision operator can be estimated as
\begin{equation}
\wint w_\parallel \collop[f_e] \sim \nu_{ei} n_e \left( \upar[e] - \upar[i] \right) \sim \nu_{ei} j_\parallel / e \sim \sqrt{\gkeps} n_e \frac{T_e}{m_i L_\parallel},
\end{equation}
which is $\sqrt{\gkeps}$ smaller than the terms in \eref{ELMAparEq}.

For the ions, we have the simpler kinetic equation
\begin{equation}
\left( \pd {}{ t} + \uE \dgR[i]{ }\right) f_i = 0,
	\label{iKE-1}
\end{equation}
so the ions are moved by the plasma flow, but do not spread thermally along field lines and neither accelerate nor collide in one Alfv\'en time. This is easily seen to be correct as all ion timescales (such as the ion streaming time, or ion-ion collision time) are small compared to the Alfv\'en time (or the electron-electron collision time).

Finally, as we know that the distribution functions are independent of gyrophase to leading order (c.f. \eref{howSmallGyr}), we have that
\begin{equation}
\Ptens = m_s \wint \gyror{\bm{w}\bm{w}} f_s =   p_{\perp s} \idmat + \left(p_{\parallel s} - p_{\perp s} \right) \bm{b}\bm{b} + \Or(\gkeps n_s T_s),
\end{equation}
where the perpendicular and parallel pressures have their usual definitions
\begin{equation}
p_{\perp s} =  \wint \frac{1}{2} m_s w_\perp^2 f_s \qquad\mathrm{and}\qquad p_\parallel =  \wint m_s w_\parallel^2 f_s,
\end{equation}
respectively, and so the pressure tensors are also gyrotropic to leading order (the off-diagonal terms are computed later in \Apref{ApVortELM}).
Hence we can now write the largest non-zero part of the momentum equation \eref{MomEq1} as
\begin{equation}
\nabla_\perp\left( \sum_s p_{\perp s} + \frac{\bm{B}_0\cdot\elmB}{4\pi} \right) = 0,
	\label{perpPressure}
\end{equation}
which determines the time-dependent part of the total field strength.

We now have equations for the distribution functions, and the two fields that comprise $\elmB$. All that remains is to derive an equation for $\pot$. 

\subsubsection{The Vorticity Equation}
\newcommand{\turbVort}{\varpi}
As is common to many long-wavelength theories of a magnetized plasma, the electrostatic potential is determined from the plasma vorticity equation. 
This is derived in \Apref{ApVortELM}, with the final result that
\begin{equation}
\begin{split}
&\dv\left\{ \sum_s \frac{n_s m_s}{B} \left[ \pd{ }{t} + \uE\dg\right]\left( \frac{\nabla p_{\perp s}}{n_s m_s \cycfreq[s]} + \frac{c \nabla \pot}{B}\right)\right\}
\\
		&\quad= \dv\left[ - \frac{j_\parallel \bm{b}}{c}\left( 1- 4\pi \frac{ p_\parallel - p_\perp}{B^2}\right) - p_\perp \curl\left(\frac{\bm{b}_0}{B}\right) - \frac{\curl \bm{b}_0}{B} \left(p_{\parallel} - p_\perp\right) \right]+\turbVort,
\end{split}
\label{vortELM}
\end{equation}
where $\turbVort$ is the possible contribution from turbulence, whose explicit expression is
	\begin{equation}
	\turbVort = - \sum_{s=i} \frac{ Z_i e}{2\cycfreq[i]} \left\{ \grad\grad\bm{:} \ensav{\wint\left[\left(\grad\turbPot\right) \bm{w}_\perp + \bm{w}_\perp \left(\grad\turbPot\right)\right] \turbf[i]} \right\}.
	\label{turbVortDef}
	\end{equation}
The parallel current in \eref{vortELM} is found from Amp\`ere's law:
\begin{equation}
j_\parallel = \frac{c}{4\pi} \bm{b}\cdot\curl\bm{B} = \frac{c}{4\pi}\left( \bm{b}_0 \cdot \curl \bm{B}_0 - \nabla_\perp^2 A_{\parallel}\right).
\label{ampLawELM}
\end{equation}
The first term of this expression is due to the shear of the confining field. This first term can be large when compared to the second, and thus the lowest order of \eref{vortELM} is
\begin{equation}
\dv\left( \bm{b} j_\parallel \right) = \frac{c}{4\pi}\left(\bm{b}_0 + \bm{b}_1 \right)\dg\left( \bm{b}_0 \cdot \curl \bm{b}_0 \right) = 0.
\label{constrainShear}
\end{equation}
Taking the time-dependent piece of this equation, we have that
\begin{equation}
\bm{b}_0 \dg \left(\bm{b}_0 \cdot \curl\bm{b}_0\right) = 0.
\end{equation}
If we assume that $\bm{B}_0$ has magnetic surfaces labelled by $\psi_0$ then this becomes a constraint on the shear of $\bm{B}_0$:
\begin{equation}
\left(\bm{b}_0 \cdot \curl\bm{b}_0\right) = K(\psi_0),
	\label{LargeShearK}
\end{equation}
and, from the time-independent piece of \eref{constrainShear}, we discover that 
\begin{equation}
\elmB \dg \psi_0 \pd{K}{\psi_0} = 0,
\end{equation}
so either $K(\psi_0)$ is constant on the spatial scale $L_\perp$ or there are exact surfaces and $K = K(\psi)$. If $K$ is approximately constant, then we have a strong, but unvarying shear across the pedestal.
If the field $\bm{B}_0$ does not have surfaces then naturally such a large shear is forbidden and the solution of these constraints is $K=0$.

Let us discuss the physics contained in this vorticity equation. The motion of the filamentary structures, related to the inertial term on the left-hand-side of \eref{vortELM} is driven by
gradients of the parallel current in $\bm{B}_0$ (kink-mode drive) and gradients in the plasma pressure (interchange-mode drive). The part of $j_\parallel$ from $\elmB$ represents the opposition 
of such motion due to field line bending. The combination of interchange drive and field-line-bending gives rise to the usual ballooning-mode physics. Thus we have all the necessary ingredients for nonlinear peeling-ballooning modes. 

We also wish to note that this vorticity equation, although in a different form, is consistent with the collisional one derived in~\citet{cattosimakovedge}. This can be seen most clearly by 
using (D2) of \citet{cattosimakovedge} in (69) \& (70) of the same paper. This gives the following vorticity equation
\begin{equation}
\begin{split}
&\dv\left\{ \sum_s \frac{n_s m_s}{B} \left[ \pd{ }{t} + \uE\dg\right]\left( \frac{\nabla p_{\perp s}}{n_s m_s \cycfreq[s]} + \frac{c \nabla \pot}{B}\right)\right\}
\\
		&\qquad= \dv\left[ - \frac{j_\parallel \bm{b}}{c}\left(1 - \frac{4\pi\pi_{ci}}{B^2}\right) - \left(p_i + p_e -\frac{1}{3}\pi_{ci}\right) \curl\left(\frac{\bm{b}}{B}\right) - \frac{\curl \bm{b}}{B} \pi_{ci} \right],
\end{split}
\label{vortCS}
\end{equation}
where we have used the identity \eref{apCurv25} from \Apref{ApVortELM} to rewrite the term involving the ion parallel viscosity. All notation is ours, except for $\pi_{ci}$ which is defined in (14) of \citet{cattosimakovedge}. Using the orderings of \citet{cattosimakovedge}, the terms involving $\pi_{ci}$ are $\Or( p_i \delta_i \Delta_i / L_\parallel B \gkeps)$ and so can be neglected compared to the terms only involving the pressure (by using the second inequality of (7) of \citet{cattosimakovedge}).

Comparing this to our vorticity equation, we see that they match if as well as taking the collisional limit of our equations (which removes the electron pressure anisotropy from our vorticity equation) one also assumes that the initial ion pressure is isotropic. Thus we include the same Alfv\'enic physics as Catto-Simakov.

We also discuss the difference of our vorticity equation with those found in reduced Braginskii models~\citep{ricci2012gbs,zeiler1997driftbraginskii}. Taking the vorticity equation in \citet{ricci2012gbs} as the most
modern and complete model currently in use, we see that we retain the diamagnetic contributions to the vorticity, but not the ion-sound contributions, which are small in our orderings.
We anticipate that this is the correct trade-off for ELM filaments as they are often observed to be both narrow and extremely rapid.

Finally, we should discuss the potential contribution from small-scale turbulence. This contribution is exactly analogous to the Reynold's stress generated by electrostatic gyrokinetic turbulence. Indeed, we will find later that the equations this turbulence obeys are reminiscent of gyrokinetics. Because of this, absent symmetry breaking effects, this stress may turn out to be small~\citep{parra2011up}. However, until calculations and simulations confirm this, we retain this term as it is formally of the required size.

\subsection{Summary of the ELM Model}
\label{secELMsummary}
The complete kinetic model for ELM-like behaviour on the fast, Alfv\'enic timescale is as follows
\vspace{2mm}
\begin{itemize}
\item Electron and ion distribution functions are determined from \eref{eKE-1} and \eref{iKE-1}, respectively
\item The vector potential for $\bm{B}_1$, $A_\parallel$, is determined from \eref{ELMAparEq}.
\item The electrostatic potential $\pot$ is determined from the vorticity equation \eref{vortELM}, in which the pressure tensors are given by moments of the distribution functions we have already calculated.
\item The turbulent contribution to the vorticity equation $\turbVort$ is defined by \eref{turbVortDef}, where the fluctuations are calculated from the equations given in \Secref{SecPedTurb}.
\end{itemize}
\vspace{2mm}
These equations should be solved in an annular domain of closed field lines, with periodic parallel boundary conditions and static Dirichlet boundary conditions for $\pot$ and $A_\parallel$ on the boundaries, such that $\uE$ and $\bm{b}$ are tangential to the boundaries. This is achieved by taking the boundaries sufficiently far from the ELM that no plasma seeks to erupt out of the domain.
Static boundary conditions are naturally appropriate as the time-variation of fields outside of the ELM region is small compared to the timescale of the ELM.
\section{Building Up: Inter-ELM pedestal Evolution}
\label{SecInterELM}
The equations presented in the preceding section describe the rapid evolution of the pedestal region during an ELM. This rapid evolution will rearrange field lines and the plasma pressure profiles 
but does not itself contain the physics that would relax these profiles after an ELM. Nor do they contain the physics that builds up the density and temperature profiles between ELMs.
For this, we need to consider what happens when the only solutions to the ELM equations are stationary. 

To achieve this, we will first, in \Secref{SecIntELMOrd} build up a set of orderings for the slow behaviour of the plasma. Then we will derive equations embodying such orderings. We will show that these equations naturally require stationary solutions to our ELM equations, which were summarised above in \Secref{secELMsummary}. This proves one major part of self-consistency.
The final piece of the puzzle will arrive in \Secref{SecPedTurb} where we show that small-scale fluctuations, consistent with slab-like electron temperature gradient-driven (ETG) or microtearing (MTM) turbulence, 
drive the inter-ELM transport through our pedestal. We will also confirm the result asserted above, that this turbulence is strong enough to provide the required transport but only affects ELM dynamics in a limited way.
\subsection{Inter-ELM Orderings}
\label{SecIntELMOrd}
We must now re-evaluate our orderings to accommodate the slow inter-ELM behaviour. 
We retain the orderings on $L_\perp/L_\parallel$, $\rho_i / L_\perp$, $\beta$, $m_e/m_i$, and collisionalities from \Secref{SecELMOrdering}.
However, we now wish to accommodate frequencies on the ion-sound timescale, $\omega \sim \vth[i]/L_\parallel$.

In order to do this, we need to slow down the nonlinear timescales due to the $\bm{E}\times\bm{B}$ motion ($\omega_{NL} \sim \uE\dg$ and due to diamagnetic effects $\omega_{NL}\sim\bm{b}\times\nabla p \dg$).
We manage this by introducing a new length scale $L_\wedge \sim \sqrt{L_\perp L_\parallel} \sim \sqrt{\gkeps} L_\parallel$. We assume that in the plane perpendicular to the magnetic field we can identify a 
fast direction, where the typical length scales are $L_\perp$, and a slow direction, where the typical length scales are $L_\wedge$.
In a pedestal this is naturally given by the radial and the in-surface perpendicular directions respectively (if one views the poloidal coordinate as the distance along a field line, then this slow direction is the \textit{toroidal} direction).

Under this assumption, we see that all the nonlinearities can be written in the form 
\begin{equation}
\bm{b}\times\nabla g \cdot \nabla f \sim \frac{g f}{L_\perp L_\wedge},
\end{equation}
where we have used the fact that the derivatives cannot both be in the fast direction.

With this estimate in hand, we see that 
\begin{equation}
\uE \dg \sim \frac{cT_e}{e B} \frac{e\pot}{T_e} \frac{1}{L_\perp L_\wedge} \sim \sqrt{\gkeps} \frac{cT_e}{e B} \frac{1}{L_\perp^2} \sim \frac{\vth[i]}{L_\parallel},
\end{equation}
as required.

No other ordering changes are required. We expect this set of equations to model the equilibrium and any possible coherent oscillation of the edge pedestal. Such coherent oscillations, with frequencies comparable to the local acoustic frequency, are thought to be important in novel modes of tokamak operation~\citep{qhmode2002,edahmode2001}.

We now revisit and justify the turbulence orderings introduced earlier. To distinguish scales relating to the fluctuations from the scales of the pedestal itself, we will use $k_\perp$ to denote
typical perpendicular wavenumbers, and $k_\parallel$ typical parallel wavenumbers.
The only length scales smaller than $L_\perp$ available to us are the gyroradii, and so we assume that the turbulence is predominantly at scales comparable to the ion gyroradius 
\begin{equation}
k_\perp \rho_i \sim 1,
\end{equation}
and because of our ordering for $\beta$ and the mass ratio, this means that turbulent scales are also comparable to the electron skin depth $d_e = c/\omega_{pe}$ :
\begin{equation}
k_\perp d_e \sim 1.
\end{equation}
Making a mixing-length estimate for the amplitude of this turbulence we see that
\begin{equation}
\frac{\turbf}{f_s} \sim \frac{e\turbPot}{T_e} \sim \frac{1}{k_\perp L_\perp} \sim \sqrt{\gkeps}.
\label{turbamplitude}
\end{equation}
This is also exactly the ordering required to balance the nonlinear turnover rate with the $\omstar$-drive:
\begin{equation}
\uE\dg \sim \frac{cT_e}{eB} \frac{e\turbPot}{T_e} k_\perp^2 \sim \sqrt{\gkeps} k_\perp \vth[i],
\end{equation}
and
\begin{equation}
\omstar \sim \frac{cT_e}{eB} \frac{k_\perp}{L_\perp} \sim \frac{\rho_i}{L_\perp} k_\perp \vth[i] \sim \uE\dg.
\end{equation}
Both of these frequencies are comparable to $\vth[i] / L_\perp$. 
If this turbulence is mainly mediated by electrons, as in ETG or MTM turbulence, then the frequency sets a typical parallel
length scale by
\begin{equation}
\omega \sim k_\parallel \vth[e].
\end{equation}
This means that we have $k_\parallel L_\parallel \sim 1/\sqrt{\gkeps} \gg 1$. This is short-parallel-wavelength, slab-like turbulence.
We will also assume that turbulence is as electromagnetic as our low-$\beta$ assumption allows:
\begin{equation}
\frac{\turbB_\perp}{B} \sim \frac{k_\parallel}{k_\perp} \sim \gkeps.
\end{equation}

We can also make a random-walk estimate of the transport from this turbulence
\begin{equation}
\tau_E^{-1} \sim \frac{D}{L_\perp^2} \sim \frac{\omega}{(k_\perp L_\perp)^2} \sim \frac{\vth[i]}{L_\parallel},
\end{equation}
where $\tau_E$ is the pedestal energy confinement time and $D$ is a typical diffusion coefficient from our turbulence.
We see that this is precisely in accord with our ordering of the inter-ELM evolution timescale.

Note that our exploration of these equations does not extend to the timescale on which the confining field evolves, $\bm{B}_0$ is fixed throughout. This is currently no great limitation, as a pedestal that is quiescent for so long that the diffusion of current becomes important is presently beyond experimental and theoretical reach. However, this must eventually be addressed in future work.

\subsection{The Slow Dynamics of a Pedestal}
\label{SecSlowPed}

We use the same variables as in the prequel, and taking care to use the new ordering to estimate the size of their time derivatives we obtain
\begin{align}
\gyroR{\dot{\bm{R}}_{s}} &= w_\parallel \bm{b}(\bm{R}_s) + \frac{c}{B} \bm{b}\times\nabla\pot + \Or\left(\gkeps^{3/2} \vth[i] \right)\\
\gyroR{\denergy} &= - Z_s e w_\parallel \left(\bm{b}\dg \pot + \frac{1}{c} \pd{A_\parallel}{t} \right) + \Or\left(\gkeps \frac{\vth[i]}{L_\parallel} T_e\right)\\
\gyroR{\dot{\gyr}} &= \cycfreq[s] + \Or(\gkeps^{1/2} \cycfreq[i]),
\end{align}
as before.

In \Apref{SlowDeriv} we prove that a consistent solution to the lowest-order electron kinetic equation, in a closed-field-line region, is a 
Maxwellian. If this is the chosen solution, then the magnetic field is frozen into a fluid flow and the magnetic topology is fixed on the timescales of interest.
Then we can assume that the field has topologically-toroidal flux surfaces (denoted by $\psi$) for all time, and we will average over them when needed.

The electrons are an isothermal (i.e. $T_e$ is constant on a flux surface) fluid, with a continuity equation for the density (c.f. \eref{SlowEContAP}):
\begin{equation}
\left(\pd{ }{t} +\uE\dg\right) n_e + \dv\left(n_e \upar[e]\bm{b}\right)+ \dv\ensav{\vint \vchi h_e } = \Psource[e],
\label{eTransp1}
\end{equation}
where the fluctuating velocity is given by
\begin{equation}
\vchi = \frac{c}{B} \bm{b}\times\nabla\gyroR{\pot - \frac{w_\parallel}{c} \turbApar},
\end{equation}
and an evolution equation for the temperature (see \eref{SlowETempAP})
\begin{equation}
\begin{split}
& \frac{3}{2}\fav{ \left( \pd { }{t} + \uE \dg\right) n_e T_e}  + \fav{\dv\ensav{\vint \left(\frac{\energy[e]}{T_e} - \frac{3}{2}\right) \vchiR \turbf[e]}} \\
		&\quad= \fav{e n_e \upar[e] \bm{b}\dg \pot} + \fav{\Esource[e]}.
\end{split}
\label{eTransp2}
\end{equation}
where $T_e = T_e(\psi)$.
The terms on the right hand side are easily interpreted as parallel compressional heating, turbulent Ohmic heating, and a possible energy source, respectively.
The fluctuating distribution function $\delta f_e$ is defined by \eref{deltaDef}, and given by the solution of \eref{electronGKE} as explained in the next section.

The ions obey a drift-kinetic equation:
\begin{equation}
\begin{split}
&\left(\pd{ }{t} + w_\parallel \bm{b}\dg + \uE\dg\right) f_i 
- Z_i e w_\parallel\bm{b}\dg\pot \pd{f_i}{\energy[i]} +\dv\ensav{ \frac{c}{B} \bm{b}\times\nabla\gyroR{\turbPot}\, h_i }
\\
		&\qquad\qquad= \collop[f_i] + \source[i].
\end{split}
\label{SlowIonKE}
\end{equation}
in which the fluctuating ion distribution function $h_i$ is defined by \eref{hDefI} and given by the solution of \eref{ionGKE} as described in \Secref{SecPedTurb}.

The fluctuating magnetic field is given by the lowest-order vorticity equation
\begin{equation}
\begin{split}
&\left(\bm{B}_0 + \bm{b}_0 \times \nabla A_\parallel \right) \dg \left( \frac{1}{B} \nabla_\perp^2 A_\parallel - \bm{b}_0 \cdot\curl\bm{b}_0\right)\\
	&\quad= 4\pi \curl\left(\frac{\bm{b}_0}{B}\right) \dg p_{\perp} 
+ \frac{4\pi}{B} \curl\bm{b}_0 \dg \left( p_\parallel - p_\perp \right).
\end{split}
\label{MagEq}
\end{equation}
Equivalently, \eref{MagEq} is simply the constraint that the ELM-timescale vorticity equation \eref{vortELM} has zero-vorticity solutions. 
The fluctuating $A_\parallel$ then determines the parallel electron flow via Amp\`ere's law
\begin{equation}
\upar[e] = -\frac{c}{4\pi e n_e} \bm{b}\cdot\curl\bm{B} + \frac{1}{n_e} \sum_{s=i} Z_s \vint w_\parallel f_s,
\end{equation}
with the sum on the right-hand side being over all ion species.
We can then use \eref{APD7}, which is
\begin{equation}
\bm{b}\dg\left( \frac{T_e}{e}\ln n_e - \pot \right) = 0,
	\label{EParSmall}
\end{equation}
to find the electrostatic potential up to a flux function $\overbar{\pot}(\psi)$.

Finally, the flux-surface average of the vorticity equation gives us the equation that determines $\overbar{\pot}$:
\begin{equation}
\begin{split}
&\fav{\dv\left\{ \sum_s \frac{n_s m_s}{B} \left[ \pd{ }{t} + \upar\bm{b}\dg + \uE\dg\right]\left( \frac{\nabla p_{\perp s}}{n_s m_s \cycfreq[s]} + \frac{c \nabla \pot}{B}\right)\right\}}
\\
		&\,= \fav{\dv\left[\sum_s \bm{b}\dg\left( \frac{\upar \grad p_{\perp s}}{B\cycfreq}\right)\right]} + \fav{ \sum_s \frac{1}{2B\cycfreq} \left[ \left(\dv\bm{b}\right) \nabla^2 - \left( \grad\bm{b}\right) \bm{:} \grad\grad\right]\qpar} \\
		&\qquad+ \fav{\turbVort},
\end{split}
\label{fsVort}
\end{equation}
where 
\begin{equation}
\qpar = \wint \frac{m_s w_\perp^2}{2} w_\parallel f_s.
\end{equation}
with the definition of turbulent vorticity $\turbVort$ from \eref{turbVortDef}.

Let us pause and comment on these equations for a moment.  These equations live a double life. Firstly, they are transport equations, with fluxes given by averages over the pedestal turbulence. Secondly, they support sound waves and their relatives like the geodesic acoustic mode (GAM).
As these are on the same timescale, the interplay between poloidally-inhomogeneous turbulence and the sound waves that such inhomogeneity can excite may result in interesting physics.
Equation \eref{MagEq} gives the response of the magnetic field to this transport, and is in effect the constraint that the field-line-bending forces always balance pressure forces so no Alfv\'en waves can be excited. 

In this system, transport occurs on the same timescale that poloidal flows are damped by parallel viscosity. Thus, this system can potentially support Stringer spin-up~\citep{stringerSpinUp,bryanStringerSpin}. This mechanism has been thought to play a role in the L-H transition~\citep{adilStringerSpin}. Indeed, investigating the behaviour of the large poloidal flows that our equations can support will be a key point of future work.

\subsection{Pedestal Turbulence}
\label{SecPedTurb}
Finally, we come to the equations governing the turbulence in the pedestal. These are derived in \Apref{APPedTurb}, by systematically expanding the fluctuating part of \eref{vfp0} in accordance with the orderings of \Secref{SecForeshadow}. This derivation is standard, and closely resembles that of gyrokinetics. Thus, we leave the derivations to \Apref{APPedTurb}, and only present the results here.

The ions obey a simple kinetic equation:
\begin{equation}
\begin{split}
&\left(\pd{}{t} + \uE\dgR[i]{ }\right) h_i + \frac{c}{B} \pb{\gyroR{\turbPot}}{h_i} \\
		&\qquad\qquad= -\left(\pd{ }{t} + \uE \dgR[i]{ }\right){Z_i e\gyroR{\turbPot}} \pd{f_i}{\energy[i]} - \frac{c}{B} \bm{b}\times\nabla\gyroR{\turbPot}\dg f_i,
	\end{split}
	\label{ionGKE}
\end{equation}
where $\uE$ is the $\bm{E}\times\bm{B}$ velocity defined with the large-scale potential $\pot$, obtained from the equations in the previous section.
Similarly, $f_i$ is the large-scale distribution function. In this context these fields are static, and they can be handled exactly the same as the Maxwellian backgrounds in local gyrokinetic flux-tube codes.
Finally, $h_i$ is the non-adiabatic part of the ion distribution function
\begin{equation}
\turbf[i] = h_i + {Z_i e} \turbPot \pd{f_i}{\energy[i]} + \frac{Z_i e}{\MeanMagB} \left( \turbPot - \gyroR{\turbPot} \right) \pd{f_i}{\magmom[i]}.
\label{hDefI}
\end{equation}
This equation is easily seen to be the same as the slab-geometry gyrokinetic equation around a non-Maxwellian background, if one were to neglect ion-sound effects (c.f. \citep{frieman1982nge,flowtome1}).

The electrons also obey a gyrokinetic equation
\begin{equation}
\begin{split}
&\left(\pd{}{t} + \uE\dgR[e]{ }\right) h_e + w_\parallel \bm{b} \dg h_e + \frac{c}{B} \pb{\gyroR{\turbChi}}{h_e} \\
		&\qquad\qquad =  - \left(\pd{}{t} + \uE\dgR[e]{ }\right)\frac{e\gyroR{\turbChi}}{T_e} f_e - \collop[h_e] - \frac{c}{B} \bm{b}\times\nabla\gyroR{\turbChi}\dg f_e,
\end{split}
\label{electronGKE}
\end{equation}
where
\begin{equation}
\turbf[e] = h_e + \frac{e\turbPot}{T_e} f_e,
	\label{hDefE}
\end{equation}
as $f_e$ is Maxwellian, and 
\begin{equation}
\turbChi = \turbPot - \frac{w_\parallel}{c} \turbApar.
\label{ChiDef}
\end{equation}

These two equations are coupled with the following two field equations. Firstly, the quasineutrality condition,
\begin{equation}
-\frac{e \turbPot}{T_e} n_e +  \sum_{s=i} Z_i e \vint \left[ \turbPot \pd{f_i}{\energy[i]} + \left(\turbPot- \gyroR{\turbPot}\right) \pd{f_i}{\magmom[i]}\right] = \sum_s Z_s e \vint \gyror{h_s},
\label{turbQN}
\end{equation}
in which the sum on the left hand side is taken over all ion species.
Secondly, the parallel (to $\bm{b}$ but not $\bm{b}+\turbB$) component of Amp\`ere's law
\begin{equation}
\frac{c}{4\pi e} \nabla_\perp^2 \turbApar = \vint w_\parallel h_e,
	\label{turbAmpLaw}
\end{equation}
where the ions do not participate due to the high frequency of the fluctuations.
Perpendicular pressure balance obtains, and allows one to calculate turbulent fluctuations in the field strength, $\turbBpar = \bm{b}\cdot\turbB$, but these do not react back upon the turbulence, nor do they contribute to the ensuing transport. This is easy to see as the drift arising from $\turbBpar$ is $\Or(\gkeps\vth[i])$ whereas the $\bm{E}\times\bm{B}$ drift due to $\turbPot$ is $\Or(\gkeps^{1/2} \vth[i])$.

We should comment upon the geometry these equations should be solved in. The turbulence they describe occurs on spatial scales that are short compared to those of the pedestal profile. Hence, as in classic gyrokinetic simulations, flux-tubes with periodic perpendicular boundary conditions~\citep{beer1995field} are the appropriate setting for analysing our equations.
However, as the turbulence is such that the parallel correlation lengths are short compared even to the connection length, shear-periodic \textit{parallel} boundary conditions are also required. Similarly, all geometric quantities that vary on the $L_\parallel$ scale along the field line (poloidally) should be evaluated at one point and not vary through the simulation domain. Importantly, this means that trapped-particle effects are irrelevant for this turbulence, as only a small fraction of particles have a bounce point in any given $1/k_\parallel$ length of plasma. The only geometry of the field that is retained is the magnetic shear. This is easily seen by expanding the operator $\bm{b}\dg$ around some radial point $r_0$
\begin{equation}
\bm{b}\dg = \bm{b}(r=r_0) \dg + (r-r_0)\pd{\bm{b}}{r} \dg +\dots \approx \bm{b}\dg + (r-r_0)\pd{\bm{b}}{r} \cdot \nabla_\perp,
\end{equation}
where, as $r-r_0 \sim \nabla_\perp^{-1}$, the second term will be comparable to the first if $k_\parallel L_s \sim 1$. Referring back to our maximal estimates for the amount of magnetic shear in the pedestal,  we see that this ordering is marginally satisfied. Thus, magnetic shear must be retained.
These simplifications should render simulation of these equations simpler than current methods for simulating pedestal-relevant turbulence.

This set of equations is also interesting for several reasons. It is a strict superset of the KREHM equations of \citet{zoccoelectrons} which is a popular reduced model for studying reconnection. The equations also contain equations used to study microtearing physics in \citet{zocco2015mtm}. In the limit of weak magnetic shear, and weak background ion gradients, our equations reduce to a model that has been shown to support strong electromagnetic ETG turbulence, due to an inverse cascade process~\citep{alexInverse}. This strongly suggests our equations support the turbulent processes believed to be most important in the pedestal~\citep{hatch2016mtm,saarelma2013pedestalstab,hillesheim2016pedestal,hatch2017jetilwped}.

\subsection{Summary}
Our inter-ELM equations consist of two timescales. Firstly the slow-timescale equations, which comprise the following:

\begin{itemize}
\item \eref{eTransp1} and \eref{eTransp2} to determine the electron density $n_e$ and temperature $T_e$ respectively,
\item \eref{SlowIonKE} to determine $f_i$,
\item \eref{MagEq} which determines $A_\parallel$,
\item \eref{EParSmall}, which determines $\pot$ up to a flux function,
\item and \eref{fsVort} that determines that flux function.
\end{itemize}

All of these equations are meant to be solved on a global annular domain containing the entire pedestal. The detailed boundary conditions will be discussed in \Secref{SecBoundaries}.

In some of these equations, there are terms that arise from averages over the turbulence. The turbulence is  governed by the following equations:
\begin{itemize}
\item the kinetic equations, \eref{ionGKE} for $h_i$ and \eref{electronGKE} for $h_e$,
\item the quasineutrality condition \eref{turbQN} for $\turbPot$,
\item and parallel Amp\`ere's law \eref{turbAmpLaw} for $\turbApar$.
\end{itemize}
There are also the relations \eref{hDefI}, \eref{hDefE}, and \eref{ChiDef} that relate these fields to the fields that arise in the flux terms in the inter-ELM equations.
As usual for coupling turbulent fluxes to a transport code, these equations should be run to a steady state and appropriately averaged to give the fluxes.

\section{Betwixt and Between: Boundary Conditions, Matching, and Existing Models}
\label{SecBoundaries}

In this section, we tackle two remaining points regarding the systems of equations derived above. Firstly, how they interact, and how they couple both to core and Scrape Off Layer models. Then, secondly, how 
they compare to existing strategies for modelling this region of the plasma. 

\subsection{The Integrated Approach: Coupling and Boundary Conditions}
\label{SecBCs}
Having derived a full multiscale approach to the pedestal, the question naturally arises of how these equations should be coupled to each other, and to other parts of the plasma.
We have a framework for physics on three disparate timescales. The fastest is the timescale of the individual turbulent fluctuations. Then comes the timescale on which ELM filaments erupt. Finally, the timescale on which transport occurs, which is the same as the acoustic timescale.

The outline of the envisioned implementation, reminiscent of the operation of multiscale core transport codes, is as follows.
Given an initial pedestal profile, one tests for stability in the ELM equations. Due to the meta-stable nature of erupting ELM filaments~\citep{CowleyExplosiveProcRoySoc,ham2016nonlinear}, a purely linear solver may be insufficient to evaluate the stability of the pedestal to ELMs.\footnote{In addition, on account of the inclusion of electron Landau resonances and diamagnetic effects, there will not be an energy principle for our ELM equations. Thus, any linear stability calculation would have to perform a full initial-value solution for a finite time anyway.}
A more robust approach is to initialise the ELM equations with the initial profile and, in addition, some specified amount of noise. One then evolves the fully nonlinear equations for a short time. Either nothing happens, if the pedestal is suitably stable, or the filaments erupt and a new equilibrium is rapidly reached. 
Once a situation is reached where no more evolution occurs on the fast timescale, the state is such that it is a valid initial condition for the inter-ELM equations.

The Inter-ELM equations are then used to evolve the profiles on the long timescale, with fluxes given by the pedestal turbulence.
We envision these equations being solved in a (topologically) annular region, with inner and outer surfaces corresponding to flux surfaces.
The boundary conditions at the inner radial location is given by matching to multiscale gyrokinetics. The core is solved as in \citet{barnes2009trinity}, with the boundary values for the transport equations being taken from our Inter-ELM pedestal profile. That profile is itself solved for with a fixed-flux boundary condition obtained from the outermost radial grid point of the gyrokinetic transport calculation.
In general, this indicates that the core transport timestep and the Inter-ELM timestep must be taken together. However, for a steady-state situation one could envision solving the Inter-ELM equations with a fixed flux given by the total heating and fuelling sources before then solving the core with a fixed value for the pedestal.
The outer radial boundary condition is given, in principle, by matching to a scrape-off-layer model just inside the separatrix. We discuss the difficulties and possibilities of such matching in \Secref{SecOuterMatch}. Absent an exact model to match to, we take the same \textit{ad hoc} approach as \citet{barnes2009trinity} and suggest fixing values for top of the scrape-off layer at the separatrix.
The pedestal turbulence, which gives rise to the fluxes, is solved locally in slab-like flux tubes, as described in detail at the end of \Secref{SecPedTurb}.

Periodically during this evolution, stability in the ELM equations is checked for. If at any point the ELM equations are unstable, then one returns to the start of this procedure and evolves on the short timescale until a new equilibrium is reached. In principle this is checked every timestep of the Inter-ELM equations (just as MHD stability should be checked every timestep of a core transport code), but hopefully heuristics can be developed to ameliorate this cost.\footnote{As an example, one simple approach would be to require a certain change in global pedestal parameters (height and width) before initialising a stability calculation. Speculatively executing both the next slow timestep and the stability calculation in parallel, throwing away whichever turns out to be irrelevant, might also save wall-clock time.} It is easy to see how this cyclical procedure would produce an ELM cycle.

Ideally, this framework would resolve the problematic issue of boundary conditions in core multiscale codes -- the solution to the core transport equations is sensitive to the fixed boundary condition at the top of the pedestal. One might hope that sensitivity studies of the equations in this paper would demonstrate an insensitivity of the pedestal top temperature to conditions at the separatrix. Of course, this is mere speculation, and one may wonder, pessimistically, if sensitivity to boundary values continues all the way to a material surface. 

Now, the fact that multiscale gyrokinetics has a subsidiary limit that provides a large enough heat flux to appropriately source the Inter-ELM equations is not obvious.
Indeed, it is not obvious that our inner boundary is even consistent with multiscale gyrokinetics. To this end we will show explicitly in the next section that these two theories can indeed be matched on to each other at the pedestal top.

\subsection{Matching at the Pedestal Top}
\label{SecGKMatch}
In this section we show that our pedestal model can be smoothly matched on to multiscale gyrokinetics. More precisely, we mean that there is a subsidiary expansion of each of these systems that results in the same set of equations. This implies that there is potentially a region in the plasma that would be well described by both multiscale gyrokinetics and the model contained in this paper.
Suggestively, we will designate this location as the top of the pedestal.

In this section, we will only detail the orderings of the subsidiary expansions. Detailed calculations showing that these orderings do indeed result in the same set of equations are performed in \Apref{APGKMatch}.
We begin with the expansion of multiscale gyrokinetics. 
This should be seen as an expansion in nearness to the pedestal. As the pedestal top is approached from the core, steepening density gradients and decreasing $\eta_i$ conspire to stabilise the ITG mode. Thus, we are looking at an expansion that should result in strongly-driven microtearing and ETG turbulence.

There are two main expansions we will need to do.
Firstly, we will need an expansion in small Mach number to match the rapidly-rotating core onto the subsonic pedestal.
Secondly, in order to begin to match the pedestal, we will need a subsidiary expansion in both the mass-ratio and in $\beta$. This will mainly affect the equations for the fluctuations, and 
closely parallels the derivation in \citep{zoccoelectrons}.

We do this as two subsidiary expansions, so the first expansion is that of small Mach number $M$. We will use a na\"ive low-Mach-number expansion, where $M \ll 1$ but the scale length of the flow is the same as the perpendicular scale length of all other mean quantities. Such an expansion simply results in the zero-flow limit of gyrokinetics.
The resulting equations are simply the equations of Section 11 of \citet{flowtome1} with any remaining terms involving the equilibrium flow neglected. This is the same as the system of equations presented in \citet{sugama1996tpa}, except in a notation that is closer to the one used in this paper.
 It is this set of equations we will now expand in $\beta$ and $m_e / m_i$.

To formalise our second subsidiary expansion, we introduce a small parameter $\apSmall$, defined by
\begin{equation}
\xi \sim \sqrt{\beta} \sim \sqrt{\frac{m_e}{m_i}} \ll 1.
\end{equation}
We then order
\begin{equation}
\frac{\delta f_s}{f_s} \sim  \frac{e\delta \pot}{T_e} \sim \frac{\realGKeps}{\apSmall},
\end{equation}
where $\realGKeps = k_\parallel / k_\perp$ is the small parameter of gyrokinetics. Other than this rebranding, we will use the notation of \citep{flowtome1} when discussing multiscale gyrokinetics.
We retain
\begin{equation}
k_\perp \rho_{i} \sim 1,
\end{equation}
from gyrokinetics and due to our ordering for $\beta_e$ we keep the electron skin depth scale,
\begin{equation}
k_\perp d_e \sim 1,
\end{equation}
as well.
With this ordering for $k_\perp$, the frequencies become 
\begin{equation}
\omega \sim \frac{cT_e}{eB} \bm{b}\times\nabla \left(\frac{e\delta \pot}{T_e}\right) \cdot \nabla \delta f_s \sim \frac{\realGKeps}{\apSmall} \cycfreq \sim k_\parallel \vth[e].
\end{equation}
As in \citep{zoccoelectrons} this expansion in $\beta$ reduces the relative importance of some electromagnetic effects.
Thus, we have
\begin{equation}
\frac{\delta B_{\perp}}{B} \sim \gkeps \qquad\mathrm{and}\qquad \frac{\delta B_\parallel}{B} \sim \xi\gkeps.
\end{equation}

To drive this increase in turbulent amplitude, we assume that the gradients in $F_{0s}$ are increased
\begin{equation}
\frac{\nabla F_{0s}}{f_s} \sim \frac{1}{a \xi^2},
\end{equation}
with $a$ the long equilibrium length scale, assumed to be comparable to the parallel connection length $q R$.
However, the parallel length scale of the fluctuations is now
\begin{equation}
k_\parallel a \sim \frac{\omega a}{\vth[e]} \sim \frac{1}{\xi}.
\end{equation}

Note that this scaling is consistent with turbulence that has its outer scale fixed (independent of drive) around $\rho_i$ such as might follow from an inverse cascade from electron-scales that is broken by ion gyroscale physics~\citep{alexInverse}. As the drive is increased, instead of the outer scale increasing, the frequency increases and the parallel length scale decreases due to causality.
Thus, this ordering is not consistent with a na\"ive scaling theory of critically-balanced ETG turbulence which might be expected to follow the orderings of \citep{barnes2011critically}.

With our ordering, the transport timescale of such turbulence is now
\begin{equation}
\tau_E^{-1} \sim \frac{\omega}{k_\perp^2} \frac{\nabla F_{0s}}{F_{0s}} \sim \frac{\omega}{\xi^2} \realGKeps^2 \sim \frac{\vth[i]}{qR} \frac{\realGKeps^2}{\xi^4} \sim \frac{\realGKeps^3\cycfreq[i]}{\xi^4}
\end{equation}

As we are allowing the various scales of the equilibrium to be ordered differently (unlike normal gyrokinetics), we must consider how to handle the magnetic shear length scale $L_s$. As our pedestal allows for strong shear, we take the ordering
\begin{equation}
L_s \sim \xi a,
\end{equation}
such that $k_\parallel L_s \sim 1$, which is consistent with electromagnetic microtearing turbulence.

Revisiting the collisionality estimate from above, $\nu_{ee} \sim \vth[e] / qR$, we see that our turbulence should be collisionless if our pedestal is trans-collisional. Hence, we will assume that the turbulence self-generates phase-space structure sufficient that the collision operator is retained in the resultant gyrokinetic equation.~\footnote{If semi-collisional or truly collisional instabilities are to be studied, they must be investigated with a collisional pedestal as the background. This is a trivial large-collisionality limit of the equations presented above.}
This assumption of steep gradients in velocity space is needed in order that dissipation can continue to balance the turbulent fluxes in the Free-Energy conservation equation of \citep{flowtome1}.

Applying this ordering to the low-Mach-number equations of gyrokinetics results in the set of equations detailed in \Apref{APGKMatch}. 
We thus turn to the subsidiary ordering of our inter-ELM equations. We do not need to revisit our ELM equations as gyrokinetics assumes that the plasma is MHD stable, and so to match on to it we make the same assumption.

The subsidiary expansion we engage in should be thought of as approaching the top of the pedestal from inside the pedestal region. Perpendicular gradients are shallower than they are in the pedestal. ITG-driven turbulence is still suppressed, but the microtearing and ETG turbulence is not as strong as it is in the pedestal proper. 
To quantify this, we introduce a second subsidiary parameter $\zeta$ to quantify our expansion of the inter-ELM equations. It is defined by
\begin{equation}
\frac{\turbf}{f} \sim \frac{e\turbPot}{T_e} \sim \zeta\sqrt{\gkeps}.
\end{equation}
This lower-amplitude turbulence naturally leads to a longer $k_\parallel$ for the fluctuations, obtained from a critical-balance estimate
\begin{equation}
k_\parallel \vth[e] \sim \frac{c}{B}\bm{b}\times\grad\turbPot \dg \sim \zeta \sqrt{\gkeps} \cycfreq[i],
\end{equation}
giving
\begin{equation}
k_\parallel \sim \frac{\zeta}{\sqrt{L_\perp L_\parallel}}.
\end{equation}
Consistent with this reduction in turbulent amplitude is a reduction in the driving gradients (a linear scaling of $\turbf/f_s$ with driving gradient is in accord with the results of \citep{barnes2011critically})
\begin{equation}
\nabla_\perp f_s \sim \zeta \frac{f_s}{L_\perp}.
\end{equation}
We will order the time derivatives of the mean quantities with the modified turbulent transport time,
\begin{equation}
\pd{f_s}{t} \sim \zeta^3 \frac{\vth[i]}{L_\parallel} f_s,
\end{equation}
but we shall keep the collision time comparable to the parallel sound time.

The mean electrostatic potential is ordered so as to allow us to match up with a low-Mach-number expansion of gyrokinetics.
We fix the amplitude of the potential at $T_e/e$ as we head towards the interior of the tokamak, but keep the scale length that of the other mean quantities:
\begin{equation}
\nabla_\perp \pot \sim \zeta\frac{\pot}{L_\perp} \qquad\mathrm{ and }\qquad \frac{e\pot}{T_e} \sim 1.
\end{equation}
This means that the shearing rate drops with decreasing $\zeta$ and so the mean flow is eliminated from our equations.

This set of orderings reproduces exactly the same set of equations as the expansion of gyrokinetics detailed above. Hence, we can confidently say that there is a physically plausible regime which is
described both by gyrokinetics and by our pedestal model, and acts as a bridge between them.
Operationally, we expect this region to be located at the top of a pedestal, but one could also envision using our model to embed a transport barrier region inside core gyrokinetics.

Having shown that we match on to the best tested model of core turbulence, in the next section we will discuss how our model matches on to Scrape-Off-Layer models which are much less developed.

\subsection{Matching on to the Scrape Off Layer}

To leave the plasma and reach such a surface, one can imagine coupling the above pedestal to a collisional Scrape Off Layer model such as Catto-Simakov or \verb#GBS#, where the boundary between closed and open field lines is completely contained in the SOL model. If the plasma is still hot enough to be weakly collisional at the separatrix then such a coupling may not be consistent, and further work is required. Exciting models for arbitrary collisionality Scrape Off Layers have recently been proposed~\citep{jorge2017gbscoll}, and matching to such models should be addressed in the future.

As yet, such an explicit matching (either collisional or collisionless) has not been performed. However, even without the explicit forms of the equations that are to be matched, we know that they must obey certain 
constraints. We know that they must conserve energy and particles, and as such we can envisage using the fluxes leaving the pedestal (as simulated by our equations) as boundary conditions for a scrape-off layer model. One could also, perhaps, use the general spectrum of fluctuations in our model to initialise some noise in a scrape-off-layer code. This procedure would, of course, be fundamentally inconsistent. Consistency requires not only a smooth matching of the fluxes at the boundary, but also a matching of the mechanisms that carry that flux. Rigorous subsidiary expansions would provide a finite-width region in which both models were valid, and we could be confident in matching across this region. We leave the detailed derivation of such expansions to future work.

\label{SecOuterMatch}

\subsection{Existing Models}
\label{SecExisting}
Individual parts of our model are directly comparable to some parts of existing models, and we have highlighted these both in the discussion of the ELM vorticity equation in \Secref{SecELMEquations} and in \Secref{SecPedTurb} where we discuss pedestal turbulence. Thus, in this section, we focus on comparing our model as a whole to other approaches for simulation and modelling of the pedestal region of a tokamak.

The most successful empirical model of the pedestal is the \verb#EPED# model~\citep{snyder2002eped,snyder2011eped}. This combines a collisional, long-wavelength model for peeling-ballooning modes~\citep{wilson2002elite}, with an assumption that the inter-ELM turbulence is given by Kinetic-Ballooning Modes (KBM), modelled heuristically by the infinite-$n$ ballooning stability limit. Our model improves upon this by including weakly-collisional kinetic effects including electron landau damping, trapped electron physics, and self-consistent diamagnetic stabilisation into the time-dependent ELM dynamics. We also have a self-consistent model of the underlying pedestal turbulence. Indeed, the assumption that the turbulence is strictly KBM in nature has recently come under scrutiny~\citep{hatch2017jetilwped} as detailed gyrokinetic analysis lends more support to the turbulence being MTM or ETG in nature.
Two-fluid simulations of ELMs with \verb#BOUT++#~\citep{xu2011boutELM}, which have been included in \verb#EPED#, can capture some of the non-ideal effects of our model, but not the electron kinetic effects. In addition, our model makes precise the potential sources of the effects of turbulence upon the ELM dynamics.

More generally, two-fluid or reduced Braginskii simulations will not capture the weakly-collisional physics that may be important in a hot pedestal.
The most complete fluid equations are those of \verb#GBS#~\citep{ricci2012gbs} or Catto-Simakov~\citep{cattosimakovedge}. The former of these has almost exclusively been used to study open-field-line regions, but the latter was formulated with the explicit intent of handling closed-field-line regions in pedestals. The fluid nature of the models means that they cannot capture kinetic effects that may affect ELM stability. In addition, these models do not consistently separate the sound and Alfv\'enic timescales, despite the low-$\beta$ nature of their models. Their ordering schemes do not touch on the possibility of small-scale fluctuations existing within the pedestal, but could be extended to include the turbulence model we have presented above. Indeed, the Catto-Simakov model, with turbulence and separating the two equilibrium timescales is the collisional limit of the equations we have presented. 

The final approach that is considered for pedestal modelling is global full-$f$ gyrokinetics~\citep{chang2006xgcped,churchill2017xgcedge}. This approach promises to include all the physics, at all scales, consistently. The potential pitfalls of this approach are twofold. Firstly, the numerical requirements are extraordinary, requiring all parts of the simulation to resolve the fastest and smallest fluctuations. This may render predictive parameter scans prohibitively expensive. Secondly, by incorporating all temporal and spatial scales on the same footing, it becomes incredibly hard to extract the physics of the interaction of these scales. Our model should be much less numerically challenging, especially as the turbulence can be parallelised over the background parameters in an approach reminiscent of current core transport codes~\citep{barnes2009trinity}. Again, because we have decoupled the filamentary dynamics from the slow sound-time dynamics and the fluctuations, we can examine each piece of the puzzle separately.

\section{Summary \& Conclusions}
\label{SecSummary}
In this paper we have derived a self-consistent first-principles model for a pedestal. 

The first part of the model is a trans-collisional ELM model, given by \eref{eKE-1}, \eref{ELMAparEq}, \eref{iKE-1}, \eref{vortELM}, and closed with \eref{elmBdef} and \eref{ampLawELM}.
This model contains the eruption of filamentary structures driven both by magnetic buoyancy forces and launched by current-driven instabilities. It is fully nonlinear and capable of dynamically evolving a filamentary structure from one equilibrium to another.

These equilibria provide the backdrop for the next part of the model. The evolution of the inter-ELM profiles. This is given by \eref{eTransp1}, \eref{eTransp2}, \eref{SlowIonKE}, \eref{MagEq}, \eref{EParSmall}, and \eref{fsVort}.
This component of the framework contains the physics of sound waves, including GAMs which are often seen in the edge, poloidal spin-up, and even systematically includes turbulent transport.

These equations are, in turn, closed by a set of equations governing the turbulence that generates that transport.
These are, explicitly, the set \eref{ionGKE}--\eref{turbAmpLaw}.
This set contains the physics of strongly-driven microtearing and ETG turbulence.

Future work must include exploring the equilibria of this system, with special focus on the stability of zero-flow solutions to spontaneous spin-up. In addition, the scaling of pedestal turbulence with drive parameters is a high priority -- if pedestal turbulence exhibits the $(R/L_T)^3$ scaling of core turbulence, then this may permit the formation of boundary layers that govern the size of ELM filaments. 


 Support for I.G.A. was provided by the Princeton Center for
Theoretical Science, for the initial work, and later the Framework grant for Strategic Energy Research (Dnr. 2014-5392) from Vetenskapsr{\aa}det. 
Discussions with G. Hammett of Princeton Plasma Physics Lab led to many of the key insights that made this work possible. 
The authors also thank P. Catto, F. Parra, P. Ricci, F. Militello, J. Connor, P. Helander, and T F\"ul\"op for fascinating and productive discussions. 
These discussions were facilitated by the generous hospitality and material support provided by the Wolfgang Pauli Institute in Vienna and Chalmers University of Technology.
This work was conducted in part within the framework of the EUROfusion Consortium and has received funding from the Euratom research and training programme 2014-2018 under grant agreement No. 633053. The views and opinions expressed herein do not necessarily reflect those of the European Commission.

\appendix
\section{Time Derivatives of the Gyrokinetic Variables}
\label{APDdt}
In this appendix we take the time derivatives of the variables used in the body of the paper.
We start with the gyrocentre position,
\begin{equation}
\dot{\bm{R}}_s = \frac{d}{dt} \left( \bm{r} + \frac{\bm{w}\times\bm{b}}{\cycfreq} \right),
\end{equation}
where the derivative is taken along unperturbed orbits.
Noting that all effects of magnetic inhomogeneity result in terms of order $\vth \rho_s / L_\parallel$, we can neglect them.
Thus this becomes
\begin{equation}
\dot{\bm{R}}_s = w_\parallel \bm{b} + \uE + \Or\left( \frac{\rho_s}{L_\parallel} \vth\right).
\end{equation}
In the fast ELM ordering, the component of $\uE$ due to $A_\parallel$ is $\Or(\gkeps^{3/2} \vth[i])$. In the inter-ELM ordering it is suppressed by another factor of $\gkeps^{1/2}$. Hence we can neglect it.
Gyroaveraging our formula is trivial, and results in
\begin{equation}
\label{APRDot}
\dot{\bm{R}}_s = w_\parallel \bhat(\bm{R}_s) + \frac{c}{B} \bhat\times\nabla\pot + \Or(\gkeps^{3/2} \vth[i]),
\end{equation}
as required.

The energy is also simple to handle. Working from the definition we immediately see that
\begin{equation}
\denergy = Z_s e \bm{E}\cdot\bm{b} - m_s \bm{w}\cdot \frac{d}{dt} \uE
\end{equation}
Gyroaveraging gives
\begin{equation}
\gyroR{\denergy} = -Z_s e \left( \bm{b}\dg \pot + \frac{1}{c} \pd{A_\parallel}{t}\right) + \Or\left( \sqrt{\gkeps} \frac{\vth[i]}{L_\parallel} T_e\right)
\label{APeDot}
\end{equation}

The magnetic moment is defined to be the exact adiabatic invariant
\begin{equation}
\begin{split}
\magmom = &\frac{m_s w_\perp^2}{2\MeanMagB} - 
\frac{m_s}{\MeanMagB} \left[\bm{w}_\perp\cdot\vdrift[s] - \frac{1}{\cycfreq[s]}\tensor{W}\bm{:}\left(w_\parallel \nabla\Meanb + \nabla\uE\right)\right] \\
&\quad+ \frac{w_\perp^2}{2\MeanMagB \cycfreq[s] } \left( w_\parallel\Meanb\cdot\curl\Meanb + \curl\uE\right) + \Or\left(\gkeps^2 \frac{m_s\vth^2}{2B}\right) .
\end{split}
\label{magmomWAP}
\end{equation}
where $\tensor{W}$ is given by
\begin{equation}
\tensor{W} = - \frac{1}{4} \left( \bm{w}_\perp\bm{w}_\perp\times\bm{b} + \bm{w}_\perp\times\bm{b} \bm{w}_\perp\right).
\end{equation}
This tensor is useful because it satisfies
\begin{equation}
\bm{w}_\perp\bm{w}_\perp = \frac{w_\perp^2}{2}\left(\idmat - \bm{b}\bm{b} \right) + \pd{\tensor{W}}{\gyr}.
\label{Wident}
\end{equation}

Finally, for $\dot{\gyr}$ we take the time derivative of
\begin{equation}
\bm{w} = w_\parallel \bm{b} + w_\perp \left(\cos \gyr\,\bm{e}_2 - \sin\gyr\,\bm{e}_1 \right),
\end{equation}
and find
\begin{equation}
\begin{split}
\frac{Z_se}{m_s} \left(E_\parallel\bm{b} + \frac{1}{c} \bm{w}\times\bm{B}\right) =& \frac{d w_\parallel}{dt} \bm{b} + w_\parallel\left( \pd{\bm{b}}{t} +  \bm{w}\dg\bm{b} \right) + \frac{d w_\perp}{dt} \frac{\bm{w}_\perp}{w_\perp} 
\\ &- \dot{\gyr} w_\perp \left( \sin \gyr\, \bm{e}_2 + \cos \gyr\, \bm{e}_1\right) + w_\perp \left(\cos \gyr\,\dot{\bm{e}}_2 - \sin\gyr\,\dot{\bm{e}}_1 \right) 
	\end{split}
\end{equation}
Taking the inner product with $\bm{w}\times\bm{b} = -w_\perp \left(\sin \gyr\, \bm{e}_2 + \cos \gyr\, \bm{e}_1\right)$ gives
\begin{equation}
w_\perp^2 \cycfreq = w_\perp^2 \dot{\gyr} + \Or\left( \vth^2 \frac{L_\perp}{L_s} \gkeps^{1/2} \cycfreq\right),
\end{equation}
where $L_s$ is the magnetic shear length. Gyroaveraging gives the required answer
\begin{equation}
\dot{\gyr} = \cycfreq(\bm{R}_s) + \Or(\gkeps \cycfreq) =  \cycfreq + \Or(\gkeps^{1/2} \cycfreq).
\end{equation}

In these variables, gyrophase derivatives have the property that
\begin{equation}
\left.\pd{ }{\gyr}\right|_{\bm{R}_s,\energy,\magmom} = \left.\pd{ }{\gyr}\right|_{\bm{R}_s,w_\parallel,w_\perp} + \left.\pd{\magmom}{\gyr}\right|_{\bm{R}_s,w_\parallel,w_\perp} \left.\pd{ }{\magmom}\right|_{\bm{R}_s,\energy,\gyr} .
\end{equation}

We will also need the leading-order gyrophase dependent pieces of the time derivatives of $\bm{R}_s$ and $\energy$, which are
\begin{equation}
\dot{\bm{R}}_s - \gyroR{\dot{\bm{R}}_{s}} = \frac{c}{B} \bm{b}\times\nabla \left(\frac{\bm{b} \times \bm{w}}{\cycfreq} \dg \pot \right) + \Or(\gkeps^{3/2} \vth[i]),
	\label{GyrRdot}
\end{equation}
and
\begin{equation}
\denergy - \gyroR{\denergy} = - m_s \bm{w}\cdot\left[ \pd{ }{t} + \left( \uE + \bm{w}\right)\dg\right] \uE +  \Or(\sqrt{\gkeps} \frac{\vth[i]}{L_\parallel} T_s).
\label{GyrEdot}
\end{equation}
respectively.

In these variables the ion distribution function is given by
\begin{equation}
f_i = f_i(\bm{R}_s,\energy[i],\magmom[i],t) + \tilde{f}_i,
\end{equation}
with $\tilde{f}_i \sim \gkeps^{3/2} f_i$.

\subsection{Kinetic Equations}
In the $(\bm{R}_s,\energy,\magmom,\gyr,t)$ variables we can write the averaged kinetic equation as
\begin{equation}
\left(\pd{ }{t} + \dot{\bm{R}}_s \dgR{ } + \denergy \pd{ }{\energy} + \dot{\gyr}\pd{ }{\gyr}\right) f_s + \ensav{\daccel \cdot\pd{f_s}{\bm{w}}} = \collop[f_s] + \source,
\label{ur-KE-AP}
\end{equation}
in which the fluctuating acceleration is 
\begin{equation}
\daccel = \frac{Z_s}{m_s} \left( \turbE + \frac{1}{c}\bm{v}\times\turbB\right).
\label{daccelDef}
\end{equation}
In all situations, the leading order term in this equation is the gyrophase derivative, from which we learn that the distribution functions are gyrophase independent to some order in $\gkeps$. 
Hence, all the kinetic equations for the mean quantities in this paper are derived by gyroaveraging \eref{ur-KE-AP} and retaining the leading order terms.

We will also need an equation for the gyrophase dependence of the ion distribution function (the gyrophase dependence of the electrons is too small to matter). 
Writing the ion distribution function as
\begin{equation}
f_i = f_i(\bm{R}_s,\energy,\magmom,t) + \tilde{f}_i,
\end{equation}
where all the gyrophase dependence is in $\tilde{f}_i$ we have
\begin{equation}
\left.\pd{f_i}{\gyr}\right|_{\bm{r},w_\parallel,w_\perp} = -\frac{\bm{w}_\perp}{\cycfreq[i]} \left.\dgR[i]{f_i}\right|_{\energy,\magmom} + \pd{\magmom}{\gyr} \left.\pd{f_i}{\magmom}\right|_{\bm{R}_i,\energy[i]} + \left.\pd{\tilde{f}_i}{\gyr}\right|_{\bm{R}_i,w_\parallel,w_\perp} + \Or(\gkeps^2 f_i),
	\label{ddGyrAP}
\end{equation}
with
\begin{equation}
\cycfreq[i] \pd{\tilde{f}_i}{\gyr} = -\left(\dot{\bm{R}}_i - \gyroR{\dot{\bm{R}}_{i}}\right)\dgR[i]{f_i} - \left(\denergy[i] - \gyroR{\denergy[i]}\right) \pd{f_i}{\energy[i]} - \ensav{\daccel[i] \cdot \pd{\turbf[i]}{\bm{w}}}.
\end{equation}

In addition, using our orderings for the inter-ELM turbulence, we can write the fluctuating acceleration as
\begin{equation}
\daccel = -\frac{Z_s}{m_s} \grad\left( \turbPot - \frac{w_\parallel}{c} \turbApar \right) - \frac{Z_s}{m_s c} \bm{b} \bm{w}_\perp \dg \turbApar + \Or\left( {\gkeps} \vth[i]\cycfreq[i] \frac{m_i}{m_s} \right),
	\label{daccelAP}
\end{equation}
which will be used in later appendices.

\section{Derivation of \eref{vortELM}}
\label{ApVortELM}
In this appendix we derive the vorticity equation to a high enough order to serve for both the ELM and inter-ELM equations.
To begin our derivation we multiply the averaged kinetic equation by $m_s \bm{v}$ and integrate over $\bm{v}$ to obtain
\begin{equation}
\begin{split}
\pd{ }{t} \sum_s m_s n_s &\bm{u}_s + \dv\left( \sum_s \vint m_s\bm{v}\bm{v} f_s \right) \\
		&=  \frac{1}{c} \bm{j}\times\bm{B} + \frac{1}{4\pi}\dv\ensav{ \turbB\turbB - \frac{1}{2} \turbB^2 \idmat},
	\end{split}
	\label{tmpVort0}
\end{equation}
where $n_s \bm{u}_s = \vint \bm{v}f_s$. In deriving this equation, we have used quasineutrality to eliminate all terms involving electric fields and the conservation properties of the collision operator.
Our next step is to write this in terms of $\bm{w}$. Defining
\begin{equation}
n_s \bm{U}_s = n_s \bm{u}_s - n_s \uE = \wint \bm{w} f_s,
\end{equation}
 we have
\begin{equation}
\begin{split}
\pd{ }{t} &\left( n_s m_s \bm{u}_s \right) + \dv\left( \vint m_s \bm{v}\bm{v} f_s \right) = \\
		&n_s m_s \left( \pd{ }{t} + \uE \dg\right) \bm{u}_s + \nabla p_{\perp s} + \dv\left[ \left( p_{\parallel s} - p_{\perp s} \right) \bm{b}\bm{b}\right] \\
		&\qquad+ \dv\pitens - \bm{U}_s\dv\left( n_s m_s \bm{U}_s \right) + n_s m_s \bm{U}_s \dg \uE,
	\end{split}
	\label{tmpLHS0}
\end{equation}
where the new viscous stress tensor is
\begin{equation}
\dv\pitens = \wint m_s \left[ \bm{w}\bm{w} - \left(w_\parallel^2 - \frac{1}{2} w_\perp^2\right) \bm{b}\bm{b} - \frac{1}{2} w_\perp^2 \idmat\right]  f_s.
\label{APpiDef}
\end{equation}
It will be convenient to rearrange \eref{tmpLHS0} slightly into
\begin{equation}
\begin{split}
\pd{ }{t} &\left( n_s m_s \bm{u}_s \right) + \dv\left( \vint m_s \bm{v}\bm{v} f_s \right) = \\
		&n_s m_s \left( \pd{ }{t} + \upar \bm{b}\dg + \uE \dg\right) \bm{u}_s + \nabla p_{\perp s} + \dv\left[ \left( p_{\parallel s} - p_{\perp s} \right) \bm{b}\bm{b}\right] + \bm{X}_s \\
	\end{split}
	\label{tmpLHS1}
\end{equation}
with the abbreviation that
\begin{equation}
\bm{X}_s = \dv\pitens - n_s m_s \upar\bm{b}\dg \bm{u}_s - \bm{U}_s\dv\left( n_s m_s \bm{U}_s \right) + n_s m_s \bm{U}_s \dg \uE.
\label{Xdef}
\end{equation}
We now proceed to partially derive our vorticity equation.
Substituting \eref{tmpLHS1} into \eref{tmpVort0} we have
\begin{equation}
\begin{split}
\sum_s &n_s m_s \left( \pd{ }{t} + \upar \bm{b}\dg + \uE \dg\right) \bm{u}_s + \nabla p_{\perp s} + \dv\left[ \left( p_{\parallel} - p_{\perp} \right) \bm{b}\bm{b}\right] + \bm{X} \\
		&=  \frac{1}{c} \bm{j}\times\bm{B} + \frac{1}{4\pi}\dv\ensav{ \turbB\turbB - \frac{1}{2} \turbB^2 \idmat},
	\end{split}
	\label{tmpVort1}
\end{equation}
again using the natural notation that $p_\parallel = \sum_s p_{\parallel s}$, etc.
Now, taking $\dv\left[\bm{b}\times(\ref{tmpVort1})/B\right]$, we obtain
\begin{equation}
\begin{split}
\dv&\left[ \sum_s \frac{n_s m_s}{B} \left( \pd{ }{t} + \upar \bm{b}\dg + \uE \dg\right) {\bm{b}\times\bm{u}_s} - \sum_s \frac{n_s m_s}{B} \upar \bm{b}\dg\bm{b} \times\bm{u}_s\right]\\
&= \dv\left[ - \frac{j_\parallel \bm{b}}{c}\left( 1- 4\pi \frac{ p_\parallel - p_\perp}{B^2}\right) - p_\perp \curl\left(\frac{\bm{b}}{B}\right) - \frac{\curl \bm{b}}{B} \left(p_{\parallel} - p_\perp\right) \right] \\
  &\qquad+ \dv\left( \frac{\bm{b}}{B} \times \bm{X} \right),
	\end{split}
	\label{tmpVort2}
\end{equation}
in which we have used \eref{apCurv1}, \eref{apCurv2}, and \eref{apCurv3} to rewrite the first line of the right-hand side.
The unadorned vector $\bm{X}$ is given by
\begin{equation}
\bm{X} = \sum_s \bm{X}_s + \frac{1}{4\pi}\dv\ensav{ \turbB\turbB - \frac{1}{2} \turbB^2 \idmat}.
\label{XRealDef}
\end{equation}

We now use the result \eref{apFlowForm} to find $\bm{U}_s$:
\begin{equation}
n_s m_s \bm{U}_s = n_s m_s \upar \bm{b} + \frac{1}{\cycfreq}\bm{b}\times\nabla p_{\perp s},
\label{pecFlowAp}
\end{equation}
which allows us to simplify $\bm{X}_s$:
\begin{equation}
\begin{split}
\bm{X}_s =& \dv\pitens - n_s m_s \upar\bm{b}\dg \left( \upar \bm{b} + \uD\right) \\
		&- \left( \upar \bm{b} + \uD\right)\dv\left( n_s m_s \upar \bm{b} + n_s m_s \uD \right) + n_s m_s \uD \dg \uE,
\label{Xtmp1}
\end{split}
\end{equation}
with the notation $\uD = \bm{b}\times\nabla p_{\perp s} / n_s m_s \cycfreq$ for the diamagnetic velocity of species $s$.

\subsection{Manipulations Leading to \eref{tmpVort2}}
\label{derivCurv}
We now provide the results alluded to above.
Firstly,
\begin{equation}
\dv\left[ \frac{\bm{b}}{B}\times\left( \bm{j}\times\bm{B} \right) \right] = -\dv\left( j_\parallel \bm{b} \right),
\label{apCurv1}
\end{equation}
in which we have used quasineutrality in the form $\dv\bm{j} = 0$.
For our next trick, we see that
\begin{equation}
\dv\left( \frac{\bm{b}}{B} \times \nabla p_\perp \right) = \curl\left( \frac{\bm{b}}{B} \right) \cdot \nabla p_\perp = \dv\left[ p_\perp \curl\left(  \frac{\bm{b}}{B} \right)\right].
\label{apCurv2}
\end{equation}
Finally, 
\begin{equation}
\dv\left[ \frac{\bm{b}}{B} \times \dv\left( a \bm{b}\bm{b}\right) \right] = \dv\left[ \frac{a}{B} \bm{b}\times\bm{b}\dg\bm{b} \right] = \dv\left[ \frac{a}{B} \left( \curl\bm{b} - \bm{b}\bm{b}\cdot\curl\bm{b}\right)\right],
	\label{apCurv25}
\end{equation}
whereupon we can use $\bm{b}\cdot\curl\bm{b} = \bm{b}\cdot\curl\bm{B} / B$ and Amp\`ere's law to obtain
\begin{equation}
\dv\left[ \frac{\bm{b}}{B} \times \dv\left( a \bm{b}\bm{b}\right) \right] = \dv\left[ \frac{a}{B} \left(\curl\bm{b} -  \frac{4\pi j_\parallel}{cB} \bm{b}\right) \right].
\label{apCurv3}
\end{equation}
Setting $a = p_\parallel - p_\perp$ gives exactly the result needed above.

\subsection{Lowest-order Flows}
Let us first calculate the perpendicular velocity of a generic ion species:
\begin{equation}
n_s\bm{u}_{\perp s} = \vint \bm{v}_\perp f_s = n_s \uE + \wint \bm{w}_\perp f_s.
\end{equation}
We then integrate by parts in $\gyr$, using $\bm{w}_\perp = \inpd{\bm{b}\times\bm{w}}{\gyr}$, to obtain
\begin{equation}
n_s \bm{u}_{\perp s} = n_s \uE - \wint \bm{b}\times\bm{w}_\perp \left.\pd{f_s}{\gyr}\right|_{\bm{r},w_\parallel,w_\perp}.
\end{equation}
Next we substitute from \eref{ddGyrAP}, noting that only the first term on the right-hand-side of \eref{ddGyrAP} is large enough to contribute, and find
\begin{equation}
n_s \bm{u}_{\perp s} = n_s \uE + \frac{1}{\cycfreq}\wint \bm{b}\times\bm{w}_\perp \bm{w}_\perp \dg f_s.
\end{equation}
Using the gyrotropy of $f_s$ and performing the gyroaverage, we obtain the final result
\begin{equation}
n_s \bm{u}_{\perp s} =  \frac{c}{B} \bhat \times\left[\nabla \left( \frac{1}{Z_s e}\vint \frac{m_s}{2} v_\perp^2 f_s \right) + n_s\nabla \pot\right],
	 \label{apFlowForm}
\end{equation}
where we have expanded the $\bm{E}\times\bm{B}$ flow to lowest order and reverted to using $\bm{v}$ instead of $\bm{w}$.
Thus the perpendicular flow is simply the sum of the diamagnetic flow and the electrostatic $\bm{E}\times\bm{B}$ flow.

\subsection{The Ion Stress Tensor}
In this section we calculate the stress tensor $\pitens$.
By using the identity \eref{Wident} in the definition \eref{APpiDef}, we can write $\pitens$ as
\begin{equation}
\pitens = \wint m_s \left( w_\parallel \bm{b} \bm{w}_\perp + w_\parallel \bm{w}_\perp \bm{b} +\pd{\tensor{W}}{\gyr}\right)  f_s.
\end{equation}
By the same manipulations as in the previous section we can show that
\begin{equation}
\wint m_s w_\parallel \bm{b} \bm{w}_\perp f_s = \bm{b} \left( \frac{m_s}{2\cycfreq} \wint w_\parallel w_\perp^2 \bm{b}\times\nabla f_s\right).
\end{equation}
Using this result, we can rewrite $\pitens$ in the following way
\begin{equation}
\pitens = \bm{b} \bm{h} + \bm{h} \bm{b} + \wint m_s \pd{\tensor{W}}{\gyr}f_s,
\label{APPitmp1}
\end{equation}
where we have abbreviated
\begin{equation}
\bm{h} = \frac{B^2}{\cycfreq} \bm{b}\times\nabla \wint w_\parallel \frac{m_s w_\perp^2}{2B^2} f_s.
\label{APhVecDef}
\end{equation}

We now proceed to the final term of \eref{APPitmp1}. Using \eref{ddGyrAP} we see that
\begin{equation}
m_s \wint \pd{\tensor{W}}{\gyr} f_s = -m_s \wint \tensor{W} \pd{f_s}{\gyr} = m_s \wint \tensor{W}\left( \frac{\bm{w}_\perp}{\cycfreq} \dg f_s - \pd{f_s}{\magmom}\pd{\magmom}{\gyr} - \pd{\tilde{f}_s}{\gyr}\right).
\label{PiPrimeTmp}
\end{equation}
The first of these terms is easily handled, 
 \begin{equation}
 \frac{m_s}{\cycfreq} \wint \,\tensor{W} \bm{w} \dg f_s(\bm{R}_s,\energy,\magmom,t) = \frac{m_s}{\cycfreq^2}\wint\, \tensor{W}\, \bm{w}\bm{w}\times\bm{b}\bm{:} \grad_\perp \grad_\perp f_s.
 \end{equation}
 We then perform the gyroaverage, to obtain
 \begin{equation}
 \begin{split}
 \frac{m_s}{\cycfreq} \wint \tensor{W} \bm{w} \dg f_s &= -\frac{m_s}{\cycfreq^2} \wint \frac{w_\perp^4}{16} \left[\grad_\perp \grad_\perp - \left( \bm{b}\times\grad \right)\left( \bm{b}\times\grad \right) \right]f_s + \Or(\gkeps^2 p_{\perp s})\\
	&=  \left[\left( \bm{b}\times\grad \right)\left( \bm{b}\times\grad \right) - \grad_\perp \grad_\perp\right]\frac{m_s}{\cycfreq^2} \wint \frac{w_\perp^4}{16}f_s + \Or(\gkeps^2 p_{\perp s}),
	\end{split}
 \label{PaPi0}
 \end{equation}
 where the operator $\left( \bm{b}\times\grad \right)\left( \bm{b}\times\grad \right)$ is given by its components as $\epsilon_{ikl}\epsilon_{jmn}b_{k} b_m \nabla_{l} \nabla_{n}$ with $\epsilon_{ijk}$ the Levi-Civita symbol. Writing the product $\epsilon_{ikl} \epsilon_{jmn}$ in terms of Kronecker deltas, we have that
 \begin{equation}
 \epsilon_{ikl}\epsilon_{jmn}b_{k} b_m \nabla_{l} \nabla_{n} = \delta_{ij} \left( \nabla^2 - \bm{b}\dg^2 \right)   - b_i b_j\nabla^2 + b_i \bm{b}\dg \nabla_j + b_j \bm{b}\dg \nabla_i - \nabla_i \nabla_j,
 \end{equation}
 whereupon \eref{PaPi0} becomes
\begin{equation}
 \frac{m_s}{\cycfreq} \wint \tensor{W} \bm{w} \dg f_s =  \left[\idmat \nabla^2 - 2 \grad_\perp \grad_\perp\right]\frac{m_s}{\cycfreq^2} \wint \frac{w_\perp^4}{16}f_s + \Or(\gkeps^2 p_{\perp s}),
 \label{PaPi1}
 \end{equation}

 Moving to the second term in \eref{PiPrimeTmp}, and substituting from \eref{magmomWAP}, we have
 \begin{equation}
 \begin{split}
 m_s &\wint \, \tensor{W} \pd{\magmom}{\gyr} \pd{f_s}{\magmom} = \\
	 &m_s \wint \gyror{\tensor{W} \,\left[ - \frac{m_s}{B} \bm{w}\times\bm{b}\cdot \vdrift + \bm{w}_\perp\bm{w}_\perp \bm{:} \left( w_\parallel
		 \grad\bm{b} + \grad\uE\right) \right] }\pd{f_s}{\magmom},
	 \end{split}
 \end{equation}
 where we have used the fact that $(\idmat - \bm{b}\bm{b}) \bm{:} (w_\parallel\grad\bm{b} + \grad\uE)$ is small.
 Explicitly performing the gyroaverages, the first term vanishes (gyroaverage of an odd power of $\bm{w}_\perp$) and the second term gives
 \begin{equation}
 m_s \wint \, \tensor{W} \pd{\magmom}{\gyr} \pd{f_s}{\magmom} = -\wint \frac{w_\perp^4}{32} \frac{m_s^2}{B\cycfreq } \left[ w_\parallel \tensor{B} + \tensor{U}  \right] \pd{f_s}{\magmom},
 \end{equation}
 where the symmetric tensors $\tensor{B}$ and $\tensor{U}$ are given by their components
 \begin{equation}
 {B}_{ij} = \epsilon_{jkl}b_l \nabla_{i} b_k + \epsilon_{jkl} b_l \nabla_k b_i +  \left( i \leftrightarrow j \right) 
	\label{BTensDef}
 \end{equation}
 and
\begin{equation}
 {U}_{ij} = \frac{c}{B} \nabla_{i} \nabla_{j} \pot - \epsilon_{ikl}b_k \nabla_{l} \left( \uE \right)_j + \left( i \leftrightarrow j \right),
 \label{UTensDef}
 \end{equation}
 respectively.
 
 Finally, the third term in \eref{PiPrimeTmp} is
 \begin{equation}
 \begin{split}
 m_s \wint \,\tensor{W}\, \pd{\tilde{f}_s}{\gyr}= - m_s\wint & \frac{\tensor{W}}{\cycfreq}\,\left[ \left(\dot{\bm{R}}_s - \gyroR{\dot{\bm{R}}_{s}}\right) \dgR{f_s} +\left(\denergy - \gyroR{\denergy}\right) \pd{f_s}{\energy}  \right]
 \\
	 &- \frac{m_s}{\cycfreq} \wint \,\tensor{W} \ensav{ \daccel\cdot\pd{\turbf}{\bm{w}}}
 \end{split}
 \end{equation}
 Using the expression \eref{GyrRdot} for $\dot{\bm{R}}_s$, we can see that the first term on the right-hand-side of this equation vanishes upon gyroaveraging. For the second term, we have
 \begin{equation}
 -\frac{m_s}{\cycfreq}\wint \tensor{W}\, \left(\denergy - \gyroR{\denergy}\right) \pd{f_s}{\energy} = \frac{m_s^2}{\cycfreq}\wint \tensor{W}\, \bm{w}_\perp\bm{w}_\perp \bm{:} \grad\uE \pd{f_e}{\energy[e]}.
 \end{equation}
 whereupon we can perform the gyroaverage as above to obtain
 \begin{equation}
 -\frac{1}{\cycfreq}\wint \tensor{W}\, \left(\denergy - \gyroR{\denergy}\right) \pd{f_s}{\energy} = -\frac{m_s^2}{\cycfreq}\wint \frac{w_\perp^4}{32} \tensor{U} \pd{f_e}{\energy[e]}.
 \end{equation}

 Collecting these intermediate results, we have that 
 \begin{equation}
 \begin{split}
 \pitens = &\bm{b}\bm{h} + \bm{h}\bm{b} + \left[\idmat \nabla^2 - 2 \grad_\perp \grad_\perp\right] \wint \frac{m_s w_\perp^4}{16 \cycfreq^2} f_s\\
  &+ \frac{\tensor{B}}{32} \wint \frac{m_s^2 w_\perp^4}{B \cycfreq} w_\parallel \pd{f_s}{\magmom} 
  + \tensor{U} \left[ \wint \frac{m_s^2 w_\perp^4}{32 \cycfreq} \left( \pd{f_s}{\energy} + \frac{1}{B} \pd{f_s}{\magmom}\right)\right] \\
  & + \frac{m_s}{\cycfreq} \wint \tensor{W}\ensav{ \daccel\cdot\pd{\turbf}{\bm{w}}}.
 \end{split}
 \label{B49-1}
 \end{equation}
One final manipulation is to integrate by parts where possible to obtain
 \begin{equation}
 \begin{split}
 \pitens = &\bm{b}\bm{h} + \bm{h}\bm{b} + \left[\idmat \nabla^2 - 2 \grad_\perp \grad_\perp\right] \wint \frac{m_s w_\perp^4}{16 \cycfreq^2} f_s\\
  &- \frac{\tensor{B}}{4\cycfreq} \wint \frac{m_s}{2} w_\perp^2 w_\parallel f_s -\frac{p_{\perp s}}{4\cycfreq} \tensor{U} - \frac{m_s}{\cycfreq}\ensav{ \wint \daccel \cdot \pd{\tensor{W}}{\bm{w}} \turbf}.
 \end{split}
 \label{B49}
 \end{equation}

 \subsection{The ELM Vorticity Equation}
 Let us now simplify the quantity $\bm{X}_s$ in the vorticity equation for the ELM ordering. Estimating the size of other terms, we see that we need the final vorticity equation
 correct to order $\Or(p_\perp / L_\parallel L_\perp B)$. This means we need $\bm{X}$ to $\Or(p_\perp / L_\parallel)$.
 
 With this ordering, \eref{Xtmp1} becomes
 \begin{equation}
\bm{X}_s = \dv\pitens - n_s m_s \upar\bm{b}\dg \left( \upar \bm{b} \right) -  \upar \bm{b}\dv\left( n_s m_s \upar \bm{b}\right) + n_s m_s \uD \dg \uE.
\label{XtmpELM}
 \end{equation}
Examining the form of $\pitens$, it can easily be seen that
 \begin{equation}
 \dv\left\{ \frac{\bm{b}}{B} \times \dv\left[ \left( \idmat \nabla^2 - 2\grad\grad\right) \lambda \right] \right\} = -\curl\left(\frac{\bm{b}}{B}\right) \dg \nabla^2 \lambda,
 \end{equation}
 and so the third term in \eref{B49} doesn't contribute to \eref{XtmpELM}.
	Similarly terms involving $\bm{h}$ or $\tensor{B}$ can be seen to be $\Or(\sqrt{\gkeps} p_\perp / L_\parallel)$ and so too small to be kept.

Hence, we can substitute from \eref{B49} into \eref{XtmpELM} to obtain
 \begin{equation}
 \begin{split}
\bm{X}_s = &-\dv\left( \frac{p_{\perp s}}{4\cycfreq} \tensor{U}\right) - n_s m_s \upar\bm{b}\dg \left( \upar \bm{b} \right) -  \upar \bm{b}\dv\left( n_s m_s \upar \bm{b}\right) \\
		&\qquad+ n_s m_s \uD \dg \uE - \frac{m_s}{\cycfreq}\ensav{ \wint \daccel \cdot \pd{\tensor{W}}{\bm{w}} \turbf} + \Or\left(\sqrt{\gkeps} \frac{p_\parallel}{L_\parallel}\right).
\end{split}
\label{XtmpELM2}
 \end{equation}

 We now use this in the definition of $\bm{X}$ to obtain
 \begin{equation}
 \begin{split}
 \frac{\bm{b}}{B}\times\bm{X} = &\frac{\bm{b}}{B} \times\left[ n_s m_s \uD \dg \uE - \dv\left( \frac{p_{\perp s}}{4\cycfreq} \tensor{U}\right)\right] - n_s m_s \upar^2 \bm{b}\times\bm{b}\dg\bm{b} \\
             & + \frac{\bm{b}}{B} \times \dv\left(\tensor{P}_{2} + \frac{1}{4\pi}\ensav{\turbB \turbB}\right) - \frac{\bm{b}}{8\pi B} \bm{b}\times\grad \ensav{\delta B^2} + \Or\left(\sqrt{\gkeps} \frac{p_\perp}{L_\parallel B}\right),
 \end{split}
 \end{equation}
 where we have used the explicit formula for $\tensor{W}$ to evaluate the term involving the fluctuations and then, as we need to keep only the leading order, substituted $\bm{v}$ for $\bm{w}$ in this expression, before using 
 \begin{equation}
 \tensor{P}_2 = \frac{m_s}{\cycfreq} \ensav{ \wint \tensor{W} \daccel\cdot \pd{\turbf}{\bm{w}}},
 \end{equation}
 to write the result concisely.

 Taking the divergence of this result, the final term vanishes and the terms involving the parallel velocity will cancel those on the left-hand side of \eref{tmpVort2}. This leaves us with the expected terms involving fluctuations, and also terms involving the diamagnetic and $\bm{E}\times\bm{B}$ velocities. 
 However, it can be shown by direct evaluation in index notation that
 \begin{equation}
 \dv\left[\frac{\bm{b}}{B} \times \dv\left( -\frac{p_{\perp s}}{4\cycfreq} \tensor{U}  \right) \right] = - \dv \left[ \frac{\bm{b}}{B} \times\left( \uD \dg \uE \right)\right] + \Or\left(\gkeps^2 \frac{p_\perp}{L_\perp^2 B}\right).
 \label{APgyrovisc}
 \end{equation}
 Thus, these terms cancel. This important result is the expression, in our formulation, of the well-known gyro-viscous cancellation~\citep{hazeltine2003plasma}.
\subsection{Turbulent Contributions to the Vorticity Equation} 
 Let us now massage the term involving the fluctuations. The tensor $\tensor{P}_{2}$ is given explicitly by,
 \begin{equation}
 \tensor{P}_2 = \frac{m_s}{\cycfreq} \ensav{ \wint \tensor{W} \daccel\cdot \pd{\turbf}{\bm{w}}},
 \end{equation}
 where we have used the fact that $\vint = \wint$ and $\partial / \partial\bm{v} = \partial / \partial\bm{w}$.
 Integrating by parts and using the definition of $\tensor{W}$ we have
 \begin{equation}
 \tensor{P}_2 = \frac{m_s}{4\cycfreq} \ensav{ \wint \left[\delta\bm{a}_{s\perp} \bm{w}\times\bm{b} + \bm{w} \daccel \times\bm{b} + \left( \leftrightarrow \right)\right] \turbf }.
 \end{equation}
 Now, we calculate the contribution to the vorticity equation in index notation
 \begin{equation}
 \begin{split}
 &\dv\left(\frac{\bm{b}}{B}\times\dv\tensor{P}_2\right)  \\
	 &\quad=\frac{m_s}{4\cycfreq} \partial_i \partial_l \ensav{ \epsilon_{ijk} b_j \wint \left[\epsilon_{kmn} {b}_n\left(\delta{a}_{\perp l} {w}_{\perp m}  + {w}_{\perp l} \delta a_{\perp m} \right) +  \left( l \leftrightarrow k \right)\right] \turbf } \\
	 &\quad=\frac{m_s}{4\cycfreq} \partial_i \partial_l \left< \wint \left[\left(\delta{a}_{\perp l} {w}_{\perp i}  + {w}_{\perp l} \delta a_{i} \right) \right.\right. \\
		 &\qquad\qquad\left.\left.\phantom{\wint} + \epsilon_{ijk}\epsilon_{lmn} b_j b_n \left(  \delta{a}_{\perp k} {w}_{\perp m}  + {w}_{\perp k} \delta a_{m} \right) \right] \turbf \right>_{\mathrm{turb}} \\
	 &\quad=\frac{m_s}{4\cycfreq} \partial_i \partial_l \ensav{ \wint \left[2\left(\delta{a}_{\perp l} {w}_{\perp i}  + {w}_{\perp l} \delta a_{i} \right) - \left( \delta_{il} - b_i b_l \right) \daccel \cdot\bm{w}_\perp \right] \turbf }.
 \end{split}
 \label{tmpTurbVort}
 \end{equation}
 We now attack the second term in the brackets here, which contains at its core the following gyroaverage
 \begin{equation}
 \nonumber
	\ensav{\gyror{\daccel\cdot\bm{w}_\perp \turbf}}.
 \end{equation}
 The first identity we will need is 
 \begin{equation}
 \daccel[i] = -\frac{Z_i e}{m_i} \grad\left( \turbPot - \frac{w_\parallel }{c} \turbApar\right) - \frac{\cycfreq[i]}{B} \bm{v}_\perp \dg\turbApar \bm{b} + \Or( \gkeps^{3/2} \vth[i] \cycfreq ),
 \end{equation}
 which can be derived from the definition of $\daccel$ by direct substitution.
 We have specialised to the ions as the electrons do not contribute to the momentum (and thus vorticity) equation.
 Inserting this result into our gyroaverage gives
\begin{equation}
\begin{split}
\ensav{\gyror{ \daccel\cdot\bm{w}_\perp \turbf}} &= \gyror{\ensav{ \turbf \bm{w}_\perp \dg \turbChi}} \\
&= \dv\ensav{\turbChi\gyror{\bm{w}_\perp  \turbf}} - \ensav{\turbChi \gyror{\bm{w}_\perp \dg \turbf}}.
\end{split}
\end{equation}
Now 
\begin{equation}
\bm{w}_\perp \dg \turbf = \cycfreq \left( \left.\pd{ }{\gyr}\right|_{\bm{R}_s,\energy,\magmom} - \left.\pd{ }{ \gyr}\right|_{\bm{r},w_\parallel,w_\perp}\right) \turbf
\end{equation}
Thus we have
\begin{equation}
\ensav{\gyror{ \daccel\cdot\bm{w}_\perp \turbf}} = \frac{Z_s e}{m_s}\dv\ensav{\turbChi\gyror{\bm{w}_\perp  \turbf}} - \frac{Z_s e}{m_s}\ensav{\turbPot\gyror{\left.\cycfreq\pd{\turbf^{(2)}}{\gyr}\right|_{\bm{R}_s}}},
	\label{tmpFoo}
\end{equation}
where we have defined the higher-order piece of $\turbf$ by
\begin{equation}
\turbf = Z_s e \turbPot \pd{f_s}{\energy} + Z_s e \left(\turbChi - \gyroR{\turbChi}\right) \pd{f_s}{\magmom} + h_s + \turbf^{(2)}.
\end{equation}

\begin{equation}
\begin{split}
\cycfreq \pd{\turbf^{(2)}}{\gyr} &= \gyroR{\pd{\turbf}{t}} - \pd{\turbf}{t} - \daccel \cdot \pd{}{\bm{v}} \left(f_s + \turbf\right)+ \gyroR{\daccel\cdot \pd{}{\bm{v}}\left(f_s + \turbf\right)}\\
		&= Z_i e \pd{ }{t}\left( \gyroR{\turbPot} - \turbPot \right) + \frac{c}{B} \bm{b}\times\ddR[i]{}\left(\gyroR{\turbPot} - \turbPot\right) \dgR[i] { } \left( f_i + h_i \right) \\
		&\qquad+ \Or( \gkeps^{3/2} \cycfreq f_i),
\end{split}
\end{equation}
where in the second line we have specialised to the ion species.
Substituting back into \eref{tmpFoo}, we see that
\begin{equation}
\ensav{\gyror{ \daccel[i]\cdot\bm{w}_\perp \turbf[i]}} = \frac{Z_s e}{m_s}\dv\ensav{\turbPot\gyror{\bm{w}_\perp  \turbf[i]}} + \Or(\gkeps^2 \vth[i]^2 \cycfreq[i]).
	\label{tmpFoo2}
\end{equation}

Combining these results together, the leading order part of \eref{tmpTurbVort} is given by
	\begin{equation}
	\turbVort \equiv \dv\left(\frac{\bm{b}}{B}\times\dv\tensor{P}_2\right)  = - \sum_{s=i} \frac{ Z_i e}{2\cycfreq[i]} \left\{ \grad\grad\bm{:} \ensav{\wint\left[\left(\grad\turbPot\right) \bm{w}_\perp + \bm{w}_\perp \left(\grad\turbPot\right)\right] \turbf[i]} \right\}
	\end{equation}

 Finally, we show that the fluctuating Maxwell stress is negligible.
 \begin{equation}
 \begin{split}
 \dv\left(\frac{\bm{b}}{B} \times\dv \ensav{\turbB\turbB}\right) &= \partial_i \epsilon_{ijk} b_j \partial_l \ensav{ \delta B_l \delta B_k} \\
				&= \epsilon_{ijk} b_j \partial_i \partial_l \ensav{ \epsilon_{lmn} b_m \partial_n \turbApar \epsilon_{krs} b_r \partial_s \turbApar } \\
				&= \epsilon_{lmn}\left( \delta_{ir} \delta_{js} - \delta_{is}\delta_{jr}\right) b_j b_m b_r \partial_i \partial_l \ensav{ \partial_n \turbApar \partial_s \turbApar },
 \end{split}
 \end{equation}
 where we have used the fact that terms involving $\turbB\cdot\bm{b}$ and those involving gradients of $\bm{b}$ are too small to contribute (these terms are $\Or(\gkeps^2 p_\perp / L_\perp^2 B)$).
 Now, using the fact that $b_r \partial_r = \bm{b}\dg \ll \nabla_\perp$, we have 
 \begin{equation}
 \begin{split}
 \dv\left(\frac{\bm{b}}{B} \times\dv \ensav{\turbB\turbB}\right) &= -\epsilon_{lmn} b_m \partial_i \partial_l \ensav{ \partial_n \turbApar \partial_s \turbApar } \\
	 &= -\epsilon_{lmn} b_m \partial_i \partial_l \left(\partial_n \ensav{ \turbApar \partial_s \turbApar } - \ensav{ \turbApar \partial_n\partial_s \turbApar } \right)\\
			 &= \epsilon_{lmn} \partial_i \partial_l \ensav{ \partial_n \partial_s \frac{1}{2} \turbApar^2} = 0.
 \end{split}
 \end{equation} 
 
 With these results, we see that we do indeed obtain the required vorticity equation \eref{vortELM}.

 \subsection{The inter-ELM Vorticity Equation}
 We now need terms in the vorticity equation to $\Or(\gkeps^{3/2} p_\perp / L_\perp^2)$. Firstly, we note that the gyroviscous calculation, and the calculation of the fluctuating Maxwell stress are already accurate to a high enough order; we do not need to revisit them.
 However, we now need to go back and handle some extra terms in both $\bm{X}_s$ and our formula for $\pitens$.
 Firstly, the terms on the left-hand side of \eref{tmpVort2} involving $\bm{b}\dg\bm{b}$ are
 \begin{equation}
 \begin{split}
 \dv&\left( n_s m_s \upar \bm{b}\dg\bm{b} \times \bm{u}_s\right)\\
	&= \dv\left[ n_s m_s \upar \bm{b}\dg\bm{b} \times \left( \upar\bm{b} + \uD + \uE \right)\right]\\
	&= \dv\left[ n_s m_s \upar^2 \bm{b}\dg\bm{b} \times\bm{b} + n_s m_s \upar \left(\bm{b}\dg\bm{b}\right) \cdot\left( \frac{\grad p_{\perp s}}{n_s m_s \cycfreq} + \frac{c\nabla\pot}{B}\right) \bm{b}\right].
	\end{split}
 \end{equation}
 We can see that, because of our ordering on $\bm{b}\dg$, only the centrifugal term is large enough to matter. This will, of course, cancel with terms in $\bm{X}_s$ as in the previous calculation.

 Returning to \eref{Xtmp1} and retaining all terms, we see that
\begin{equation}
 \begin{split}
 \frac{\bm{b}}{B}\times\bm{X}_s = &\frac{\bm{b}}{B} \times\left(\dv\pitens\right) - \frac{n_s m_s}{B} \upar^2 \bm{b}\times\left(\bm{b}\dg\bm{b}\right) - \frac{n_s m_s}{B} \upar \bm{b}\times\left(\bm{b}\dg \uD\right) \\
	 &\quad+ \frac{\grad_\perp p_{\perp s}}{n_sm_sB\cycfreq} \dv\left(n_s m_s \upar \bm{b} + n_s m_s \uD\right) + \frac{n_s m_s}{B}\bm{b}\times\left(\uD\dg\uE\right),
 \end{split}
 \end{equation}
 in which we have used the definition of the diamagnetic velocity.
 Expanding the third and fourth terms in this expression it becomes
\begin{equation}
 \begin{split}
 &\frac{\bm{b}}{B}\times\bm{X}_s = 
 \frac{\bm{b}}{B} \times\left(\dv\pitens\right) - \frac{n_s m_s}{B} \upar^2 \bm{b}\times\left(\bm{b}\dg\bm{b}\right) \\
	 &\qquad- \frac{n_s m_s}{B} \upar \left[ \frac{\left(\bm{b}\dg\bm{b}\right) \bm{b}\dg p_{\perp s}}{n_s m_s \cycfreq} + \bm{b}\dg\left(\frac{\grad p_{\perp s}}{n_s m_s\cycfreq}\right) \right]  \\
	 &\qquad+ \frac{\grad_\perp p_{\perp s}}{n_sm_sB\cycfreq} \left[ n_s m_s \upar \dv\bm{b} + \bm{b}\dg \left(n_s m_s \upar\right) + \dv\left(\frac{\bm{b}\times\grad p_{\perp s}}{\cycfreq} \right)\right] \\
	 &\qquad+ \frac{n_s m_s}{B}\bm{b}\times\left(\uD\dg\uE\right).
 \end{split}
 \end{equation}
 Eliminating terms that are smaller than $\Or(\gkeps^{3/2} p_{\perp s} / B L_\perp^2)$ and collecting terms together we obtain
\begin{equation}
 \begin{split}
 &\frac{\bm{b}}{B}\times\bm{X}_s = 
 \frac{\bm{b}}{B} \times\left(\dv\pitens\right) - \frac{n_s m_s}{B} \upar^2 \bm{b}\times\left(\bm{b}\dg\bm{b}\right) \\
	 &\qquad +\bm{b}\dg \left( \frac{\upar \grad p_{\perp s}}{B\cycfreq} \right) + \frac{n_s m_s}{B}\bm{b}\times\left(\uD\dg\uE\right).
 \end{split}
 \end{equation}

 Ultimately we must handle the remaining terms in $\dv\pitens$.
 It can be shown that:
 \begin{equation}
 \dv\left[\frac{\bm{b}}{B}\times\dv\left(\bm{h}\bm{b} + \bm{b}\bm{h}\right)\right] = -\bm{B}\dg\left[ \frac{1}{\cycfreq[s]} \nabla^2\left( \frac{1}{2B^2} \wint m_s w_\parallel w_\perp^2 f_s\right)\right] + \Or\left(\gkeps^2 \frac{p_{\perp s}}{BL_\perp^2}\right).
 \end{equation}
 Similarly,
 \begin{equation}
 \begin{split}
 \dv&\left[ \frac{\bm{b}}{B}\times\dv\left( -\frac{\tensor{B}}{4\cycfreq} \wint \frac{m_s}{2} w_\parallel w_\perp^2 f_s\right)\right]\\
  &= \frac{m_s c}{2 Z_s e} \left[ \left(\dv\bm{b}\right) \nabla^2 - \left( \grad\bm{b}\right) \bm{:} \grad\grad\right] \wint \frac{m_s}{2B^2} w_\parallel w_\perp^2 f_s.
  \end{split}
 \end{equation}
 Thus, the vorticity equation in the inter-ELM ordering, accurate to $\Or(\gkeps^{3/2} p_\perp / L_\perp^2 B)$ is
\begin{equation}
\begin{split}
&\dv\left\{ \sum_s \frac{n_s m_s}{B} \left[ \pd{ }{t} + \upar\bm{b}\dg + \uE\dg\right]\left( \frac{\nabla p_{\perp s}}{n_s m_s \cycfreq[s]} + \frac{c \nabla \pot}{B}\right)\right\}
\\
		&\qquad= \dv\left[ - \frac{j_\parallel \bm{b}}{c}\left( 1- 4\pi \frac{ p_\parallel - p_\perp}{B^2}\right) - p_\perp \curl\left(\frac{\bm{b}}{B}\right) - \frac{\curl \bm{b}}{B} \left(p_{\parallel} - p_\perp\right) \right]
		\\
			&\qquad\qquad - \bm{B}\dg \left( \frac{\nabla^2 \qpar}{B^2\cycfreq} \right) + \dv\left[\bm{b}\dg\left( \frac{\upar \grad p_{\perp s}}{B\cycfreq}\right)\right] \\
			&\qquad\qquad+ \frac{1}{2B\cycfreq} \left[ \left(\dv\bm{b}\right) \nabla^2 - \left( \grad\bm{b}\right) \bm{:} \grad\grad\right]\qpar + \dv\left(\frac{\bm{b}}{B}\times\dv \tensor{P}_{2}\right),
\end{split}
\label{SlowVortAP}
\end{equation}
in which we have abbreviated
\begin{equation}
\qpar = \wint \frac{m_s w_\perp^2}{2} w_\parallel f_s.
\end{equation}
We postpone the flux-surface-averaging of this equation to \Apref{APFavVort}.
\subsection{Gyroaverages}
In the previous section we needed to gyroaverage terms like $\tensor{W} \bm{w} \bm{w}$. These all stem from the following identity
\begin{equation}
\gyror{ w_{\perp i} w_{\perp j}w_{\perp k}w_{\perp l}} = \frac{w_\perp^4}{8} \left( \delta^{\perp}_{i j}\delta^{\perp}_{k l} + \delta^{\perp}_{i k}\delta^{\perp}_{j l} +\delta^{\perp}_{i l}\delta^{\perp}_{j k}\right),
\label{w4th}
\end{equation}
where
\begin{equation}
\delta^{\perp}_{ij} = \delta_{ij} - b_i b_j,
\end{equation}
with $\delta$ the usual Kronecker delta. The tensor structure of the right-hand-side of \eref{w4th} follows from the absence of a preferred perpendicular direction. The coefficient in front can be calculated by picking a particular component and using e.g. $\gyror{\cos^4 \gyr} = 3/8$.
We can then perform gyroaverages like
\begin{equation}
\gyror{\tensor{W} \bm{w}_\perp \bm{w}_\perp } = \frac{-1}{4} \epsilon_{jmn} b_n \gyror{ w_{\perp i} w_{\perp m}w_{\perp k}w_{\perp l}} + \left( i \leftrightarrow j \right),
\end{equation}
which lead immediately to \eref{BTensDef} and \eref{UTensDef}.

\section{Flux-Surface-Averaged Vorticity}
\label{APFavVort}
Here we take the flux-surface average of the slow-timescale vorticity equation \eref{SlowVortAP} to obtain \eref{fsVort}.
The usual expressions for the flux-surface average are needed (see \citet{DhaeseleerFlux}), namely

\begin{equation}
\fav{g} = \frac{1}{V'} \oint \frac{d\ell d\alpha}{B},
\end{equation}
where $g$ is an arbitrary function, $V'$ is the area of the flux surface in question, and $\psi, \alpha, \ell$ are a set of Clebsch coordinates for the field $\bm{B}$, and 
\begin{equation}
\fav{\dv{\bm{Y}}} = \frac{1}{V'} \pd{ }{\psi} V'\fav{\bm{Y}\dg\psi},
\end{equation}
with $\bm{Y}$ an arbitrary vector field.
With these identities, we see immediately that
\begin{equation}
\fav{ \dv\left[\frac{j_\parallel \bm{b}}{c} \left( 1 - 4\pi \frac{ p_\parallel - p_\perp }{ B^2} \right)\right]} = 0,
\end{equation}
and
\begin{equation}
\begin{split}
\fav{\dv\left( \frac{p_\parallel - p_\perp}{B} \curl \bm{b} \right)} &= \fav{\dv{ \left[\bm{b} \times \nabla \left( \frac{p_\parallel - p_\perp }{B}\right) \right]}} \\
		&= \frac{1}{V'} \pd{ }{\psi} V' \oint d\ell d\alpha \pd{ }{\alpha} \left( \frac{p_\parallel - p_\perp}{B}\right) \\
		&= 0,
	\end{split}
	\label{favcurlb}
\end{equation}
where in both we have used only the properties of the flux surfaces, including $\alpha$-periodicity, but not used any ordering assumptions. These are identities that hold to all orders in $\gkeps$.

Finally we tackle the term
\begin{equation}
\fav{\dv\left( p_{\perp} \curl\left(\frac{\bm{b}}{B}\right)\right)}.
\label{finalFav}
\end{equation}
To analyse this term we need the following expression for the perpendicular current, derived from the exact momentum equation under the inter-ELM orderings
\begin{equation}
\bm{j}_\perp = \frac{c}{B} \bm{b}\times\grad p_\perp + c \frac{p_\parallel - p_\perp}{B} \left( \curl\bm{b} - \bm{b}\bm{b}\cdot\curl\bm{b}\right) + \Or\left( \sqrt{\gkeps} \frac{c}{B} \frac{p_\perp}{L_\parallel}\right),
\end{equation}
which can be trivially rewritten as
\begin{equation}
\bm{j} = j_\parallel \bm{b}\left[ 1 + \frac{4\pi}{B^2} \left(p_\parallel - p_\perp\right)\right] + \frac{c}{B} \bm{b}\times\grad p_\perp + c \frac{p_\parallel - p_\perp}{B} \left( \curl\bm{b} \right) + \Or\left( \sqrt{\gkeps} \frac{c}{B} \frac{p_\perp}{L_\parallel}\right).
\label{jExp}
\end{equation}
Returning to \eref{finalFav} we use the identity
\begin{equation}
\bm{b}\times\nabla B = B{\curl\bm{b}} - \frac{4\pi}{c} \bm{j},
\end{equation}
which follows from Amp\`ere's law to obtain
\begin{equation}
\fav{\dv\left( p_{\perp} \curl\left(\frac{\bm{b}}{B}\right)\right)} = 
\fav{\dv\left[ p_\perp \left( \frac{\curl\bm{b}}{B} + \frac{\bm{b}\times\grad B}{B^2}\right) \right]} = \fav{\dv\left( \frac{4 \pi}{c} p_\perp \bm{j}\right)},
\end{equation}
where we have used the same results that lead to \eref{favcurlb} to eliminate terms containing $\fav{ \dv\left( p_\perp \curl\bm{b} / B \right)}$.

Now, substituting \eref{jExp} into this result, we immediately obtain
\begin{equation}
\fav{\dv\left( p_{\perp} \curl\left(\frac{\bm{b}}{B}\right)\right)} = \fav{j_\parallel \bm{b} \dg\left(\frac{4\pi p_\perp}{cB^2}\right)} + \Or\left( \frac{{\gkeps} p_\perp}{B L_\perp L_\parallel}\right),
\end{equation}
where we retain the $j_\parallel$ term because of our ordering for the shear length as $\sqrt{L_\perp L_\parallel}$.
However, from our solution \eref{LargeShearK} for the large part of $j_\parallel$, we see that this entire expression becomes
\begin{equation}
\fav{\dv\left( p_{\perp} \curl\left(\frac{\bm{b}}{B}\right)\right)} = \frac{4\pi K(\psi)}{c} \fav{ \bm{B} \dg\left(\frac{4\pi p_\perp}{cB^2}\right)} = 0,
\end{equation}
and so the entire term is small.

Using these identities, the flux-surface average of \eref{SlowVortAP} is
\begin{equation}
\begin{split}
&\fav{\dv\left\{ \sum_s \frac{n_s m_s}{B} \left[ \pd{ }{t} + \upar\bm{b}\dg + \uE\dg\right]\left( \frac{\nabla p_{\perp s}}{n_s m_s \cycfreq[s]} + \frac{c \nabla \pot}{B}\right)\right\}}
\\
		&\qquad= \fav{\dv\left[\bm{b}\dg\left( \frac{\upar \grad p_{\perp s}}{B\cycfreq}\right)\right]} + \fav{\frac{1}{2B\cycfreq} \left[ \left(\dv\bm{b}\right) \nabla^2 - \left( \grad\bm{b}\right) \bm{:} \grad\grad\right]\qpar} \\
		&\qquad\qquad+ \fav{\dv\left(\frac{\bm{b}}{B}\times\dv \tensor{P}_{2}\right)},
\end{split}
\label{FavSlowVortAP}
\end{equation}
again, with the abbreviation for $\qpar$.
\section{Derivation of The Ballooning Equation}
In this appendix we demonstrate that our equations for ELM dynamics reduce in the appropriate limit to the Ballooning Equation of \citet{tang1980kbm}.
The equations that we will linearise are \eref{eKE-1} and \eref{iKE-1} for the distribution functions, along with \eref{ELMAparEq} and \eref{vortELM}.
For the background magnetic field, we will assume that it has topologically toroidal flux surfaces labelled by $\psi$ -- the entire equilibrium field is contained 
in $\bm{B}_0$, there is no equilibrium $A_\parallel$.
We will linearise around Maxwellian equilibria for both ions and electrons, with the density and temperature of the equilibria being flux functions.
Consistently, we will assume that there is no equilibrium $\pot$.

We chose a form for our fluctuating quantities that parallels that in \citet{tang1980kbm}, but simplifies our notation. We will assume we are linearising in a local flux tube~\cite{beer1995field} and so
every fluctuating quantity can be expressed in a Fourier series perpendicular to the field line, with a wave-vector $\bm{k}_\perp$. The wave vector is assumed to be large so that $k_\perp L_\perp \gg 1$, and we really are looking at a local perturbation. This is appropriate for a Ballooning mode as we know that the most unstable mode occurs for $n \rightarrow \infty$, where $n$ is the toroidal mode number.
Naturally, we assume exponential time dependence with frequency $\omega$.
We also define the following auxiliary frequencies to parallel the definitions in \citet{tang1980kbm}:
\begin{eqnarray}
\omega_{*s} &=& \frac{c T_s}{Z_s e B} \bm{b}\times\bm{k}_\perp \dg \ln n_s, \\
\omega_{*s}^{T} &=& \omega_{*s} \left[ 1+ \eta_s \left(\frac{m_s v^2}{2 T_s} - \frac{3}{2} \right)\right],
\end{eqnarray}
with $\eta_s$, $\omega_{*p}$, $\omega_{\kappa}$ and $\omega_{B}$ defined exactly as in \citet{tang1980kbm}.

Let us start by linearising and solving the ion kinetic equation\eref{iKE-1} to obtain:
\begin{equation}
\delta f_i = - \frac{\omega_{*i}^T}{\omega} \frac{Z_i e \delta\pot}{T_i} f_i.
\end{equation}
Hence, the perturbed ion pressure is isotropic and given by
\begin{equation}
\delta p_i = - p_i \frac{\omega_*}{\omega} \left(1+\eta_i\right) \frac{Z_i e \delta \pot}{T_i}.
\end{equation}

We now linearise the electron kinetic equation \eref{eKE-1}, resulting in
\begin{equation}
\begin{split}
- i \omega \delta f_e + &w_\parallel \bm{b} \dgR[e]{\delta f_e} + \frac{i c}{B} \bm{b}\times\bm{k} \dg f_{0e} \delta \pot
- e w_\parallel \left( \bm{b}\dg \delta \pot + \frac{1}{c} \pd{\delta A_\parallel}{t} \right) \frac{f_{0e}}{T_e}\\
		&- w_\parallel \bm{b}\times\nabla\delta A_\parallel \dg f_{0e} = \lincol[\delta f_e],
	\end{split}
	\label{eKE-lin}
\end{equation}
where $\lincol$ is the collision operator linearised about the Maxwellian equilibria. 
To solve this equation we need to do two things. Firstly, as is usual in linear calculations we introduce a new field $\xi$ defined by
\begin{equation}
\frac{1}{c} \pd{\delta A_\parallel}{t} = \bm{b}\dg \xi.
\end{equation}
N.B. this representation is not in general possible, but because we are dealing with perturbations that have non-zero $n$, neither $\delta A_\parallel$ nor $\xi$ can have non-zero toroidal averages, and so this representation is both unique and well defined.
Secondly, in order to make contact with Equation (3.40) of \citet{tang1980kbm}, we will need to assume that the electron transit and frequencies are large compared to the frequencies of interest, and to self-consistently neglect trapped particle effects we take $\nu_{ee} \sim \vth[e]/L_\parallel \gg \omega$. 
Using this subsidiary ordering, we can immediately solve \eref{eKE-lin} to find
\begin{equation}
\delta f_e = \left[\frac{e\delta \pot}{T_e} + \frac{e\xi}{T_e} \left(1- \frac{\omega_{*e}^{T}}{\omega}\right) \right] f_e + \Or\left(\frac{\omega}{k_\parallel \vth[e]}\right).
\end{equation}
Unfortunately, this solution satisfies the linearisation of \eref{ELMAparEq} identically. Thus, we need to go to the next order in $\omega \left/ k_\parallel \vth[e]\right.$ to obtain a relationship between $\delta \pot$ and $\xi$. This is most rapidly done by simply integrating \eref{eKE-lin} over all velocities and using the fact that any parallel velocity in $\delta f_e$ must vanish:
\begin{equation}
-i \omega \delta n_e + \frac{ic}{B} \bm{b}\times\bm{k} \dg n_e  \delta \pot = 0.
\end{equation}
Using the lowest-order solution for $\delta n_e$ above, we obtain
\begin{equation}
\left( \delta \pot + \xi\right) \left( 1- \frac{\omega_{*e}}{\omega} \right) \frac{e}{T_e} = 0.
\end{equation}
Thus, 
\begin{equation}
\delta \pot = -\xi.
\end{equation}
This is just the usual condition of $\delta E_\parallel = 0$, which always obtains when the electron diamagnetic corrections to Ohm's law can be neglected due to rapid electron motion.

Turning now to the vorticity equation, we can drop all terms involving pressure anisotropy, and thus we have
\begin{equation}
\begin{split}
\dv&\left[ \frac{n_i m_i}{B} \pd{ }{t}\left( \frac{\nabla \delta p_i}{n_i m_i \cycfreq[i]} + \frac{c}{B} \nabla \delta\pot \right) + \frac{c n_i m_i}{B^2} \bm{b}\times\bm{\nabla}\delta\pot\dg \left(\frac{\nabla p_i}{n_i m_i \cycfreq[i]} \right)\right]\\
&\qquad 	= \dv\left[ - \frac{j_\parallel \bm{b}}{c} - \left( \delta p_i + \delta p_e \right) \curl \left(\frac{\bm{b}}{B}\right) \right].
	\end{split}
	\label{linVort}
\end{equation}

Now, for consistency with the \citet{tang1980kbm} approach, we will furthermore assume $\rho_i^2 / L_n L_T \ll \omega / \cycfreq[i]$ which allows us to drop the final term on the left-hand side of \eref{linVort}.
This is in fact a consequence of our local assumption -- if $\omega \sim \omega_{*}$ then 
\begin{equation*}
\frac{\omega L_n L_T }{ \rho_i^2 \cycfreq[i]} \sim k_\perp L_T \gg 1.
\end{equation*}
We will also lean on the local approximation on the right-hand side of \eref{linVort} and assume that $k_\perp L_s \ll 1$, where $L_s$ is the shear length; thus eliminating the kink drive (i.e. $\delta\bm{b}\dg j_{\parallel 0}$) from \eref{linVort}.

With these approximations we obtain
\begin{equation}
\begin{split}
  \frac{i \omega n_i T_i}{B \cycfreq[i] }& k_\perp^2 \left[ 1 - \frac{\omega_{*i}}{\omega} \left( 1 + \eta_i \right) \right] \frac{Z_ie \xi}{T_i}
 = \\
	&\frac{1}{4\pi} \bm{B} \dg\left( \frac{k_\perp^2 c i }{\omega B} \bm{b}\dg \xi \right) - i \frac{m_i n_i\cycfreq[i]}{B} \left( 2\omega_{\bm{\kappa}} + \frac{\beta_i}{2} \omega_{*p}\right) \frac{\omega_{*p}}{\omega} \frac{Z_i e\xi}{T_i} ,
 \end{split}
\label{ballEq0}
\end{equation}
where we have used
\begin{equation}
\bm{k}_\perp \cdot \curl \left( \frac{\bm{b}_0}{B_0}\right) = \frac{m_i \cycfreq[i]}{B T_i} \left( 2 \omega_{\bm{\kappa}} + \frac{\beta_i}{2} \omega_{*p}\right).
\end{equation}
Rearranging \eref{ballEq0} and multiplying by $4\pi T_i \omega / i Z_i e \cycfreq[i] B$, we obtain
\begin{equation}
\begin{split}
  \frac{1}{B}\bm{b} \dg&\left( b_i B\bm{b}\dg \xi \right)
 = \\
	& -\frac{ \omega^2 }{\vA^2} b_i \left[ 1 - \frac{\omega_{*i}}{\omega} \left( 1 + \eta_i \right) \right] \xi - \frac{1}{\vA^2} \left( 2\omega_{\bm{\kappa}} + \frac{\beta_i}{2} \omega_{*p}\right) \omega_{*p}\xi,
 \end{split}
\label{ballEq}
\end{equation}
where we have used the fact that $\bm{b}\dg T_i = 0$ and introduced the notation $b = k_\perp^2 \rho_i^2 / 2$.

Inserting explicit geometric expressions for $\bm{b}\dg$, and dropping the $\beta_i$ term, \eref{ballEq} matches the first line of (3.40) of \citet{tang1980kbm}. The second line of the ballooning equation in \citet{tang1980kbm} is small in our ordering, because $ \omega_* / \omega_{\bm{\kappa}} \sim L_n / R  \ll 1$. Taking the JET-ILW pedestals from \citet{hatch2017jetilwped} as typical, then $R/L_n \gtrsim 50$ in the pedestal region.

Hence we see that both formulations capture the key physics of ballooning modes, with our model retaining more electron and finite-$\beta$ physics, and the model of \citet{tang1980kbm} retaining plasma compressibility effects (the final term in 3.6 and 3.40 of that paper).

\section{Derivations for the Inter-ELM Equations}
\label{SlowDeriv}
In this appendix, we perform the derivations required for the Inter-ELM equations. 

\subsection{Electron behaviour on the Slow Timescale}
\label{SlowElectrons}
In this section we discuss the possible solutions to the lowest-order electron kinetic equation, under the inter-ELM orderings.
Taking \eref{vfp} for the electrons, and applying the inter-ELM orderings, we see that the largest terms are those involving electron velocities:
\begin{equation}
w_\parallel \bm{b}\dg f_e + ew_\parallel\bm{b}\dg \pot \pd{f_e}{\energy}  = \collop[f_e].
\label{SDtmp0}
\end{equation}
We multiply this by $1+\ln f_e$, and integrate over all velocities to obtain
\begin{equation}
\dv\left( \bm{b} \vint w_\parallel f_e \ln f_e\right)  = \vint \ln f_e \collop[f_e],
\label{SDtmp1}
\end{equation}
where we use the fact that $f_e \rightarrow 0$ implies $f_e \ln f_e \rightarrow 0$ at high energies.
We will now {\textit{assume}} that the magnetic field has good flux surfaces, at least to some approximation.
Then, averaging over these surfaces, and applying the $H$-theorem for the nonlinear collision operator, we obtain Maxwell-Boltzmann electrons
\begin{equation}
f_e = \frac{N_e(\psi)}{\pi^{3/2} \vth[e]^3} e^{- \infrac{\energy[e] - e\pot}{T_e(\psi)}},
\end{equation}
where $N_e$ and $T_e$ are flux functions. The parameter $N_e$ is related to the electron density $n_e$ by
\begin{equation}
N_e(\psi) = n_e e^{-\infrac{e\pot}{T_e}}.
\end{equation}
We can rewrite this equation as
\begin{equation}
 \bm{b}\dg\left( \frac{T_e}{e} \ln n_e - \pot \right) = 0.
 \label{APD7}
\end{equation}

Note that because the induced parallel electric field is small compared to the electrostatic field there is a field-line preserving flow~\citep{newcombLinesOfForce,flowtome2-electrons}. Thus, the assumption of good magnetic surfaces is consistent -- if we assume that they exist initially for some time then they will be preserved by this flow.

Going to the next order in $\sqrt{\gkeps}$, we expand $f_e = \feN{0} + \feN{1} +\cdots$ where $\feN{1} \sim \sqrt{\gkeps} \feN{0}$ and so forth. Using this expansion, we obtain the kinetic equation 
\begin{equation}
\begin{split}
\left(\pd{ }{t} + \uE \dgR[e]{ } \right)& \feN{0} + w_\parallel \bm{b}\dg \feN{1} + e w_\parallel \left( \bm{b}\dg \pot^{(1)} + \pd{A_\parallel}{t} \right)\pd{\feN{0}}{\energy[e]}\\&
	+e w_\parallel\bm{b}\dg \pot \pd{\feN{1}}{\energy} + \gyroR{\ensav{ \daccel[e]\cdot\pd{\turbf[e]}{\bm{v}}}}  = \lincol[\feN{1}] + \source[e].
	\end{split}
\label{SlowAP-ke1}
\end{equation}

Integrating this over all velocities we obtain
\begin{equation}
\left( \pd{ }{t} + \uE \dg \right) n_e + \dv\left( \bm{b} \upar[e] n_e \right) + \dv\ensav{\vint \vchiR \turbf[e] } = \Psource[e],
\label{SlowEContAP}
\end{equation}
where we have used \eref{ElectronTurbAP} for the fluctuations, and introduced
\begin{equation}
\upar[e] = \vint w_\parallel \feN{1},
\end{equation}
because there is no electron parallel velocity in $\feN{0}$.

If we multiply \eref{SlowAP-ke1} by $( \infrac{\energy[e]}{T_e} - \infrac{3}{2})$ and integrate over all velocities we find that
\begin{equation}
\begin{split}
\frac{3}{2}&\left( \pd { }{t} + \uE \dg\right) n_e T_e  + \dv\ensav{\vint \left(\frac{\energy[e]}{T_e} - \frac{3}{2}\right) \vchiR \turbf[e]} \\
	&= e n_e \upar[e] \bm{b}\dg\pot +  \ensav{ e\left(\bm{b}\dg\turbPot + \frac{1}{c}\pd{\turbApar}{t}\right) \delta\upar[e] } - \dv\left( q_{\parallel e}^{(1)} \bm{b}\right) + \Esource[e],
\end{split}
\end{equation}
where again we have used \eref{ElectronTurbAP} for the fluctuations, and $q_{\parallel e}^{(1)}$ is a moment of $\feN{1}$ which is not solved for. As $T_e$ is a flux function, we average this equation over the flux surfaces to finally obtain
\begin{equation}
\begin{split}
& \frac{3}{2}\fav{ \left( \pd { }{t} + \uE \dg\right) n_e T_e}  + \fav{\dv\ensav{\vint \left(\frac{\energy[e]}{T_e} - \frac{3}{2}\right) \vchiR \turbf[e]}} \\
		&\quad= \fav{e n_e \upar[e] \bm{b}\dg \pot} -\fav{\ensav{ e \delta E_{\parallel} \delta\upar[e]} } +\fav{\Esource[e]} .
\end{split}
\label{SlowETempAP}
\end{equation}

\subsection{Pedestal Gyrokinetic Turbulence}
\label{APPedTurb}
In this section, we derive the equations governing the turbulence in the pedestal.
To obtain the equations for the inter-ELM turbulence, we take the fluctuating part of the fundamental kinetic equation, written in Catto-transformed variables
\begin{equation}
\left(\pd{ }{t} + \dot{\bm{R}}_s \dgR{ } + \denergy \pd{ }{\energy} + \dot{\gyr}\pd{ }{\gyr}\right)\turbf 
+ \daccel \cdot\left( \pd{f_s}{\bm{v}} + \pd{\turbf}{\bm{v}}\right) = \lincol[\turbf].
\label{fluct-KE}
\end{equation}
The leading order of this equation is
\begin{equation}
\cycfreq \pd{\turbf}{\gyr} = -\daccel \cdot \pd{f_s}{\bm{v}} =  \frac{Z_s e}{m_s}\bm{v}_\perp \dg \turbPot \pd{f_s}{\energy} + \frac{Z_s e}{m_s} \bm{v}_\perp \dg\turbPot \pd{f_s}{\magmom},
\end{equation}
whose solution can be written as
\begin{equation}
\turbf = h_s(\bm{R}_s,\energy,\magmom,t) + Z_se \turbPot \pd{f_s}{\energy} + Z_se \left( \turbPot - \gyroR{\turbPot}\right) \pd{f_s}{\magmom},
\end{equation}
where we have split out part of the gyrophase-independent piece to more closely mirror other derivations.

Substituting this back into \eref{fluct-KE} and gyroaveraging gives
\begin{equation}
\begin{split}
&\left[\pd{ }{t} + \left(w_\parallel \bm{b}(\bm{R}_s) + \uE \right) \dgR{ } \right] h_s + \gyroR{\daccel \cdot\pd{\delta f_s}{\bm{v}}} \\
		&\qquad= Z_s e \pd{f_s}{\energy} \left( \pd{ }{t} + \uE\dgR{ }\right) \gyroR{\turbChi } - \frac{c}{B}\bm{b}\times\grad\gyroR{\turbChi} \dg f_s \\
&\qquad\qquad+ \gyroR{\lincol[h_s]} + \Or({\gkeps} \cycfreq h_s),
	\end{split}
\label{gkeAP}
\end{equation}
where we have used explicit forms for the time derivatives, written $f_s = f_s(\bm{R}_s,\energy,\magmom,t)$ and used the chain rule.
By exactly the same manipulations as we use in the next section for the nonlinear term in the transport equations, we can show that
\begin{equation}
\gyroR{\daccel \cdot\pd{\delta f_s}{\bm{v}}} = \frac{c}{B} \bm{b}\times\grad\gyroR{\turbChi} \dgR{h_s}.
\end{equation}
All that remains to derive the gyrokinetic equations in \Secref{SecPedTurb} is to specialise the above results to electrons and ions.
The field equations follow immediately from the fluctuating parts of quasineutrality and Amp\`ere's law.

\subsection{Turbulent transport in the pedestal}
To derive the terms involving the fluctuations in the inter-ELM equations, we need the following results.
Firstly, the nonlinear term in the averaged kinetic equation is
\begin{equation}
\gyroR{\ensav{ \daccel\cdot\pd{\turbf}{\bm{v}}}},
\end{equation}
and so in order to perform the gyroaverage, we need to convert the velocity derivative into a derivative of the gyrokinetic variables. 
We require this term up only to $\Or( \vth[i] f_s / L_\parallel)$ and will systematically neglect any higher-order terms.
We first do this for the non-adiabatic part of the distribution function:
\begin{equation}
\begin{split}
&\gyroR{\ensav{\daccel\cdot{\pd{h_s}{\bm{v}}}}} =\\
		&\qquad\ensav{ \gyroR{ \daccel \cdot \pd{\bm{R}_s}{\bm{v}} \dgR{h_s} + m_s \daccel \cdot \bm{v} \pd{h_s}{\energy} + \frac{m_s}{B} \daccel \cdot \bm{v}_\perp \pd{h_s}{\magmom}}}.
\end{split}
\label{fooTmp1}
\end{equation}
Using the explicit form of the fluctuating acceleration, we see that
\begin{equation}
\gyroR{\daccel \cdot\bm{v}} = -\frac{Z_se}{m_s} w_\parallel \gyroR{ \bm{b}\dg\turbPot },
\end{equation}
and
\begin{equation}
\gyroR{\daccel \cdot \bm{v}_\perp} = 0,
\end{equation}
where we have used $\gyroR{\bm{v}_\perp\dg \delta G} = 0$ for any fluctuating field $\delta G$ that is independent of gyrophase at fixed $\bm{r}$ (see (A.4) of \citep{flowtome1}).

These results are then substituted back into \eref{fooTmp1} to arrive at
\begin{equation}
\begin{split}
&\gyroR{\ensav{\daccel\cdot{\pd{h_s}{\bm{v}}}}} = \ensav{  \frac{1}{\cycfreq} \gyroR{\daccel} \cdot \bm{b}\times \pd{h_s}{\bm{R}_s} } -\frac{Z_se}{m_s} w_\parallel \ensav{\gyroR{ \bm{b}\dg\turbPot} \pd{h_s}{\energy}}.
\end{split}
\end{equation}
If we also use
\begin{equation}
\gyroR{\daccel} = -\frac{Z_s e}{m_s} \grad \gyroR{ \turbPot - \frac{1}{c}{w_\parallel}\turbApar},
\end{equation}
then we obtain
\begin{equation}
\begin{split}
&\gyroR{\ensav{\daccel\cdot{\pd{h_s}{\bm{v}}}}} = \ensav{  \vchiR \dgR{h_s} } -\frac{Z_s}{m_s} w_\parallel \ensav{\gyroR{ \bm{b}\dg\turbPot }}.
\end{split}
\label{APTurbTranspH}
\end{equation}
To find a full expression for the original nonlinear term, we need to also handle the Boltzmann-like pieces of $\turbf$.

For the electrons this is simple, and we obtain
\begin{equation}
\begin{split}
\gyroR{\ensav{ \daccel[e]\cdot\pd{\turbf[e]}{\bm{v}}}} = \ensav{  \vchi \dgR{h_e} }+\frac{e}{m_e} w_\parallel \ensav{ \bm{b}\dg\turbPot  \pd{h_s}{\energy}},
		\end{split}
		\label{ElectronTurbAP}
\end{equation}
where gyroaverages have been removed as the electron gyroradius is smaller than all scales of interest.
The first term will vanish when summing over the signs of $w_\parallel$ and so does not appear in any resulting equations.

For the ions, we can neglect terms involving $w_\parallel$ as they are small in the mass-ratio compared to the same terms for electrons. Thus we have
\begin{equation}
\gyroR{\ensav{ \daccel[i]\cdot\pd{\turbf[i]}{\bm{v}}}} = \ensav{  \frac{c}{B} \bm{b}\times\grad\gyroR{\turbPot} \dgR[i]{h_i} },
\end{equation}
because all the non-adiabatic pieces only contain terms proportional to $\turbPot$, which vanish under the turbulence average.
\section{Matching to Multiscale Gyrokinetics}
\label{APGKMatch}
In this appendix we detail how our inter-ELM transport equations can be matched at the top of the pedestal to multiscale gyrokinetics.
We will demonstrate that there exists a subsidiary expansion of multiscale gyrokinetics and a different subsidiary expansion
of our inter-ELM equations that both result in the same system of equations.
In the main text we have explained the concepts behind these subsidiary orderings, and provided the orderings themselves. Thus, all that remains is to actually perform the subsidiary expansions.

\subsection{Subsidiary Expansion of Multiscale Gyrokinetics}
In this section we now apply the ordering from \Secref{SecGKMatch} to the low-Mach number equations of \citep{flowtome1} (henceforth in this appendix we will refer to this as paper I, and refer to equations in that reference with a prefix I-). Consistently with the ordering discussed in \Secref{SecGKMatch}, we drop all flow-related terms, either because they are small, or because their shear is small and so they can be removed by a Galilean transformation.
Taking (I-248) and retaining the leading order in $\apSmall$, the gyrokinetic equation for ions is, 
\begin{equation}
\begin{split}
&\pd{h_i}{t} + \frac{c}{B} \bm{b}\times\nabla\gyroR{\delta\pot} \cdot\ddR{h_i} - \gyroR{\lincol[h_i]} \\
&\qquad=\frac{Z_i e F_{0i}}{T_i}\pd{\gyroR{\delta\pot}}{t} - \frac{c}{B} \bm{b}\times\nabla\gyroR{\delta\pot}\dg\psi\pd{F_{0s}}{\psi},
	\end{split}
\label{LM-gke-i}
\end{equation}
where we have used the definition (I-250) of $\gkpot$ and the orderings for the magnetic field to replace $\gkpot$ with $\delta\pot$ in this equation.
For electrons, we have
\begin{equation}
\begin{split}
&\pd{h_e}{t} + \left[ v_\parallel \left(\Meanb+\delta\bm{b}\right) + \frac{c}{B} \bm{b}\times\nabla{\delta\pot} \right]\cdot\ddR{h_e} - {\lincol[h_e]} \\
&\qquad=-\frac{e F_{0e}}{T_e}\pd{\gkpot}{t} -\pd{F_{0s}}{\psi} \frac{c}{B} \bm{b}\times\nabla\gkpot\dg\psi,
	\end{split}
\label{LM-gke-e}
\end{equation}
where here we have been able to drop all gyroaverages because $k_\perp \rho_e \sim \xi \ll 1$. 
Because of the ordering on $\delta B_\parallel$, $\gkpot$ is now given by the simpler form
\begin{equation}
\gkpot = \delta\pot - \frac{v_\parallel}{c} \delta A_\parallel.
\end{equation}
As the flow is sub-sonic, we have replaced $w_\parallel \approx v_\parallel$ in these expressions.
In addition, we have the field equations (I-251) and (I-149)  
\begin{equation}
 \sum_s \frac{Z_s^2 e^2 n_s\delpot}{T_s}   = \sum_s Z_s e \wint  \gyror{h_s}
\end{equation}
and
\begin{equation}
-\nabla_\perp^2 \delAp = \frac{4\pi}{c} \sum_s Z_s e \wint v_\parallel \gyror{h_s}.
\end{equation}
which determine $\delta\pot$ and $\delta A_\parallel$ respectively.

Turning now to the transport equations, we first have (I-252), for particle transport
\begin{equation}
\phantom{\frac{3}{2}}\frac{1}{V'}\left.\pd{}{t}\right|_\psi V'{n_s} + \frac{1}{V'} \pd{}{\psi} V'\fav{ \ParticleFlux } = \fav{\Psource},
\label{dndt-lowmach}
\end{equation}
in which we have not dropped any terms, because the timescale of the sources and the evolution of the equilibrium is defined by the turbulent transport time. So all terms are of the same order in $\apSmall$ by definition. 
However, evaluating the terms in the particle flux (I-170) gives
\begin{equation}
\fav{\ParticleFlux} = \fav{\ensav{\wint \gyror{h_s\,\vchi}\dg\psi}},
\end{equation}
where we have dropped both collisional transport terms as they are $\Or( \nu_{ss} \rho_s^2 |\nabla\psi| / L_\perp)$, and we have dropped the electric field term as $\fav{\bm{E}\cdot\bm{B}}$ is small in the mass ratio (see Appendix C of \citet{flowtome2-electrons}).
Secondly, we have the pressure evolution equation (I-254):
\begin{equation}
\begin{split}
\frac{3}{2}\frac{1}{V'}\ddtpsi V'p_s &+ \frac{1}{V'} \pd{}{\psi}  V' \fav{\HeatFlux} = \\
&\CompHeat + \TurbPow +  \fav{\Esource},
\end{split}
\label{dpdt-lowmach}
\end{equation}
where again we have dropped the flow related terms, including the viscous heating, because the shearing rate is small, and the Ohmic heating is dropped because $\fav{\bm{E}\cdot\bm{B}}$ is small in the mass-ratio. Similarly, the collisional energy exchange term is small.
The heat flux is now
\begin{equation}
\fav{\HeatFlux} = \fav{\ensav{\wint \energy \gyror{h_s \vchi}\dg\psi}},
\end{equation}
and the turbulent heating is given by (I-259):
\begin{equation}
\TurbPow = Z_s e \fav{ \ensav{\wint \gyror{h_s\left( \pd{}{t} + \bm{u}\dg \right) \gkpot} }} = \TurbColl - \TurbInj.
\end{equation}
Examining the size of the terms in this formula, we immediately see that the turbulent heating occurs on a rate
\begin{equation}
\TurbPow \sim \frac{\gkeps^3}{\xi^3} \cycfreq[i],
\end{equation}
one order in $\xi$ too small to appear in our transport equation. This is due to the fact that the rate of perpendicular transport is 
enhanced by a power of $\xi$ due to the decreasing distance $L_\perp$ over which transport has to occur, whilst both terms benefit from the 
increased amplitude and frequency of the turbulence.

At this point we have to make an assumption on the background magnetic field. 
We will assume that $\psi$ doesn't change by $\Or(1)$ on the transport timescale -- this is for consistency with our $\beta$ ordering.
Thus, we order
\begin{equation}
\pd{\psi}{t} \sim \apSmall \frac{\psi}{\tau_E}.
\end{equation}
Applying this ordering means that we drop the compressional heating term $\CompHeat$ from \eref{dpdt-lowmach}. 
Thus, there are no heat sources in our transport equation save for the explicit source.

\subsection{Subsidiary Expansion of Inter-ELM Equations} 
Now we turn to obtaining the same equations from the inter-ELM system.
We apply our subsidiary ordering first to \eref{EParSmall} to find
\begin{equation}
\bm{b}\dg\pot = - \frac{T_e}{e} \bm{b}\dg \ln n_e = \Or\left( \zeta \frac{\pot}{L_\parallel}\right),
	\label{APFluxPot}
\end{equation}
so, to lowest order in $\zeta$, $\pot = \overbar{\pot}(\psi)$.
Then the ion kinetic equation \eref{SlowIonKE}, to lowest order in $\zeta$, becomes
\begin{equation}
w_\parallel \bm{b}\dg f_i = \collop[f_i].
\end{equation}
The general solution of this equation is, by the usual arguments, a stationary Maxwellian, with no parallel flow (i.e. $\upar[i] \ll \vth[i]$).
In addition, the ion densities and temperatures are functions only of $\psi$. 

Examining parallel Amp\`ere's law, we see that 
\begin{equation}
e\left( n_e \upar[e] - \sum_{s=i} Z_i n_i \upar[i]\right) = \Or\left( e n_e \vth[i] \frac{\beta\rho_i}{L_s}\right),
\end{equation}
and so, as the shear length $L_s$ gets longer as $\zeta^{-1}$, we have that the electron parallel flow is also small in $\zeta$ relative to the ion sound speed. Again, the electron density is now a flux function to lowest order in $\zeta$. This in fact holds to two orders in $\zeta$ as we can iterate \eref{APFluxPot} to show $\pot$ is a flux-function to $\Or(\zeta^2)$, assume axisymmetry and repeat the above argument.

As the ions are Maxwellian, we can integrate \eref{SlowIonKE} over all velocities to find a continuity equation equivalent to \eref{eTransp1} for the electrons. Flux-surface averaging any one of these continuity equations, we obtain
\begin{equation}
\frac{1}{V'} \ddtpsi V' n_s  + \frac{1}{V'}\pd{}{\psi} \fav{\ensav{\wint \gyror{h_s\,\bm{V}_\pot}\dg\psi}}  = \fav{\Psource},
\end{equation}
where we have used identities from Section 3.4 of I for the flux-surface averages.
This is the same transport equation as we obtained from gyrokinetics. 

Multiplying \eref{SlowIonKE} by $\energy - 3T_s/2$, and integrating gives us the following heat transport equation
\begin{equation}
\begin{split}
\frac{3}{2}\fav{ \left( \pd { }{t} \right) n_i T_i}  + \fav{\dv\ensav{\vint \left(\frac{\energy[i]}{T_i} - \frac{3}{2}\right) \vchiR \turbf[i]}} = \fav{\Esource[i]},
\end{split}
\end{equation}
which, upon using the usual identities for the flux-surface average becomes the gyrokinetic heat transport equation.
Similarly, taking \eref{eTransp2} and using our solution for the mean potential and the electron parallel velocity, we arrive at the same equation. Thus, all heat transport equations match.

However, we still need to show that the equations governing $\turbPot$ and $h_s$ are also the same. 
Using our Maxwellian solution for $f_i$ in \eref{ionGKE}, we have that
\begin{equation}
\begin{split}
\pd{h_i}{t} + \frac{c}{B} \pb{\gyroR{\turbPot}}{h_i} = \frac{Z_ie}{T_i}\pd{\gyroR{\turbPot}}{t} - \frac{c}{B} \bm{b}\times\nabla\gyroR{\turbPot}\cdot \nabla\psi \pd{f_i}{\psi},
\end{split}
\end{equation}
where we have used the fact that the differences between the energy variable used here and the one defined by (I-241) are all negligible in $\zeta$. Similarly the distinction between the exact $\magmom$ that we use here and the lowest-order one used in Paper I can also be ignored. We have also used the fact that we can transform the mean flow out of this equation due to its low shear.
For the electrons there is no such difficulty, there are no flow terms in \eref{electronGKE} and it is already in exactly the same form as the equation derived from gyrokinetics.
The field equations also trivially match in this limit. Thus, we have proved what we set out to. The two subsidiary expansions given in \Secref{SecGKMatch} result in identical sets of equations for the pedestal top region.


\begin{thebibliography}{63}
\expandafter\ifx\csname natexlab\endcsname\relax\def\natexlab#1{#1}\fi

\bibitem[Abel \& Cowley(2013)]{flowtome2-electrons}
{\sc Abel, I.~G. \& Cowley, S.~C.} 2013 Multiscale gyrokinetics for rotating
  tokamak plasmas {II}: {Reduced} models for electron dynamics. {\em New\ J.\
  Phys.\/} {\bf 15}, 023041.

\bibitem[Abel {\em et~al.\/}(2013)Abel, Plunk, Wang, Barnes, Cowley, Dorland \&
  Schekochihin]{flowtome1}
{\sc Abel, I.~G., Plunk, G.~G., Wang, E., Barnes, M.~A., Cowley, S.~C.,
  Dorland, W. \& Schekochihin, A.~A.} 2013 Multiscale gyrokinetics for rotating
  tokamak plasmas : {Fluctuations} and transport. {\em Rep. Prog. Phys\/} {\bf
  76}, 116201.

\bibitem[Barnes {\em et~al.\/}(2011)Barnes, Parra \&
  Schekochihin]{barnes2011critically}
{\sc Barnes, M., Parra, F.~I. \& Schekochihin, A.~A.} 2011 Critically balanced
  ion temperature gradient turbulence in fusion plasmas. {\em Phys.\ Rev.\
  Lett.\/} {\bf 107}~(11), 115003.

\bibitem[Barnes {\em et~al.\/}(2010)Barnes, Abel, Dorland, G{\"o}rler, Hammett
  \& Jenko]{barnes2009trinity}
{\sc Barnes, M.~A., Abel, I.~G., Dorland, W., G{\"o}rler, T., Hammett, G.~W. \&
  Jenko, F.} 2010 Direct multiscale coupling of a transport code to gyrokinetic
  turbulence codes. {\em Phys.\ Plasmas\/} {\bf 17}, 056109.

\bibitem[Beer {\em et~al.\/}(1995)Beer, Cowley \& Hammett]{beer1995field}
{\sc Beer, M.~A., Cowley, S.~C. \& Hammett, G.~W.} 1995 Field-aligned
  coordinates for nonlinear simulations of tokamak turbulence. {\em Phys.\
  Plasmas\/} {\bf 2}, 2687.

\bibitem[Berkowitz \& Gardner(1959)]{berkowitz1959asymptoticparticle}
{\sc Berkowitz, J. \& Gardner, C.~S.} 1959 On the asymptotic series expansion
  of the motion of a charged particle in slowly varying fields. {\em Comm. Pure
  Appl. Maths\/} {\bf 12}, 501.

\bibitem[Beurskens {\em et~al.\/}(2011)Beurskens, Osborne, Schneider, Wolfrum,
  Frassinetti, Groebner, Lomas, Nunes, Saarelma, Scannell, Snyder, Zarzoso,
  Balboa, Bray, Brix, Flanagan, Giroud, Giovannozzi, Kempenaars, Loarte, de~la
  Luna, Maddison, Maggi, McDonald, Pasqualotto, Saibene, Sartori, Solano, Walsh
  \& Zabeo]{beurskens2011pedestals}
{\sc Beurskens, M. N.~A., Osborne, T.~H., Schneider, P.~A., Wolfrum, E.,
  Frassinetti, L., Groebner, R., Lomas, P., Nunes, I., Saarelma, S., Scannell,
  R., Snyder, P.~B., Zarzoso, D., Balboa, I., Bray, B., Brix, M., Flanagan, J.,
  Giroud, C., Giovannozzi, E., Kempenaars, M., Loarte, A., de~la Luna, E.,
  Maddison, G., Maggi, C.~F., McDonald, D., Pasqualotto, R., Saibene, G.,
  Sartori, R., Solano, Emilia~R., Walsh, M. \& Zabeo, L.} 2011 H-mode pedestal
  scaling in {DIII-D}, {ASDEX} upgrade, and {JET}. {\em Phys.\ Plasmas\/} {\bf
  18}, 056120.

\bibitem[Burrell {\em et~al.\/}(2002)Burrell, Austin, Brennan, DeBoo, Doyle,
  Gohil, Greenfield, Groebner, Lao, Luce, Makowski, McKee, Moyer, Osborne,
  Porkolab, Rhodes, Rost, Schaffer, Stallard, Strait, Wade, Wang, Watkins, West
  \& Zeng]{qhmode2002}
{\sc Burrell, K~H, Austin, M~E, Brennan, D~P, DeBoo, J~C, Doyle, E~J, Gohil, P,
  Greenfield, C~M, Groebner, R~J, Lao, L~L, Luce, T~C, Makowski, M~A, McKee,
  G~R, Moyer, R~A, Osborne, T~H, Porkolab, M, Rhodes, T~L, Rost, J~C, Schaffer,
  M~J, Stallard, B~W, Strait, E~J, Wade, M~R, Wang, G, Watkins, J~G, West, W~P
  \& Zeng, L} 2002 Quiescent {H-mode plasmas in the DIII-D tokamak}. {\em
  Plasma\ Phys.\ Control.\ Fusion\/} {\bf 44}, A253.

\bibitem[Chang {\em et~al.\/}(2006)Chang, Ku, Adams, D’Azevedo, Chen,
  Cummings, Ethier, Greengard, Hahm, Hinton, Keyes, Klasky, Lee, Lin,
  Nishimura, Parker, Samtaney, Stotler, Weitzner, Worley, Zorin \& the
  {CPES}~Team]{chang2006xgcped}
{\sc Chang, C.S., Ku, S., Adams, M., D’Azevedo, G., Chen, Y., Cummings, J.,
  Ethier, S., Greengard, L., Hahm, T.S., Hinton, F., Keyes, D., Klasky, S.,
  Lee, W.~W., Lin, Z., Nishimura, Y., Parker, S., Samtaney, R., Stotler, D.,
  Weitzner, H., Worley, P., Zorin, D. \& the {CPES}~Team} 2006 Integrated
  particle simulation of neoclassical and turbulence physics in the tokamak
  pedestal/edge region using {XGC}. In {\em Proceedings of the 21st IAEA Fusion
  Energy Conference\/}, pp. {TH/P6--14}. IAEA.

\bibitem[Chang {\em et~al.\/}(2017)Chang, Ku, Loarte, Parail, Köchl,
  Romanelli, Maingi, Ahn, Gray, Hughes, LaBombard, Leonard, Makowski \&
  Terry]{chang2017divertor}
{\sc Chang, C.S., Ku, S., Loarte, A., Parail, V., Köchl, F., Romanelli, M.,
  Maingi, R., Ahn, J.-W., Gray, T., Hughes, J., LaBombard, B., Leonard, T.,
  Makowski, M. \& Terry, J.} 2017 Gyrokinetic projection of the divertor
  heat-flux width from present tokamaks to iter. {\em Nucl. Fusion\/} {\bf
  57}~(11), 116023.

\bibitem[Churchill {\em et~al.\/}(2017)Churchill, Chang, Ku \&
  Dominski]{churchill2017xgcedge}
{\sc Churchill, R~M, Chang, C~S, Ku, S \& Dominski, J} 2017 Pedestal and edge
  electrostatic turbulence characteristics from an {XGC1} gyrokinetic
  simulation. {\em Plasma\ Phys.\ Control.\ Fusion\/} {\bf 59}, 105014.

\bibitem[Connor {\em et~al.\/}(1998)Connor, Hastie, Wilson \&
  Miller]{connor-edge-stability}
{\sc Connor, J.~W., Hastie, R.~J., Wilson, H.~R. \& Miller, R.~L.} 1998
  Magnetohydrodynamic stability of tokamak edge plasmas. {\em Phys.\ Plasmas\/}
  {\bf 5}, 2687.

\bibitem[Cowley {\em et~al.\/}(2015)Cowley, Cowley, Henneberg \&
  Wilson]{CowleyExplosiveProcRoySoc}
{\sc Cowley, S.~C., Cowley, B., Henneberg, S.~A. \& Wilson, H.~R.} 2015
  Explosive instability and erupting flux tubes in a magnetized plasma. {\em
  Proc.\ R.\ Soc.\ A\/} {\bf 471}, 20140913.

\bibitem[Dickinson {\em et~al.\/}(2012)Dickinson, Roach, Saarelma, Scannell,
  Kirk \& Wilson]{dickinson2012prl}
{\sc Dickinson, D., Roach, C.~M., Saarelma, S., Scannell, R., Kirk, A. \&
  Wilson, H.~R.} 2012 Kinetic instabilities that limit $\ensuremath{\beta}$ in
  the edge of a tokamak plasma: A picture of an {$H$}-mode pedestal. {\em
  Phys.\ Rev.\ Lett.\/} {\bf 108}, 135002.

\bibitem[Dickinson {\em et~al.\/}(2013)Dickinson, Roach, Saarelma, Scannell,
  Kirk \& Wilson]{dickinson2013microtearing}
{\sc Dickinson, D, Roach, C~M, Saarelma, S, Scannell, R, Kirk, A \& Wilson,
  H~R} 2013 Microtearing modes at the top of the pedestal. {\em Plasma\ Phys.\
  Control.\ Fusion\/} {\bf 55}, 074006.

\bibitem[D’haeseleer {\em et~al.\/}(1991)D’haeseleer, Hitchon, Callen \&
  Shohet]{DhaeseleerFlux}
{\sc D’haeseleer, W.~D., Hitchon, W. N.~G., Callen, J.~D. \& Shohet, J.~L.}
  1991 {\em Flux Coordinates and Magnetic Field Structure\/}. Springer-Verlag,
  Berlin.

\bibitem[Freidberg(2014)]{IdealMHD}
{\sc Freidberg, J.~P.} 2014 {\em Ideal MHD\/}. Cambridge University Press.

\bibitem[Frieman \& Chen(1982)]{frieman1982nge}
{\sc Frieman, E.~A. \& Chen, L.} 1982 {Nonlinear gyrokinetic equations for
  low-frequency electromagnetic waves in general plasma equilibria}. {\em
  Phys.\ Fluids\/} {\bf 25}, 502.

\bibitem[Goldston(2012)]{goldstonScaling}
{\sc Goldston, R.~J.} 2012 Heuristic drift-based model of the power scrape-off
  width in low-gas-puff {H}-mode tokamaks. {\em Nucl. Fusion\/} {\bf 52}~(1),
  013009.

\bibitem[Ham {\em et~al.\/}(2016)Ham, Cowley, Brochard \&
  Wilson]{ham2016nonlinear}
{\sc Ham, C.~J., Cowley, S.~C., Brochard, G. \& Wilson, H.~R.} 2016 Nonlinear
  stability and saturation of ballooning modes in tokamaks. {\em Phys.\ Rev.\
  Lett.\/} {\bf 116}, 235001.

\bibitem[Hammett {\em et~al.\/}(1993)Hammett, Beer, Dorland, Cowley \&
  Smith]{hammett1993developments}
{\sc Hammett, G.~W., Beer, M.~A., Dorland, W., Cowley, S.~C. \& Smith, S.~A.}
  1993 Developments in the gyrofluid approach to tokamak turbulence
  simulations. {\em Plasma\ Phys.\ Control.\ Fusion\/} {\bf 35}, 973.

\bibitem[Hassam {\em et~al.\/}(1991)Hassam, Antonsen, Drake \&
  Liu]{adilStringerSpin}
{\sc Hassam, A.~B., Antonsen, T.~M., Drake, J.~F. \& Liu, C.~S.} 1991
  Spontaneous poloidal spin-up of tokamaks and the transition to the {$H$}
  mode. {\em Phys.\ Rev.\ Lett.\/} {\bf 66}, 309.

\bibitem[Hatch {\em et~al.\/}(2016)Hatch, Kotschenreuther, Mahajan, Valanju,
  Jenko, Told, Görler \& Saarelma]{hatch2016mtm}
{\sc Hatch, D.R., Kotschenreuther, M., Mahajan, S., Valanju, P., Jenko, F.,
  Told, D., Görler, T. \& Saarelma, S.} 2016 Microtearing turbulence limiting
  the {JET-ILW} pedestal. {\em Nucl. Fusion\/} {\bf 56}, 104003.

\bibitem[Hatch {\em et~al.\/}(2017)Hatch, Kotschenreuther, Mahajan, Valanju \&
  Liu]{hatch2017jetilwped}
{\sc Hatch, D.R., Kotschenreuther, M., Mahajan, S., Valanju, P. \& Liu, X.}
  2017 A gyrokinetic perspective on the {JET-ILW} pedestal. {\em Nucl.
  Fusion\/} {\bf 57}, 036020.

\bibitem[Hazeltine \& Meiss(2003)]{hazeltine2003plasma}
{\sc Hazeltine, R.D. \& Meiss, J.D.} 2003 {\em {Plasma Confinement}\/}. Dover
  Publications.

\bibitem[Hillesheim {\em et~al.\/}(2016)Hillesheim, Dickinson, Roach, Saarelma,
  Scannell, Kirk, Crocker, Peebles, Meyer \& the
  MAST~Team]{hillesheim2016pedestal}
{\sc Hillesheim, J~C, Dickinson, D, Roach, C~M, Saarelma, S, Scannell, R, Kirk,
  A, Crocker, N~A, Peebles, W~A, Meyer, H \& the MAST~Team} 2016
  Intermediate-$k$ density and magnetic field fluctuations during inter-{ELM}
  pedestal evolution in {MAST}. {\em Plasma\ Phys.\ Control.\ Fusion\/} {\bf
  58}, 014020.

\bibitem[Jackson(1960)]{jackson1960twostream}
{\sc Jackson, E.~Atlee} 1960 Drift instabilities in a maxwellian plasma. {\em
  Phys.\ Fluids\/} {\bf 3}, 786.

\bibitem[{Jorge} {\em et~al.\/}(2017){Jorge}, {Ricci} \&
  {Loureiro}]{jorge2017gbscoll}
{\sc {Jorge}, R., {Ricci}, P. \& {Loureiro}, N.~F.} 2017 {A full-F
  Drift-Kinetic Analytical Model for SOL Plasma Dynamics at Arbitrary
  Collisionality}. {\em ArXiv e-prints\/} .

\bibitem[Kagan \& Catto(2008)]{kagan2008apg}
{\sc Kagan, G. \& Catto, P.~J.} 2008 {Arbitrary poloidal gyroradius effects in
  tokamak pedestals and transport barriers}. {\em Plasma\ Phys.\ Control.\
  Fusion\/} {\bf 50}, 085010.

\bibitem[Kagan \& Catto(2010)]{kagan2010neoclassical}
{\sc Kagan, G. \& Catto, P.~J.} 2010 Neoclassical ion heat flux and poloidal
  flow in a tokamak pedestal. {\em Plasma\ Phys.\ Control.\ Fusion\/} {\bf 52},
  055004.

\bibitem[Kotschenreuther {\em et~al.\/}(2017)Kotschenreuther, Hatch, Mahajan,
  Valanju, Zheng \& Liu]{mikek2017pedestalfail}
{\sc Kotschenreuther, M., Hatch, D.R., Mahajan, S., Valanju, P., Zheng, L. \&
  Liu, X.} 2017 Pedestal transport in {H}-mode plasmas for fusion gain. {\em
  Nucl. Fusion\/} {\bf 57}, 064001.

\bibitem[Kruskal(1958)]{kruskal1958gyration}
{\sc Kruskal, M.} 1958 The gyration of a charged particle. {\em Project
  Matterhorn Report PM-S-33\/} .

\bibitem[Ku {\em et~al.\/}(2006)Ku, Chang, Adams, Cummings, Hinton, Keyes,
  Klasky, Lee, Lin, Parker \& the CPES~team]{ku2006gkped}
{\sc Ku, S, Chang, C-S, Adams, M, Cummings, J, Hinton, F, Keyes, D, Klasky, S,
  Lee, W, Lin, Z, Parker, S \& the CPES~team} 2006 Gyrokinetic particle
  simulation of neoclassical transport in the pedestal/scrape-off region of a
  tokamak plasma. {\em J.~Phys.: Conf.~Ser.\/} {\bf 46}, 87.

\bibitem[Milanese {\em et~al.\/}(2017)Milanese, Schekochihin, Dorland \&
  Loureiro]{alexInverse}
{\sc Milanese, L, Schekochihin, A~A, Dorland, W \& Loureiro, N~F} 2017 Private
  communications.

\bibitem[Militello \& Fundamenski(2011)]{militello2011multimachine}
{\sc Militello, F \& Fundamenski, W} 2011 Multi-machine comparison of drift
  fluid dimensionless parameters. {\em Plasma\ Phys.\ Control.\ Fusion\/} {\bf
  53}, 095002.

\bibitem[Newcomb(1958)]{newcombLinesOfForce}
{\sc Newcomb, W.~A.} 1958 Motion of magnetic lines of force. {\em Ann. Phys.\/}
  {\bf 3}, 347.

\bibitem[Parra {\em et~al.\/}(2011)Parra, Barnes \& Peeters]{parra2011up}
{\sc Parra, F.I., Barnes, M. \& Peeters, A.G.} 2011 Up-down symmetry of the
  turbulent transport of toroidal angular momentum in tokamaks. {\em Phys.\
  Plasmas\/} {\bf 18}, 062501.

\bibitem[Parra \& Catto(2008)]{parra2008lgt}
{\sc Parra, F.~I. \& Catto, P.~J.} 2008 {Limitations of gyrokinetics on
  transport time scales}. {\em Plasma\ Phys.\ Control.\ Fusion\/} {\bf 50},
  065014.

\bibitem[Rasmussen {\em et~al.\/}(2016)Rasmussen, Nielsen, Madsen, Naulin \&
  Xu]{rasmussen2016lhtransition}
{\sc Rasmussen, J~Juul, Nielsen, A~H, Madsen, J, Naulin, V \& Xu, G~S} 2016
  Numerical modeling of the transition from low to high confinement in
  magnetically confined plasma. {\em Plasma\ Phys.\ Control.\ Fusion\/} {\bf
  58}, 014031.

\bibitem[Ricci {\em et~al.\/}(2012)Ricci, Halpern, Jolliet, Loizu, Mosetto,
  Fasoli, Furno \& Theiler]{ricci2012gbs}
{\sc Ricci, P, Halpern, F~D, Jolliet, S, Loizu, J, Mosetto, A, Fasoli, A,
  Furno, I \& Theiler, C} 2012 Simulation of plasma turbulence in scrape-off
  layer conditions: the {GBS} code, simulation results and code validation.
  {\em Plasma\ Phys.\ Control.\ Fusion\/} {\bf 54}, 124047.

\bibitem[Rogers \& Drake(1999)]{rogers1999diamagEdge}
{\sc Rogers, B.~N. \& Drake, J.~F.} 1999 Diamagnetic stabilization of ideal
  ballooning modes in the edge pedestal. {\em Phys.\ Plasmas\/} {\bf 6}, 2797.

\bibitem[Rosenbluth \& Taylor(1969)]{bryanStringerSpin}
{\sc Rosenbluth, M.~N. \& Taylor, J.~B.} 1969 Plasma diffusion and stability in
  toroidal systems. {\em Phys.\ Rev.\ Lett.\/} {\bf 23}, 367.

\bibitem[Saarelma {\em et~al.\/}(2013)Saarelma, Beurskens, Dickinson,
  Frassinetti, Leyland, Roach \& Contributors]{saarelma2013pedestalstab}
{\sc Saarelma, S., Beurskens, M.N.A., Dickinson, D., Frassinetti, L., Leyland,
  M.J., Roach, C.M. \& Contributors, EFDA-JET} 2013 {MHD} and gyro-kinetic
  stability of {JET} pedestals. {\em Nucl. Fusion\/} {\bf 53}, 123012.

\bibitem[Schneider {\em et~al.\/}(2006)Schneider, Bonnin, Borrass, Coster,
  Kastelewicz, Reiter, Rozhansky \& Braams]{solps2006}
{\sc Schneider, R., Bonnin, X., Borrass, K., Coster, D.~P., Kastelewicz, H.,
  Reiter, D., Rozhansky, V.~A. \& Braams, B.~J.} 2006 Plasma edge physics with
  {B2-Eirene}. {\em Contrib. Plasma Phys.\/} {\bf 46}, 3.

\bibitem[Scott(1997)]{scott2000dalf}
{\sc Scott, B.~D.} 1997 Three-dimensional computation of drift alfv{\'e}n
  turbulence. {\em Plasma\ Phys.\ Control.\ Fusion\/} {\bf 39}, 1635.

\bibitem[Scott(2007)]{scott2007edge}
{\sc Scott, B.~D.} 2007 Tokamak edge turbulence: background theory and
  computation. {\em Plasma\ Phys.\ Control.\ Fusion\/} {\bf 49}, S25.

\bibitem[Shi {\em et~al.\/}(2017)Shi, Hammett, Stoltzfus-Dueck \&
  Hakim]{shi2017gksol}
{\sc Shi, E.~L., Hammett, G.~W., Stoltzfus-Dueck, T. \& Hakim, A.} 2017
  Gyrokinetic continuum simulation of turbulence in a straight open-field-line
  plasma. {\em J.\ Plasma Phys.\/} {\bf 83}.

\bibitem[Simakov \& Catto(2003)]{cattosimakovedge}
{\sc Simakov, A.~N. \& Catto, P.~J.} 2003 Drift-ordered fluid equations for
  field-aligned modes in low-{$\beta$} collisional plasma with equilibrium
  pressure pedestals. {\em Phys.\ Plasmas\/} {\bf 10}, 4744.

\bibitem[Snipes {\em et~al.\/}(2001)Snipes, LaBombard, Greenwald, Hutchinson,
  Irby, Lin, Mazurenko \& Porkolab]{edahmode2001}
{\sc Snipes, J~A, LaBombard, B, Greenwald, M, Hutchinson, I~H, Irby, J, Lin, Y,
  Mazurenko, A \& Porkolab, M} 2001 The quasi-coherent signature of enhanced
  {$D\alpha$ H-mode in Alcator C-Mod}. {\em Plasma\ Phys.\ Control.\ Fusion\/}
  {\bf 43}, L23.

\bibitem[Snyder {\em et~al.\/}(2007)Snyder, Burrell, Wilson, Chu,
  Fenstermacher, Leonard, Moyer, Osborne, Umansky, West \&
  Xu]{snyder2007qhmode}
{\sc Snyder, P.B., Burrell, K.H., Wilson, H.R., Chu, M.S., Fenstermacher, M.E.,
  Leonard, A.W., Moyer, R.A., Osborne, T.H., Umansky, M., West, W.P. \& Xu,
  X.Q.} 2007 Stability and dynamics of the edge pedestal in the low
  collisionality regime: physics mechanisms for steady-state elm-free
  operation. {\em Nucl. Fusion\/} {\bf 47}, 961.

\bibitem[Snyder {\em et~al.\/}(2011)Snyder, Groebner, Hughes, Osborne,
  Beurskens, Leonard, Wilson \& Xu]{snyder2011eped}
{\sc Snyder, P.B., Groebner, R.J., Hughes, J.W., Osborne, T.H., Beurskens, M.,
  Leonard, A.W., Wilson, H.R. \& Xu, X.Q.} 2011 A first-principles predictive
  model of the pedestal height and width: development, testing and iter
  optimization with the eped model. {\em Nucl. Fusion\/} {\bf 51}, 103016.

\bibitem[Snyder {\em et~al.\/}(2002)Snyder, Wilson, Ferron, Lao, Leonard,
  Osborne, Turnbull, Mossessian, Murakami \& Xu]{snyder2002eped}
{\sc Snyder, P.~B., Wilson, H.~R., Ferron, J.~R., Lao, L.~L., Leonard, A.~W.,
  Osborne, T.~H., Turnbull, A.~D., Mossessian, D., Murakami, M. \& Xu, X.~Q.}
  2002 Edge localized modes and the pedestal: A model based on coupled
  {peeling--ballooning} modes. {\em Phys.\ Plasmas\/} {\bf 9}, 2037.

\bibitem[Stringer(1969)]{stringerSpinUp}
{\sc Stringer, T.~E.} 1969 Diffusion in toroidal plasmas with radial electric
  field. {\em Phys.\ Rev.\ Lett.\/} {\bf 22}, 770.

\bibitem[Sugama {\em et~al.\/}(1996)Sugama, Okamoto, Horton \&
  Wakatani]{sugama1996tpa}
{\sc Sugama, H., Okamoto, M., Horton, W. \& Wakatani, M.} 1996 {Transport
  processes and entropy production in toroidal plasmas with gyrokinetic
  electromagnetic turbulence}. {\em Phys.\ Plasmas\/} {\bf 3}, 2379.

\bibitem[Tang {\em et~al.\/}(1980)Tang, Connor \& Hastie]{tang1980kbm}
{\sc Tang, W.~M., Connor, J.~W. \& Hastie, R.~J.} 1980 Kinetic-ballooning-mode
  theory in general geometry. {\em Nucl. Fusion\/} {\bf 20}, 1439.

\bibitem[{The ASDEX Team}(1989)]{ASDEXHMode}
{\sc {The ASDEX Team}} 1989 The {H-Mode of ASDEX}. {\em Nucl. Fusion\/} {\bf
  29}, 1959.

\bibitem[Wagner {\em et~al.\/}(1982)Wagner, Becker, Behringer, Campbell,
  Eberhagen, Engelhardt, Fussmann, Gehre, Gernhardt, Gierke, Haas, Huang,
  Karger, Keilhacker, Kl\"uber, Kornherr, Lackner, Lisitano, Lister, Mayer,
  Meisel, M\"uller, Murmann, Niedermeyer, Poschenrieder, Rapp, R\"ohr,
  Schneider, Siller, Speth, St\"abler, Steuer, Venus, Vollmer \&
  Y\"u]{wagner1982hmode}
{\sc Wagner, F., Becker, G., Behringer, K., Campbell, D., Eberhagen, A.,
  Engelhardt, W., Fussmann, G., Gehre, O., Gernhardt, J., Gierke, G.~v., Haas,
  G., Huang, M., Karger, F., Keilhacker, M., Kl\"uber, O., Kornherr, M.,
  Lackner, K., Lisitano, G., Lister, G.~G., Mayer, H.~M., Meisel, D., M\"uller,
  E.~R., Murmann, H., Niedermeyer, H., Poschenrieder, W., Rapp, H., R\"ohr, H.,
  Schneider, F., Siller, G., Speth, E., St\"abler, A., Steuer, K.~H., Venus,
  G., Vollmer, O. \& Y\"u, Z.} 1982 Regime of improved confinement and high
  beta in neutral-beam-heated divertor discharges of the asdex tokamak. {\em
  Phys.\ Rev.\ Lett.\/} {\bf 49}, 1408.

\bibitem[Wilson {\em et~al.\/}(2002)Wilson, Snyder, Huysmans \&
  Miller]{wilson2002elite}
{\sc Wilson, H.~R., Snyder, P.~B., Huysmans, G. T.~A. \& Miller, R.~L.} 2002
  Numerical studies of edge localized instabilities in tokamaks. {\em Phys.\
  Plasmas\/} {\bf 9}, 1277.

\bibitem[Xu {\em et~al.\/}(2011)Xu, Dudson, Snyder, Umansky, Wilson \&
  Casper]{xu2011boutELM}
{\sc Xu, X.Q., Dudson, B.D., Snyder, P.B., Umansky, M.V., Wilson, H.R. \&
  Casper, T.} 2011 Nonlinear elm simulations based on a nonideal
  peeling–ballooning model using the {BOUT++} code. {\em Nuclear Fusion\/}
  {\bf 51}~(10), 103040.

\bibitem[Zeiler {\em et~al.\/}(1997)Zeiler, Drake \&
  Rogers]{zeiler1997driftbraginskii}
{\sc Zeiler, A., Drake, J.~F. \& Rogers, B.} 1997 Nonlinear reduced
  {Braginskii} equations with ion thermal dynamics in toroidal plasma. {\em
  Phys.\ Plasmas\/} {\bf 4}, 2134.

\bibitem[Zocco {\em et~al.\/}(2015)Zocco, Loureiro, Dickinson, Numata \&
  Roach]{zocco2015mtm}
{\sc Zocco, A, Loureiro, N~F, Dickinson, D, Numata, R \& Roach, C~M} 2015
  Kinetic microtearing modes and reconnecting modes in strongly magnetised slab
  plasmas. {\em Plasma\ Phys.\ Control.\ Fusion\/} {\bf 57}, 065008.

\bibitem[Zocco \& Schekochihin(2011)]{zoccoelectrons}
{\sc Zocco, A. \& Schekochihin, A.~A.} 2011 Reduced fluid-kinetic equations for
  low-frequency dynamics, magnetic reconnection and electron heating in
  low-beta plasmas. {\em Phys.\ Plasmas\/} {\bf 18}, 102309.

\bibitem[Zohm(1996)]{zohm1996elms}
{\sc Zohm, H} 1996 Edge localized modes ({ELM}s). {\em Plasma\ Phys.\ Control.\
  Fusion\/} {\bf 38}, 105.

\end{thebibliography}
\end{document}